\documentclass{article}
\usepackage[utf8]{inputenc}
\usepackage{float}
\usepackage{amsbsy}
\usepackage{amsmath}
\usepackage{amsthm}
\usepackage{amssymb}
\usepackage{epsfig}
\usepackage{multicol}
\usepackage{multirow}
\usepackage{adjustbox}
\usepackage{color}
\usepackage{bm}
\usepackage{bbm}
\usepackage{threeparttable}
\usepackage{subfigure}
\usepackage{rotating}
\usepackage{array}
\usepackage{verbatim}
\usepackage{fullpage}
\usepackage{setspace}
\usepackage{makecell}
\usepackage{hyperref}
\usepackage[ruled]{algorithm2e}
\usepackage{natbib}
\usepackage{graphicx}
\usepackage{booktabs}
\usepackage{authblk}
\usepackage{xr}
\usepackage{geometry}

\newtheorem{theorem}{Theorem}

\newcommand{\var}{\mbox{VAR}}

\def\tilde{\widetilde}

\def\n{\noindent}

\def\var{\mbox{var}}

\def\({\Big{(}}
\def\){\Big{)}}
\def\|{\Big{|}}
\def\tx{\boldsymbol{x}}

\def\bbeta{\boldsymbol{\beta}}
\def\bbetad{\boldsymbol{\beta}_{n,0}^\dagger}
\newcommand{\bg}{\begin{eqnarray}}
\newcommand{\ed}{\end{eqnarray}}
\newcommand{\bgn}{\begin{eqnarray*}}
\newcommand{\edn}{\end{eqnarray*}}

\title{Assessing Estimation Uncertainty under Model Misspecification}
\author{Rong Li$^1$, Yichen Qin$^2$, Yang Li$^{1*}$ \\
1: School of Statistics, Renmin University of China, Beijing, China. \\
2: Department of Operations, Business Analytics, and Information Systems, University of Cincinnati, Cincinnati, Ohio, USA.}
\date{}

\begin{document}

\maketitle

\begin{quotation}
\noindent {\it Abstract:}
Model misspecification is ubiquitous in data analysis because the data-generating process is often complex and mathematically intractable. 
Therefore, assessing estimation uncertainty and conducting statistical inference under a possibly misspecified working model is unavoidable. 
In such a case, classical methods such as bootstrap and asymptotic theory-based inference frequently fail since they rely heavily on the model assumptions. 
In this article, we provide a new bootstrap procedure, termed local residual bootstrap, to assess estimation uncertainty under model misspecification for generalized linear models.
By resampling the residuals from the neighboring observations, we can approximate the sampling distribution of the statistic of interest accurately.
Instead of relying on the score equations, the proposed method directly recreates the response variables so that we can easily conduct standard error estimation, confidence interval construction, hypothesis testing, and model evaluation and selection. 
It performs similarly to classical bootstrap when the model is correctly specified and provides a more accurate assessment of uncertainty under model misspecification, offering data analysts an easy way to guard against the impact of misspecified models.
We establish desirable theoretical properties, such as the bootstrap validity, for the proposed method using the surrogate residuals. Numerical results and real data analysis further demonstrate the superiority of the proposed method.

\vspace{9pt}
\par
\end{quotation}\par

\section{Introduction}\label{sec:intro}

Estimation uncertainty measurements such as standard errors and confidence intervals play important roles in statistical inference. To evaluate estimation uncertainty, we traditionally rely on closed-form solutions of the asymptotic sampling distribution or bootstrap. 
However, when the assumed model is misspecified, neither approach is ideal.
Meanwhile, model misspecification is ubiquitous in practice because the true data-generating process is unknown and complicated, and the assumed model often simplifies the reality.
Therefore, evaluating estimation uncertainty under model misspecification is a pressing task. 
In this article, we propose a new bootstrap method for standard error estimation, confidence interval construction, hypothesis testing, and model selection under model misspecification.

Suppose the data are generated by an unknown distribution $\mathbb{P}^*$. 
We usually assume $\mathbb{P}^*$ belongs to a parametric distribution family $\mathcal{P}= \left\{ \mathbb{P}_\theta: \theta \in \Theta \right\}$ with $\mathbb{P}^* = \mathbb{P}_{\theta_0}\in \mathcal{P}$. 
The target parameter $\theta_0$ can be consistently estimated by maximum likelihood estimation (MLE).
However, when the distribution family is misspecified and $\mathbb{P}^* \notin \mathcal{P}$, it results in a potentially biased probability limit of the estimator and a different asymptotic variance \citep{Lee2016Asymptotic}. 
The former is called the pseudo-true parameter $\theta^\dagger$, which can be understood as the projection of $\mathbb{P}^*$ onto $\mathcal{P}$. 
The latter is called the pseudo-true standard error, which measures estimation uncertainty under the misspecified model.
We focus on evaluating the uncertainty in estimating the pseudo-true parameter by providing its pseudo-true standard error estimate and confidence interval.

Model misspecification has been studied extensively in the literature. \cite{White1982Maximum} establishes the asymptotic properties of MLE for misspecified models under general settings.
For the misspecified generalized linear models (GLM), \cite{Fahrmexr1990Maximum} shows the asymptotic variance tends to be overestimated, yielding conservative confidence intervals and tests.
Although standard bootstrap estimates are available for
generalized linear models \citep{Moulton1991Bootstrapping, Friedl1997Variance, Claeskens2003A},
they often fail when the model is misspecified.
Targeting the pseudo-true standard error, \cite{Bose2003Generalized} and \cite{Chatterjee2005Generalized} show the consistency of weighted bootstrap in a general setting by working with the estimating equations. 
\citet{Spokoiny2015Bootstrap} show that multiplier bootstrap procedure can work well in constructing confidence set under mild model assumption violation.
In addition, \citet{Kline2012Higher} examines the higher order properties of the wild bootstrap \citep{Wu1986Jackknife, Liu1988Bootstrap, Mammen1993Bootstrap} in linear regressions with $\mathbb{E}(\epsilon|X) \neq 0$ under heteroscedasticity and other cases.
The aforementioned methods work well mostly under certain types of misspecification, 
whereas their performance highly depend on the type and the degree of misspecification.
Meanwhile, they often deal with the score equations and avoid recreating the bootstrapped response variable, limiting their applications in other tasks such as model selection and evaluation.

To bridge this gap, we propose a new bootstrap method termed local residual bootstrap (LRB) for generalized linear models under various types of model misspecification.
Our method resamples residuals locally under the neighborhood constraint instead of globally as in the classical residual bootstrap. 
Since the residuals carry the important information about the lack of fit, 
resampling residuals locally preserves this information as much as possible and recreates the data faithfully to the data-generating process. 
If the assumed model is correct, the residuals follow roughly identical distributions, and resampling locally and globally yields similar results. 
If the model is misspecified, the residuals are no longer exchangeable, thus resampling locally will show its advantages over resampling globally.

The idea of resampling locally is originally used in the time series analysis and nonparametric statistics. 
\citet{Paparoditis1999Local} approximate the distribution of the statistics of interest by locally resampling the periodogram ordinates \citep{Silva2006local}.
\citet{Arteche2009Bootstrap} locally resample residuals instead of the periodogram ordinates in the log-periodogram regression \citep{Arteche2017Strategy}. 
\cite{Shi1991Local} draws bootstrap samples with kernel weights locally in nonparametric regression \citep{Gozalo1997}. 
\cite{Gonccalves2004Maximum, Goncalves2005Bootstrap} establish the consistency of the moving block bootstrap in dynamic models and linear regressions.
In this article, local resampling is adopted for the first time in the generalized linear model setting.

In this article, we contribute to the literature in the following aspects. 
We provide a general approach to estimate the pseudo-true standard errors, to construct confidence intervals for the pseudo-true parameter, and to select models, all under model misspecification.
The proposed method can be paired with surrogate residuals \citep{Liu2018Residuals}, sign-based residuals \citep{Li2012A}, Pearson residuals, deviance residuals, and many others.
It is designed for generalized linear models and can be extended to linear models.
The desirable theoretical properties are established for commonly used generalized linear models and linear models with various types of residuals.

The proposed method can guard against different degrees of misspecification and various types of misspecification, such as missing covariates, misspecified model forms, mixed populations, and heteroscedasticity.
When the model is correctly specified, the proposed method performs approximately equally well as the parametric bootstrap.
If the model is mildly, moderately, or even seriously misspecified, the proposed method provides accurate assessment of the estimation uncertainty. 
To prevent severe model misspecification where the pseudo-true parameters are significantly distorted, the proposed method can evaluate the model performance and select the least misspecified model from a class of models with high accuracy.
Therefore, regardless of the severity and the types of model assumption violations, the proposed method serves as an insurance for inference under misspecification with a very small premium and significant protection.

To the best of our knowledge, this article is the first attempt in the literature to recreate/bootstrap response variable for generalized linear models under misspecification. 
In contrast, most existing bootstrap methods for generalized linear models, such as weighted bootstrap and one-step bootstrap, focus on estimating equations and bypass the response variables.
The advantage of recreating the response variables is that, as long as the statistic is a function of the data,
we can approximate its sampling distribution by plugging in the bootstrap data. 
Recreating the response variable is also valuable for model evaluation and selection, such as cross-validation and bootstrap model selection based on prediction accuracy.
In addition, the proposed method also focuses on fixed design which is often overlooked in the literature.

The rest of the paper is organized as follows. Section \ref{sec:prelim} introduces the framework, and Section \ref{sec:uncertainty} presents the local residual bootstrap, establishes its theoretical properties, such as the bootstrap validity, and discusses implementation details such as neighborhood selection. 
We adopt the proposed method in model selection in Section \ref{sec:selection}. 
Extensive simulations and a real data analysis are presented in Sections \ref{sec:simu} and \ref{sec:realdata}. 
Finally, we conclude in Section \ref{sec:conclusions}  and relegate the proofs and additional numerical results to the appendix.


\section{Preliminary}\label{sec:prelim}

\subsection{Framework}

Suppose we observe the sample $\{y_i, \textbf{x}_i\}_{i=1}^n$, where $y_i$ is the independent observation of the response variable $Y_i$ conditional on the uniformly bounded predictors $\textbf{x}_i=(x_{i1}, \dots, x_{ip})^T$. The response vector $\textbf{y}=(y_1, \dots, y_n)^T$ has a true unknown density function 
$g_n(\textbf{y}) = \prod_{i=1}^n g_{n,i}(y_i)$,
where $g_{n,i}(\cdot)$ depends on $\textbf{x}_i$. Let $\textbf{X}=(\textbf{x}_1, \dots, \textbf{x}_n)^T$ be an $n \times p$ fixed design matrix with each column standardized. Since $g_n$ is unknown, for modeling purposes, we choose a family of generalized linear models as our assumed model:
\[
f_n(\textbf{y},\boldsymbol{\beta})=\prod_{i=1}^n f(y_i, \theta_i)=\prod_{i=1}^n \exp \left\{ y_i\theta_i-b(\theta_i) + c(y_i) \right\},
\]
where $\theta_i$ is a function of $\textbf{x}_i^T\boldsymbol{\beta}$ and $b(\cdot)$ is a function of known form. Under the assumed model, we have $\mathbb{E}(Y_i|\theta)=\mu_i =h(\textbf{x}_i^T\boldsymbol{\beta})=b'(\theta_i)$ and $\mbox{Var}(Y_i|\theta)= V_i = V(\mu_i)=b''(\theta_i)$, and 
$h^{-1}$ is the link function.

Such a model family may not contain the true model $g_n(\textbf{y})$.
In that case, we can estimate the parameter using the quasi-maximum likelihood estimator (QMLE),
$\hat{\boldsymbol{\beta}}_n = \arg\max_{\boldsymbol{\beta} \in \mathbb{R}^p} \mathcal{L}_n(\textbf{y}, \boldsymbol{\beta})$,
where $\mathcal{L}_n(\textbf{y}, \boldsymbol{\beta}) = \sum_{i=1}^n \log f(y_i, \theta_i)$, and $\hat{\boldsymbol{\beta}}_n$ is a consistent estimator for $\boldsymbol{\beta}^\dagger_{n,0} = \arg\max_{\boldsymbol{\beta} \in \mathbb{R}^p} \mathbb{E}_g \mathcal{L}_n(\textbf{y}, \boldsymbol{\beta})$ \citep{White1982Maximum}.
In the literature, $\boldsymbol{\beta}^\dagger_{n,0}$ is often referred to as the pseudo-true parameter \citep{Lee2016Asymptotic, Komunjer2005Quasi} and $f_n(\textbf{y},\boldsymbol{\beta}^\dagger_{n,0})$ is the distribution in the assumed model family that has the smallest Kullback-Leibler distance to  $g_n(\textbf{y})$. 
If the model is correctly specified, that is, there exists a $\boldsymbol{\beta}_0$ such as $g_n(\textbf{y}) = f_n(\textbf{y}, \boldsymbol{\beta}_0) \in \left\{ f_n(\textbf{y}, \boldsymbol{\beta}): \boldsymbol{\beta} \in \mathbb{R}^p \right\}$, then we have 
$\boldsymbol{\beta}^\dagger_{n,0} = \boldsymbol{\beta}_0$.

In real data analysis, model misspecification is often unavoidable because the underlying process may be complicated and cannot be easily expressed in the mathematical form. Under model misspecification, a more realistic goal is to find the best approximation to the data-generating process, i.e., the pseudo-true parameter. 
More importantly, although the assumed model is only an approximation and simplification of the data-generating process, it is often the only vehicle to carry any statistical inference. 
In the sense that ``all models are wrong", most of inferences are essentially conducted for pseudo-true value.

Therefore, we are interested in the standard error of $\hat{\boldsymbol{\beta}}_n$, defined as $\psi = \sqrt{\mathbb{E}_{g}(\hat{\boldsymbol{\beta}}_n- \boldsymbol{\beta}^\dagger_{n,0})^2}$, 
where $\mathbb{E}_{g}$ is over the true distribution, since it measures the estimation uncertainty associated with $\hat{\boldsymbol{\beta}}_n$. In the literature, $\psi$ is referred to as the pseudo-true standard error. 
Our focus is to accurately estimate the pseudo-true standard error and conduct statistical inference such as confidence intervals for the pseudo-true parameter under model misspecification.

\subsection{Residuals as a Diagnostic Tool}

Residuals have long been used to detect model misspecification. Conceptually, we have residuals = data - fit. When the assumed model is correctly specified, the residuals are patternless and randomly distributed around zero. When the model is misspecified, the residuals carry important information about how the model assumptions are violated. Here, we introduce several types of residuals.

For generalized linear models, the Pearson residual is defined as $r_i=(y_i-\hat{\mu}_i)/V(\hat{\mu}_i)^{1/2}$, where $\hat{\mu}_i$ and $V(\hat{\mu}_i)$ are the estimated mean and variance of $Y_i$ under the assumed model.
The deviance residual is defined as $r_i = \mbox{sign}(y_i -\hat{\mu}_i)\sqrt{2[y_i \log(y_i/\hat{\mu}_i)-(y_i -\hat{\mu}_i)]}$.
The sign-based statistic (SBS) residual \citep{Li2012A} is defined as $r_i =\hat{P}(Y_i<y_i|\textbf{x}_i)-\hat{P}(Y_i>y_i|\textbf{x}_i)$ where $\hat{P}(\cdot)$ is the estimated probability according to the model.
The surrogate residual \citep{Liu2018Residuals}, which is primarily designed for binary data or ordinal data within the class of generalized linear models,
is defined as $r_i = s_i - \textbf{x}_i^T \hat{\boldsymbol{\beta}}_n$. Here, $s_i$ is considered as a ``surrogate'' of $y_i$ and follows a truncated distribution: $s_i \sim z_i|\alpha_{j-1} \leq z_i<\alpha_j$, if  $y_i=j$ for $j=0,1,...,\infty$ where $z_i$ is a latent variable follows the rules that if $\alpha_{j-1} \leq z_i < \alpha_j$ then $y_i=j$. 
The surrogate residual follows a standard normal distribution when the model is correctly specified and deviates from that when the model is misspecified.
For linear regression, the residual is $r_i = y_i - \textbf{x}_i^T \hat{\boldsymbol{\beta}}_n$. 
When the model is correctly specified, the residuals are scattered around zero following a normal distribution.

To introduce our main approach, we use probit models and surrogate residuals as a running example, but our results apply to other models and residuals too.


\section{Estimation Uncertainty under Misspecification}\label{sec:uncertainty}

\subsection{Local Residual Bootstrap}

In this section, we introduce our main method, local residual bootstrap, to assess estimation uncertainty under model misspecification. 
Here we assume $\textbf{X}$ is nonrandom, in which case the standard resample method is the residual bootstrap \citep{Freedman1981Bootstrapping}.
Our motivation is that since the residuals contain important information about model misspecification, we bootstrap the residual locally within a neighborhood constraint such that the conditional distribution of the residuals given the covariates is preserved. Specifically, given data $\{y_i, \textbf{x}_i \}_{i=1}^n$, we can obtain the estimate $\hat{\boldsymbol{\beta}}_n$ and residuals $\left\{ r_i \right\}_{i=1}^n$ under the assumed model. Let $d_{ij} = \lVert \textbf{x}_{i}-\textbf{x}_{j} \rVert$ be the covariate Euclidean distance between the $i$-th and $j$-th observations. Define the neighborhood index set of the $i$-th observation as $N_i = \left\{ j: d_{ij} \mbox{ are among the smallest } l \mbox{ distances of } \{d_{ik}\}_{k=1}^n \right\}$, where $l$ is referred to as the neighborhood size. For the $i$-th observation, we first obtain its bootstrapped residual $r_i^*$ by sampling from the neighborhood residual set $\left\{ r_j: j \in N_i \right\}$, then generate the bootstrapped response $y_i^*$ based on $r_i^*$ and $\hat{\boldsymbol{\beta}}_n$, and finally obtain the bootstrap estimate based on the bootstrap sample $\{ y_i^*, \textbf{x}_i \}_{i=1}^n$. Using the bootstrap estimate, we can assess estimation uncertainty by calculating the pseudo-true standard error and confidence interval for the pseudo-true parameter. The proposed procedure is summarized in Algorithm \ref{algo:LRB}.

\begin{spacing}{2.0}
\begin{algorithm}[!t]
\caption{Local residual bootstrap procedure}\label{algo:LRB}
\LinesNumbered
\KwIn{data $\{ y_i, \textbf{x}_i \}_{i=1}^n$}
\KwOut{pseudo-true standard error estimate and confidence interval}
Obtain the parameter estimate $\hat{\boldsymbol{\beta}}_n$ under the assumed model and its residuals $r_i$ for $i=1, \dots, n.$ \\
Obtain the covariate distance matrix ${\bf D}= [d_{ij}]_{n \times n}$, where $d_{ij} = \lVert \textbf{x}_{i}-\textbf{x}_{j} \rVert$.

\For{$b=1,\dots,B$}{

  \For{$i = 1, \dots, n$}{
   Generate the bootstrapped residual $r_i^*$ by sampling from $\left\{ r_j: j \in N_i \right\}$, where $N_i=\left\{ j: d_{ij} \mbox{ are among the smallest } l \mbox{ distances of } \{d_{ik}\}_{k=1}^n \right\}$; \\
    Generate the bootstrapped response $y_i^*$ according to $r_i^*$, ${\bf x}_i$ and $\hat{\boldsymbol{\beta}}_n$. 
    }
    Obtain the bootstrap estimate $\hat{\boldsymbol{\beta}}_n^{*(b)}$ based on the bootstrap sample $\{ y_i^*, \textbf{x}_i \}_{i=1}^n$.
    }
   
   The pseudo-true standard error estimate is $\hat{\psi}=\sqrt{\sum_{b=1}^B (\hat{\boldsymbol{\beta}}_n^{*(b)}- \overline{\widehat{\boldsymbol{\beta}}_n^*})^2/B}$, where $\overline{\widehat{\boldsymbol{\beta}}_n^*} = \sum_{b=1}^B \hat{\boldsymbol{\beta}}_n^{*(b)}/B$. 
   
   The confidence interval for the pseudo-true parameter is given according to bootstrap estimate.
\end{algorithm}
\end{spacing}

Algorithm \ref{algo:LRB} can accommodate various classes of models such as generalized linear models and linear regressions by letting the generating process of $y_i^*$ in Line 6 take different forms. For example, for probit models with surrogate residuals, $y_i^*$ is generated by $y_i^*=0$ if $s_i^* \leq 0$ and $y_i^*=1$ otherwise, where $s_i^*=r_i^*+\textbf{x}_i^T\hat{\boldsymbol{\beta}}_n$. For probit models with Pearson residuals, we let $y_i^*=\hat{\mu}_i+V(\hat{\mu}_i)^{1/2}r_i^*$. For probit models with SBS residuals, $y_i^*$ is generated by $y_i^*=0$ if $r_i^* \leq 0$ and $y_i^*=1$ otherwise. Meanwhile, for linear regressions with regular residuals, we let $y_i^*=\textbf{x}_i^T\hat{\boldsymbol{\beta}}_n + r_i^*$. 
The algorithm is applicable to various models as long as the residuals are well defined. 

Local residual bootstrap also includes the classical residual bootstrap as a special case by setting $N_i=\{1,...,n\}$ in Line 5, in which case it samples globally from all the residuals as opposed to a subset. 
Alternatively, we could resample the residuals in the neighborhood with different weights, e.g.,  larger weight for $r_j$ when $\textbf{x}_j$ is closer to $\textbf{x}_i$, because the residual distributions of the neighboring observations should be more similar. 

The main idea of local residual bootstrap is illustrated as follows:
\[
\begin{array}{ccccc}
\mbox{Original data} &y_i  \stackrel{\mbox{decompose}}{\longrightarrow} & \textbf{x}_i^T \hat{\boldsymbol{\beta}}_n & + & r_i  \\  
& & \begin{sideways}  =  \end{sideways}  & & \stackrel{\mbox{locally resample}}{\downarrow}  \\ 
\mbox{Bootstrap data} &y_i^*  \stackrel{\mbox{recreate}}{\longleftarrow} & \textbf{x}_i^T\hat{\boldsymbol{\beta}}_n  & + & r_i^*|\textbf{x}_i \end{array}
\]

After fitting the assumed model to the data, the original response variable can be conceptually decomposed into two parts: the part that can be explained by the fitted model, i.e., model fit, and the part that cannot, i.e., residuals. 
When the model is misspecified, the distributions of these residuals usually indicate how the data deviate from the model assumptions. For example, any patterns, gaps, and non-linearity in the residuals may indicate the assumed model fails to fit the data in these regards. By resampling these residuals locally, we preserve their distributions conditional on the covariates and ensure the bootstrapped residuals have approximately the same conditional distributions. 
We estimate such a conditional distribution empirically by nearest neighbors, and other methods are also available, such as kernel-based approach.
Lastly, combining the model fit with the bootstrapped residuals, we can recreate the bootstrapped response that shares the same probabilistic behaviors as the original response. In addition, the bootstrapped estimate $\hat{\boldsymbol{\beta}}_n^*$ can accurately approximate the true sampling distribution of $\hat{\boldsymbol{\beta}}_n$. By contrast, the classical residual bootstrap resamples all the residuals uniformly and hence erases the trend/pattern in them. The bootstrapped residual in this case only preserves the marginal distribution of the original residuals but fails to retain the conditional distribution given the covariates.

Under the surrogate residuals, when the model is correctly specified, all the residuals follow the same distribution.
Therefore, local residual bootstrap yields the same results as in the classical residual bootstrap, and is asymptotically equivalent to the parametric bootstrap.

We present an example to illustrate our proposed method below.

\begin{figure}[!t]
  \centering
  \includegraphics[width=0.8\textwidth, angle=0]{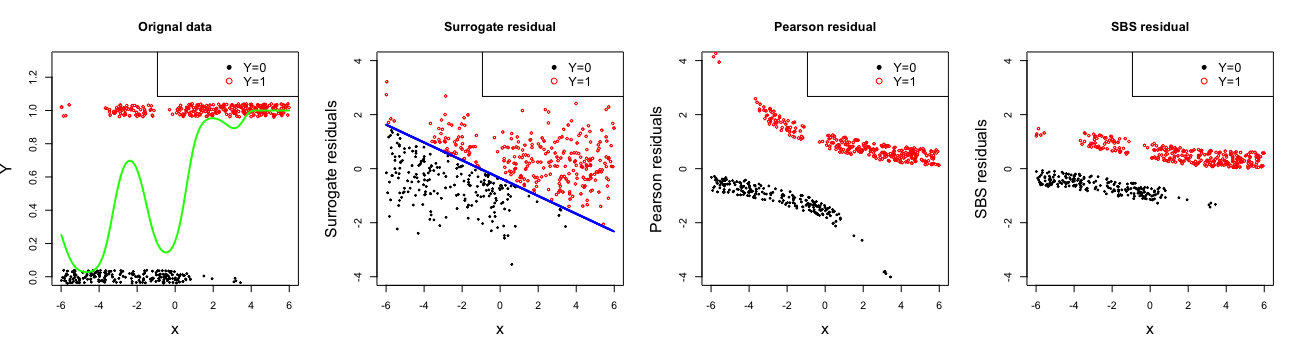} \\
  \includegraphics[width=0.8\textwidth, angle=0]{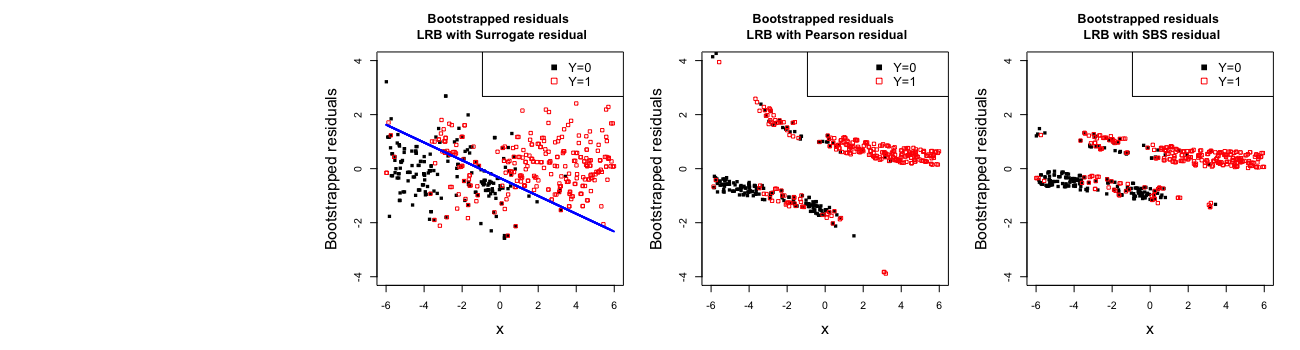} \\
  \includegraphics[width=0.8\textwidth, angle=0]{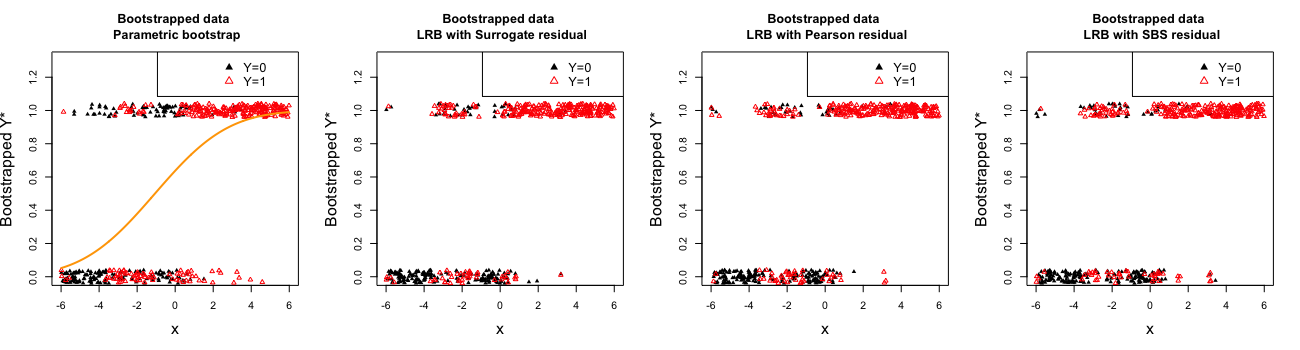} \\
  \caption{Illustration of Example 1. In the top row, the original data are in the left panel, while the surrogate, Pearson, and SBS residuals are in the right panels. The blue line is $-(\hat{\beta}_0 + \hat{\beta}_1 x)$. In the middle row, the bootstrapped residuals are shown. In the bottom row, the bootstrapped responses by the parametric bootstrap are in the left panel and the bootstrapped responses by local residual bootstrap are in the right panels. The orange line indicates the fitted probit model.
  The responses and residuals are jittered and colored according to their original observations' labels, with black indicating $y_i = 0$ and red $y_i =1$.
  The surrogate residuals are lower bounded by $-(\hat{\beta}_0 + \hat{\beta}_1 x)$ (in blue) when $y_i=1$ and upper bounded when $y_i=0$.
  }\label{Fig:resid_case1}
\end{figure}

{\bf Example 1:} For the data $\{ y_i, x_i \}_{i=1}^{500}$ presented in the upper left panel of Figure \ref{Fig:resid_case1}, we fit a probit model $\Phi^{-1} ( P(Y_i=1) ) = \beta_0 + \beta_1 x_i$.
This is clearly misspecified because the synthetic data is generated from a complex nonlinear model with an irregular curve. 
It deviates from a monotone smooth ``S" shaped curve from an assumed univariate probit model.
We compare local residual bootstrap and the parametric bootstrap on this data in Figure \ref{Fig:resid_case1}.
The first row of the figure presents the (jittered) response variable and the surrogate/Pearson/SBS residuals.
Even if we don't know the true data generating process, the misspecification can clearly identified from residuals.
The surrogate residuals are not evenly distributed, and the Pearson residuals and SBS residuals are not symmetrical. There are gaps in the distribution of residuals, especially in surrogate residuals. All of these patterns in the residuals can suggest the model misspecification.
The second row presents the bootstrapped residuals from local residual bootstrap ($l=4$), which closely mimic the original residuals. 
The third row presents the bootstrapped response by the parametric bootstrap and the local residual bootstrap. 
As we can see, the bootstrapped response by our method effectively retains the patterns in the original responses. 
In addition, the red and black dots are mixed up around the boundaries of these gaps, meaning these observations' labels are flipped by the bootstrap. 
On the contrary, the parametric bootstrap fails to preserve these patterns. 

In addition to parametric bootstrap, another common alternative is nonparametric bootstrap, which is suitable for misspecification when the predictors are assumed to be random. 
But in the fix design as we consider in this article, it introduces additional randomness for predictors and overestimates the uncertainty.
For example, it may worsen the unbalanced nature of the data and cannot produce new observations, which is prone to error when dealing with the $(p+1)$-dimensional joint distribution. 
By constrast, the proposed bootstrap empirically estimates univariate distribution of residuals. 
We also consider a hybrid of our local residual bootstrap and nonparametic bootstrap by resampling the responses instead of the residuals under the neighborhood constraint, and call it local response bootstrap. 
It is a variation of nonparametric bootstrap in the fixed design, utilizing the relationship between the predictors and the response without imposing any structure of model.
However, it is inferior to local residual bootstrap according to our numerical results, and the detailed explanation can be found in Section \ref{sec:localresponse} and Section \ref{sec:add_simu} in the appendix. 

\subsection{Theoretical Properties}

In this section, we present the theoretical results of the proposed method. 
Let $\textbf{A}_n(\boldsymbol{\beta}) = -\mathbb{E}_g(\partial^2 \mathcal{L}_n/\partial\boldsymbol{\beta} \partial \boldsymbol{\beta}^T)$ and $\textbf{B}_n(\boldsymbol{\beta}) = \mathbb{E}_g (\partial \mathcal{L}_n/\partial\boldsymbol{\beta} (\partial \mathcal{L}_n/ \partial\boldsymbol{\beta})^T)$, and let $\tilde{\textbf{B}}_n=\textbf{B}_n(\boldsymbol{\beta}_{n,0}^\dagger)$ and $\tilde{\textbf{A}}_n=\textbf{A}_n(\boldsymbol{\beta}_{n,0}^\dagger)$. 
Denote $\lambda_{\min}(\cdot)$ and $\lambda_{\max}(\cdot)$ as the smallest and largest eigenvalues of a matrix and $\Vert \cdot \Vert$ as the Euclidean norm. 
Following \citet{Lv2014Model}, we have the conditions below.

\begin{enumerate}
\setlength{\itemsep}{0.4ex}

\item[(C1)]
$b(\theta)$ is twice differentiable and  $b'(\theta)$ is monotonic. $\textbf{X}$ has full column rank.

\item[(C2)]
$\lambda_{\min}(\tilde{\textbf{B}}_n)\rightarrow \infty$ as $n \rightarrow \infty$. There is some positive constant $c>0$,  such that
$\min_{\boldsymbol{\beta} \in H_n(\delta)} \lambda_{\min}\{ \textbf{T}_n(\boldsymbol{\beta})\} \ge c$ for any $\delta>0$ and all sufficiently large $n$, where $\textbf{T}_n(\boldsymbol{\beta})=\tilde{\textbf{B}}_n^{-1/2} \textbf{A}_n(\boldsymbol{\beta})\tilde{\textbf{B}}_n^{-1/2}$, and $H_n(\delta)$ is the shrinking neighborhood
$H_n(\delta) = \left\{ \boldsymbol{\beta} \in \mathbb{R}^p: \Vert \tilde{\textbf{B}}_n^{1/2}(\boldsymbol{\beta}-\boldsymbol{\beta}_{n,0}^\dagger) \Vert \leq \delta \right\}$.

\item[(C3)]
For any $\delta >0$, $\max_{\boldsymbol{\beta} \in H_n(\delta)} \Vert \textbf{T}_n(\boldsymbol{\beta}) - \tilde{\textbf{T}}_n \Vert = o(1)$, where $\tilde{\textbf{T}}_n = \textbf{T}_n(\boldsymbol{\beta}_{n,0}^\dagger)$.

\item[(C4)]
$\max_{1 \leq i \leq n} \mathbb{E}|Y_i-\mathbb{E}(Y_i)|^3=O(1)$ and $\sum_{i=1}^n (\partial \theta_i/\partial \boldsymbol{\beta}^T \tilde{\textbf{B}}_n^{-1} \partial \theta_i/\partial \boldsymbol{\beta})^{3/2}=o(1)$.
\end{enumerate}

Condition C1 is the regularity assumption, which implies the strict concavity of $\mathcal{L}_n$ and ensures the uniqueness of the QMLE $\hat{\boldsymbol{\beta}}_n$ as well as the pseudo-true parameter $\boldsymbol{\beta}^\dagger_{n,0}$. 
C2 guarantees the consistency of $\hat{\boldsymbol{\beta}}_n$, where $\lambda_{\min}(\tilde{\textbf{B}}_n)\rightarrow \infty$ is the divergence condition, which is indispensable for asymptotic inference.
When the variance of the response is bounded away from 0 and $\infty$ and the link function is the natural link, this condition degenerates into the design matrix $\lambda_{\min}(\textbf{X}^T\textbf{X}) \rightarrow \infty$ as $n \rightarrow \infty$.
The bounded eigenvalue of $\textbf{T}_n(\boldsymbol{\beta})$ concerns the relationship between the true data-generating model and assumed model family.
With the natural link, $\tilde{\textbf{A}}_n$ is the covariance matrix of $\textbf{X}^T\textbf{Y}$ under the `best' model among the assumed parametric family, while $\tilde{\textbf{B}}_n$ is the covariance matrix under the true model. 
If the assumed model is correctly specified, $\tilde{\textbf{B}}_n=\tilde{\textbf{A}}_n$ and $\tilde{\textbf{T}}_n=\textbf{I}_p$.
C3 restricts the continuity of $\textbf{T}_n(\boldsymbol{\beta})$ in terms of $\boldsymbol{\beta}$ in the shrinking neighborhood of the pseudo-true parameter.
This function basically represents the divergence between the covariance structures given by the assumed misspecified generalized linear models and the covariance structure under the true model.
C4 is a typical moment condition for establishing asymptotic normality. When the link function is the natural link, $\sum_{i=1}^n (\partial \theta_i /\partial \boldsymbol{\beta}^T \tilde{\textbf{B}}_n^{-1} \partial \theta_i/\partial \boldsymbol{\beta})^{3/2}$ amounts to  $\sum_{i=1}^n (\textbf{x}_i^T \tilde{\textbf{B}}_n^{-1} \textbf{x}_i)^{3/2}$ in C4. 
Condition C3 and C4 are required for the asymptotic normality of $\hat{\boldsymbol{\beta}}_n$ with model misspecification,
and are classical conditions commonly used in the literature, as shown by \citet{Fahrmexr1990Maximum} and \citet{Lv2014Model}. We now present the bootstrap validity for the proposed method.

\begin{theorem}\label{Theorem1}
Under C1 and C2, using local residual bootstrap with either surrogate or Pearson residuals, suppose $l_n \rightarrow \infty$ and $ l_n/n \rightarrow 0$ as $n \rightarrow \infty$, then the distribution of bootstrapped residual $r_i^*$ converges weakly to the distribution of the surrogate or Pearson residual under the pseudo-true model $r^\dagger_i$.
\end{theorem}

Theorem \ref{Theorem1} implies that when the neighborhood size $l_n$ is selected properly, our bootstrapped residuals behave similarly to the surrogate or Pearson residuals under the pseudo-true model, which reflect how the pseudo-true model deviates from the data-generating process.
Here, a properly selected $l_n$ means that as $n \to \infty$, the neighborhood contains a sufficient number of observations to sample from and it is tight enough that the distributions of residuals in the neighborhood are similar.
Asymptotically, it implies that every observation has approximately $l_n$ i.i.d. copies in the original sample, and this number $l_n$ goes to infinity slower than the total sample size $n$. 
This guarantees the Euclidean distance between the observations in the neighborhood is neglectable under the large sample size. 
Such a condition has been commonly used in the literature, such as \citet{Biau2010Rate, Shi1991Local, Biau2010Layered}.

Because the bootstrapped response is generated according to the bootstrapped residual, Theorem \ref{Theorem1} guarantees that the distribution of $y_i^*$ can approximate the distribution of the original $y_i$. Based on the well-approximated distribution of the bootstrapped response, the bootstrapped sampling distribution of the pseudo-true estimation can approximate the true sampling distribution of $\hat{\boldsymbol{\beta}}_n$, as illustrated in Theorem \ref{Theorem2}.

\begin{theorem}\label{Theorem2}
Under C1--C4, using local residual bootstrap with surrogate or Pearson residuals, suppose $l_n \rightarrow \infty$ and $ l_n/n \rightarrow 0$ as $n \rightarrow \infty$, 
then the conditional distribution of 
$\sqrt{n}(\hat{\boldsymbol{\beta}}^*_n - \hat{\boldsymbol{\beta}}_n)$ 
converges weakly to the same limit distribution as
$\sqrt{n}(\hat{\boldsymbol{\beta}}_n - \boldsymbol{\beta}_{n,0}^\dagger)$.

\end{theorem}

Theorem \ref{Theorem2} indicates that the bootstrapped distribution converges to the true distribution of the pseudo-true estimator under model misspecification. This is the foundation of standard error estimation, confidence interval construction, and hypothesis testing for the pseudo-true parameter under model misspecification.

\begin{theorem}\label{Theorem-v}
Under the same conditions in Theorem \ref{Theorem2}, suppose $l_n \rightarrow \infty$ and $ l_n/n \rightarrow 0$ as $n \rightarrow \infty$, then 
${\rm Var}^*(\sqrt{n}(\hat{\boldsymbol{\beta}}^*_n-\hat{\boldsymbol{\beta}}_n))$ converges to the asymptotic variance of $\sqrt{n}(\hat{\boldsymbol{\beta}}_n - \boldsymbol{\beta}_{n,0}^\dagger)$ in probability.
\end{theorem}

We let $\mbox{Var}^*(\cdot)$ denote the variance with respect to the bootstrap distribution conditional on the original sample. Theorem \ref{Theorem-v} shows that the bootstrap variance estimator is consistent to the asymptotic variance of the pseudo-true estimator.

Note that the above theorems are established for a wide range of generalized linear models, which includes the probit
model, logistic model, Poisson model, ordinal model, gamma model, inverse Gaussian model 
where either the surrogate or Pearson residuals are well defined.
The proofs for the general cases are provided in the appendix.
In practice, we suggest using the surrogate residual if it is available, such as binary/ordinal regressions, because of its desirable properties. 
Other residuals, such as Pearson residual, quantile residuals \citep{Scudilio2020Adjusted}, and functional residuals \citep{Lin2022Model} can be applied similarly when the surrogate residuals are not available.

\subsection{Neighborhood Construction}\label{sec:neighborhood}

In this section, we discuss the selection of neighborhood size and the identification of the neighbors.
First, the neighborhood size plays an important role since it governs the amount of the information we borrow from the neighboring observations. If the neighborhood size is too large, the distributions of the residuals within the neighborhood become too different. If it is too small, there are insufficient residuals with which to sample. Both extremes cause the bootstrapped residuals to be inconsistent with the original residuals in distribution. In the extreme case of $l=1$, the bootstrapped residual is the same as in the original residual and the bootstrapped response is equal to the original response. When $l=n$, local residual bootstrap becomes the classical residual bootstrap; hence, it ignores the information about model misspecification hidden in the residuals. The following example is provided to illustrate the effect of the neighborhood size.

\begin{figure}[!h]
  \centering
 \includegraphics[width=0.4\textwidth, angle=0]
 {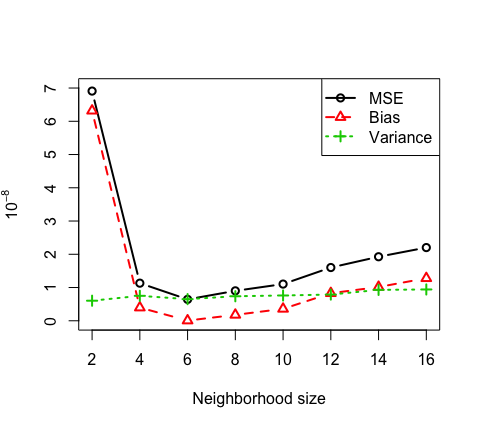}
 \includegraphics[width=0.4\textwidth, angle=0]
 {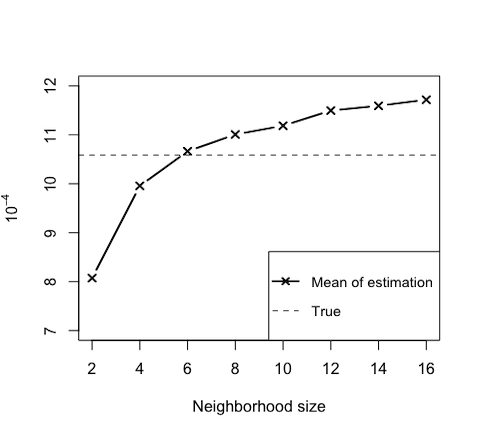}
 \caption{Left panel: the performance of standard error estimation by local residual bootstrap at different neighborhood sizes. The black solid line is the MSE, $\mathbb{E}(\hat{\psi} - \psi)^2$. The red dashed line is the bias, $(\mathbb{E}(\hat{\psi})-\psi)^2$. The green dotted line is the variance, $\mathbb{E}(\hat{\psi}-\mathbb{E}(\hat{\psi}))^2$. The mean of standard error estimation (50 replications). Right: the mean of the estimation of the pseudo-true standard error at different neighborhood sizes. The dotted horizontal line is the pseudo-true standard error.} 
\label{Fig:delta_ratio}
\end{figure}

We adopt the setting SC1 in Section \ref{sec:simu} in main text. Figure \ref{Fig:delta_ratio} plots the mean squared error (MSE), bias, and variance of the estimation of the pseudo-true standard error as the function of the neighborhood size. As we can see, as the neighborhood size increases, the estimation variance tends to increase slowly and the bias decreases first and then increases, with the MSE following the same pattern as the latter. As shown in the right panel of Figure \ref{Fig:delta_ratio}, the expectation of the pseudo-true standard error estimate increases with the neighborhood size, approaches the true value first, and then moves away. The MSE is minimized when the neighborhood size is around 6, which is the optimal size. The forgiven region of the optimal size is wide since the MSE is relatively low as long as the neighborhood size is not too far from the optimal.

Following the intuition of \cite{Hall1995On} and \cite{MacKinnon2006Bootstrap}, we select the neighborhood size to minimize mean squared error while balancing variance and bias. 
According to the calculation from Section 2.3 in \citet{Hall1995On}, the mean squared error of the estimated standard error by block bootstrapping is given by $\mathbb{E}(\hat{\psi}-\psi) \sim n^{-2}(C_1l^{-2}+C_2n^{-1}l)$, where $C_1$, $C_2$ are positive constants. The first component derives from the squared bias, while the second from the variance. 
Therefore, the asymptotically optimal neighborhood size is obtained by minimizing $\mathbb{E}(\hat{\psi}-\psi)$ over $l$. 
By setting the first derivative to zero, the optimal value achieves at $l=C_0n^{1/3}$ where $C_0=(2C_1/C_2)^{1/3}$.
Since $C_0$ is unknown in practice, we propose an iterative procedure for selecting the neighborhood size \citep{Hall1996On, Paparoditis2010Residual}.
Given the initial neighborhood size $\hat{l}_n^{(0)}$, we compare the standard error estimate $\hat{\psi}$ from the entire data set to the estimate $\hat{\psi}_{kq}, ~ k=1,\dots,K, q=1,\dots,Q$ from $K$ subsamples of size $m$ at various neighborhood sizes $l^\prime_q$s, and pick the optimal neighborhood size $\hat{l}_m$. We iterate the procedure with updated $\hat{l}_n^{(1)}=(n/m)^{1/3}\hat{l}_m$. 
The detailed process is summarized in Algorithm \ref{Algorithm2}.

\begin{spacing}{2.0}
\begin{algorithm}[!t]
	\caption{Neighborhood size selection}	\label{Algorithm2}
	\LinesNumbered
	\KwIn{data $\{ y_i, \textbf{x}_i \}_{i=1}^n$ and a grid of neighborhood sizes $\{l^\prime_1, \dots, l^\prime_Q\}$}
	\KwOut{the selected neighborhood size $\hat{l}_n$}
	\For{$k=1,\dots, K$}{

		Obtain the subsample $\{ y_i^{(k)}, \textbf{x}_i^{(k)} \}_{i=1}^m$ by resampling $m$ observations without replacement from the original data.\\
		\For{$q= 1, \dots, Q$}{
			Using the neighborhood size $l^\prime_q$, obtain the standard error estimate $\hat{\psi}_{kq}$ on the subsample $\{ y_i^{(k)}, \textbf{x}_i^{(k)} \}_{i=1}^m$.
		}
	}
	
	Initialize $t=0$, $\Delta > \delta > 0$, and the neighborhood size for the entire data set $\hat{l}_n^{(0)}=\lceil n^{1/3} \rceil$. \\
	
	\While{$\Delta > \delta$}{
		Using the neighborhood size $\hat{l}_n^{(t)}$, obtain the standard error estimate $\hat{\psi}^{(t)}$ based on the entire data set $\{ y_i, \textbf{x}_i \}_{i=1}^n$. \\
		Choose the optimal subsample neighborhood size $\hat{l}_m^{(t+1)}=l^\prime_{q^*}$, where $q^*=\arg\min_{1 \leq q \leq Q} \sum_{k=1}^K (\hat{\psi}_{kq} - \hat{\psi}^{(t)})^2/K$. \\
		Obtain the neighborhood size of the entire data set $\hat{l}_n^{(t+1)}=(n/m)^{1/3}\hat{l}_m^{(t+1)}$. \\
		Let $\Delta = |\hat{l}_n^{(t+1)} - \hat{l}_n^{(t)}|$ and $t=t+1$.
	}
	\Return the selected neighborhood size $\hat{l}_n=\hat{l}_n^{(t)}$.
\end{algorithm}
\end{spacing}

We also include an example illustrating the usage of the algorithm. We simulate a data set of $n=2000$ under setting SC1 and apply this algorithm. Table \ref{Neigh_example} details the selection process. We take $\hat{l}_n^{(0)}=[n^{1/3}]=13$ as the initial neighborhood size. The first column shows the candidate neighborhood sizes for each subsample ($K=20$), which has $m=0.9n=1800$ observations sampled from the entire data set without replacement. The second column is the MSE of the estimation ($\times 10^{-7}$) in the subsample at various neighborhood sizes $l^\prime_q, q=1,\dots,8$. The minimum of those values is $\hat{l}_m^{(1)}=4$, which suggests taking $\hat{l}_n^{(1)}=(1/0.9)^{1/3}\hat{l}_m^{(1)}=4.14 \simeq 4$. Using the updated $\hat{l}_n^{(1)}$, the MSE ($\times 10^{-7}$) is at its minimum at $\hat{l}_m^{(1)}=4$ in the third column. The process converges after two iterations, allowing us to select the neighborhood size as $\hat{l}_n=4$.

\begin{table}[!htp]
	\caption{\label{Neigh_example} MSE for various neighborhood sizes. The first column is the candidate neighborhood sizes, the second column is the MSE ($\times 10^{-7}$) based on the initial neighborhood size $\hat{l}_n^{(0)}$, and the third column is the MSE ($\times 10^{-7}$) based on the updated neighborhood size from the next iteration. } 
	\centering
	\begin{tabular}{c c c} 
		\toprule
		&$\hat{l}_n^{(0)}=13$ & $\hat{l}_n^{(1)}=4$ \\
		\midrule
		$l^\prime_1=2$ & 14.393 & 16.313 \\
		$l^\prime_2=4$ & 0.871*  & 0.694* \\
		$l^\prime_3=6$ & 2.643  & 2.053\\
		$l^\prime_4=8$ & 4.179  & 3.364\\
		$l^\prime_5=10$ & 3.379 & 2.709\\
		$l^\prime_6=12$ & 3.545 & 2.844\\
		$l^\prime_7=14$ & 4.656 & 3.844\\
		$l^\prime_8=16$ & 5.342 & 4.387 \\
		\bottomrule
	\end{tabular}
\end{table}

In addition to selecting the neighborhood size, identifying the neighbors is also an important task. 
Besides the Euclidean distance, other distances can also be used, such as Mahalanobis distance, kernel weighted distance and $L_k (k\leq 1)$ distance. Meanwhile, we can also consider clustering observations and resample from the same cluster.
It is also natural to define the neighborhood according to $\textbf{X}\hat{\boldsymbol{\beta}}_n$, which can be convenient for large $p$. When there are categorical predictors, the neighborhood set can be extended in a similar fashion, for example, $N_i = \left\{ j: x_{ik}=x_{jk}, \forall k \right\}$. If there are both continuous and categorical predictors, the neighborhood set can be the intersection of the neighborhood sets by the continuous and categorical predictors. All these methods are viable choices under various settings.
Note that in this article we have restricted to equal size neighborhood, we can certainly allow varying neighborhood sizes by thresholding the distance matrix.



\section{Bootstrap Model Selection under Misspecification}\label{sec:selection}

The previous section presents a solution to the uncertainty estimation under mildly or moderately misspecified models.
Under severely misspecified models, the pseudo-true parameters are hard to interpret.
Therefore, in this section, we present a bootstrap model selection procedure based on the proposed method to identify the least misspecified model from a class of models and to prevent severe misspecification.

Given a data set, although multiple models are available for conducting inference, we usually look for the true/optimal model. 
However, since the data-generating process is complex, all candidate models are essentially approximations to the true model with varying degrees of misspecification.
Therefore, more realistically, we are instead interested in finding the least misspecified model, under which we can assess the estimation uncertainty.
Model selection under misspecification has been extensively discussed based on testing \citep{Vuong1989Likelihood, Rivers2002Model} or information criterion \citep{Lv2014Model, Yu2018Asymptotic, Hsu2019Model}.
Bootstrap model selection has unique advantages benefiting from the computational saving and applicability of our proposed bootstrap. 
In practice, to avoid reusing the data, we may conduct the selection procedure with half the data and make the inference with the other half.
                                            
Given a data set $\left\{ y_i, \textbf{x}_i \right\}_{i=1}^n$, suppose we have a candidate model set $\mathcal{M}$, which may or may not contain the true model $\xi_0$. Our goal is that if $\xi_0 \in \mathcal{M}$, we can identify $\xi_0$; and if $\xi_0 \notin \mathcal{M}$, which is more common in practice, we can rank the candidate models for practical use. 
We achieve this goal by evaluating each candidate model's in-sample average loss and out-of-sample prediction error.

Consider the in-sample average loss as 
$L_n(\xi) = \frac{n^{-1} \sum_{i=1}^n  \left[ \mathbb{E}(y_i)-\hat{\mu}_i(\textbf{x}_i, \hat{\boldsymbol{\beta}}_n) \right]^2}{\mbox{Var}(y_i)}$ under model $\xi$,
where $\hat{\boldsymbol{\beta}}_n$ is the QMLE estimate under model $\xi$.
We also consider the out-of-sample prediction error as $
\Gamma_n(\xi) = n^{-1} \sum_{i=1}^n \left[ \mathbb{E}(y'_i)-\hat{\mu}_i(\textbf{x}_i, \hat{\boldsymbol{\beta}}_n) \right]^2 / \mbox{Var}(y'_i)$, where $y'_i$ is a future response at $\textbf{x}_i$, which is independent of $y_i$, $i=1, \cdots, n$.
Since both $\mathbb{E}(L_n(\xi))$ and $\mathbb{E}(\Gamma_n(\xi))$ are unknown in practice, we can estimate them by bootstrap, respectively.
\begin{equation}\label{eq:bootmse}
\hat{L}_n(\xi)=\mathbb{E}^* \sum_{i=1}^n \frac{ \left[ y_i - \hat{\mu}_i(\textbf{x}_i, \hat{\boldsymbol{\beta}}^*_n) \right]^2}{n V(\hat{\mu}_i(\textbf{x}_i, \hat{\boldsymbol{\beta}}_n))},
\end{equation}
\begin{equation}\label{eq:bootpred}
    \hat{\Gamma}_n(\xi) = \mathbb{E}^* \left[
    \sum_{i=1}^n \frac{\left[ y_i-\hat{\mu}_i(\textbf{x}_i, \hat{\boldsymbol{\beta}}_n) \right]^2 }{nV(\hat{\mu}_i(\textbf{x}_i, \hat{\boldsymbol{\beta}}_n))} 
    +  \sum_{i=1}^n \frac{\left[ y_i-\hat{\mu}_i(\textbf{x}_i, \hat{\boldsymbol{\beta}}_n^*) \right]^2 }{nV(\hat{\mu}_i(\textbf{x}_i, \hat{\boldsymbol{\beta}}_n))} 
    - \sum_{i=1}^n \frac{\left[ y_i^*-\hat{\mu}_i(\textbf{x}^*_i, \hat{\boldsymbol{\beta}}^*_n) \right]^2}{nV(\hat{\mu}_i(\textbf{x}^*_i, \hat{\boldsymbol{\beta}}^*_n))} \right],
\end{equation}
where  $V(\hat{\mu}_i)$ is the variance function and $\hat{\boldsymbol{\beta}}^*_n$ is the bootstrapped estimate under model $\xi$. $y_i^*$ and $\textbf{x}_i^*$ are the bootstrapped response and predictors, respectively ($\textbf{x}_i^* = \textbf{x}_i$ in residual bootstrap).
Here, $\mathbb{E}^*$ is the expectation with respect to the bootstrap resampling, for which we could use local residual bootstrap. 
Therefore, we select the model that minimizes (\ref{eq:bootmse}) or (\ref{eq:bootpred}) over $\xi \in \mathcal{M}$ as our final model, that is, $\hat{\xi}_L =\arg\min_{\xi \in \mathcal{M}} \hat{L}_n(\xi)$ and $\hat{\xi}_\Gamma = \arg\min_{\xi \in \mathcal{M}} \hat{\Gamma}_n(\xi)$. 
The similar bootstrap model selection procedure can be found in \cite{Shao1996Bootstrap, Efron1983Estimating} and Section 8.2 of \cite{Shao2012Jackknife}. 

Here, $L_n(\xi)$ and $\Gamma_n(\xi)$ are used as the model selection criteria.
Their estimates $\hat{L}_n(\xi)$ and $\hat{\Gamma}_n(\xi)$ require the bootstrap estimate $\hat{\boldsymbol{\beta}}_n^*$, while $\hat{\Gamma}_n(\xi)$ also requires the bootstrap response $y_i^*$, which is only accessible through the bootstrap that regenerates the response variable.
By contrast, many other bootstrap procedures, e.g., one-step bootstrap, are not applicable to this out-of-sample prediction error approach because they do not recreate the response.

Given a data set, suppose the candidate model set does not contain the true model. 
Then we need to select the least misspecified model according to $L_n(\xi)$ and $\Gamma_n(\xi)$, i.e., the minimizer of $L_n(\xi)$ and $\Gamma_n(\xi)$ over $\mathcal{M}$.
In this case, the traditional bootstrap often fails to estimate $\hat{L}_n(\xi)$ and $\hat{\Gamma}_n(\xi)$ for $\xi \in \mathcal{M}$, because it cannot approximate the distributions of $\hat{\boldsymbol{\beta}}_n$ and response under misspecification. 
Therefore, the minimizers of $\hat{L}_n(\xi)$ and $\hat{\Gamma}_n(\xi)$ are rarely the optimal model choice for the data and the rank of the candidate models is also inaccurate.
On the contrary, local residual bootstrap can mimic the true data-generating process faithfully under misspecified models, offering accurate estimates for $\hat{L}_n(\xi)$ and $\hat{\Gamma}_n(\xi)$.
Hence, we not only select a model as the `best' approximation to the true data-generating process, but also rank candidate models according to the in-sample average loss and out-of-sample prediction error.


\section{Simulation Studies}\label{sec:simu}

\subsection{Standard Errors and Confidence Intervals}\label{sec:simu1}

Using simulations, we compare the standard error estimation and confidence interval coverage rate of the proposed method with other methods. 
Two types of confidence intervals are considered, namely, the asymptotic normal theory-based confidence interval $\textup{CI}_{\mbox{nor}} = \hat{\boldsymbol{\beta}}_n \pm \Phi^{-1}(1-\alpha/2) \hat{\psi}$ and bootstrap percentile confidence interval $\mbox{CI}_{\mbox{per}}$.
We conduct the local residual bootstrap with Pearson residual (LRB-Pearson), SBS residual (LRB-SBS), and surrogate residual (LRB-Surrogate), along with parametric bootstrap, pairwise bootstrap, wild bootstrap, multiplier bootstrap, and local response bootstrap. 
The wild bootstrap is defined as $y_i^*=\textbf{x}_i^T\hat{\boldsymbol{\beta}}_n + e_i^*$, where $e_i^*=w_i(y_i-\textbf{x}_i^T\hat{\boldsymbol{\beta}}_n)$ and $w_i$ is the `Rademacher' weights with $P(w_i=1)=P(w_i=-1)=0.5$. 
The multiplier bootstrap generates $\hat{\boldsymbol{\beta}}_n^* = \arg \max_{\boldsymbol{\beta} \in \mathbb{R}^p} \sum_{i=1}^n w_i \log f(y_i, \theta_i) $, where $w_i \sim \exp(1)$. The local response bootstrap is a variation of local residual bootstrap detailed in the appendix. For all these bootstrap methods, we focus on the standard error and confidence interval of the slope estimate of the first covariate.

We consider the twelve scenarios in total (SC1 -- SC12) of model misspecification. These scenarios include multivariate/univariate regressions with missing predictors, misspecified mean structure, mixed population, categorical predictors, over-dispersion for the case of 
binary/poisson/gamma/ordinal regressions (SC1 -- SC9), and missing interaction terms, heteroscedasticity, mixed population for the case of linear regressions (SC10--SC12).
Due to space limit, we present the results of SC1 here and the rest of results in Section \ref{sec:add_simu} of the appendix.
We use the ratio of the estimated standard error to the true standard error and the coverage rate for the confidence intervals to evaluate performance. 
The pseudo-true parameter and pseudo-true standard error are obtained by $10^4$ Monte Carlo replications. We set $B=500$ and the coverage rate is calculated by 100 replications.

\n
{\bf  SC1: Probit/Logistic/Ordinal Models for Univariate Predictor with Missing Term}

We simulate data $\{ y_i, x_i \}_{i=1}^n$ according to $G^{-1} ( P(Y_i = 1) ) =  \beta_0 + \beta_1x_i +\beta_2x_i^2$, where $x_i \sim \mbox{Unif}(-6,6)$ and $(\beta_0, \beta_1, \beta_2)=(12, 2, -2)$. The assumed model is $G^{-1} ( P(Y_i = 1) ) =  \beta_0 + \beta_1 x_i$, which ignores the quadratic term and is misspecified. We consider two link functions for $G$: probit model with normal link and logistic model with logit link. We also consider the ordinal regression case in which the data-generating process is $\Phi^{-1}( P(Y_i \leq j) )= \alpha_j + \beta_1 x_i + \beta_2 x_i^2$ for $j=1,2,3,4$. Let $x_i \sim \mbox{Unif}(1,7)$ and $(\alpha_1, \alpha_2, \alpha_3, \beta_1, \beta_2)=(-16, -12, -8, 8, -1)$. The assumed model is $\Phi^{-1}( P(Y_i \leq j) )= \alpha_j + \beta_1 x_i$, which again misses the quadratic term. The neighborhood size is $l=10$.

The results with $n=2000$ and $n=500$ are shown in Table \ref{Tab:cover_ci_s1} and Table \ref{Tab:cover_ci_s1_2} (in appendix), respectively. 
Table \ref{Tab:cover_ci_s1} lists the coverage rates of the pseudo-true parameter and ratio of standard error estimation, showing that all three types of local residual bootstrap provide accurate standard error estimations and effective coverage probabilities. On the contrary, the parametric bootstrap, pairwise bootstrap, wild bootstrap and multiplier bootstrap overestimate the standard errors and generate conservative confidence intervals. 
Considering the width of bootstrapped confidence interval, the proposed bootstrap achieves more desirable coverage rates with narrower confidence intervals. This results further verify that the alternatives tend to overestimate the standard errors and generate conservative confidence intervals in this setting. 
In particular, the bootstrap percentile confidence interval for the wild bootstrap cannot contain the pseudo-true parameter because its bootstrap estimates are biased in this setting. 
The pairwise bootstrap overestimates the standard errors because it introduces additional randomness for predictors under this fix design. The similar results can be found in the Table \ref{Tab:cover_ci_s1_2} (in appendix) with smaller sample size.

\begin{figure}[t]
\centering
\includegraphics[width=0.75\textwidth]{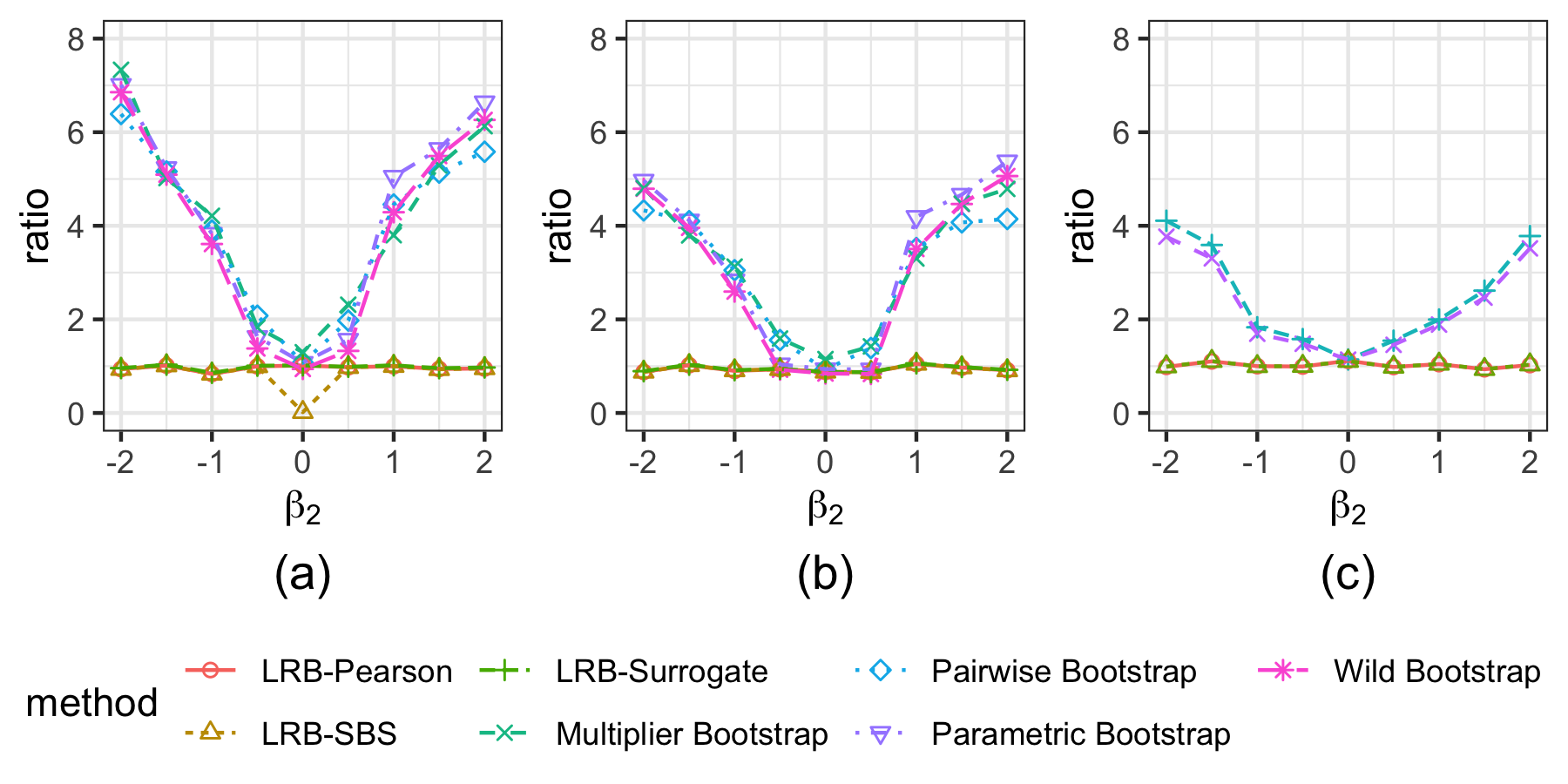}
\caption{The ratios of the estimated standard error to the true standard error under different values of $\beta_2$ in the true data-generating process under SC1. The probit, logistic, and ordinal models are presented in Panels (a), (b), and (c) respectively. We only consider four bootstrap methods in the ordinal regression because the Pearson residual is not well defined.
}\label{Fig:s1}
\end{figure}

We further vary $\beta_2$ in the data-generating process; Figure \ref{Fig:s1} plots the ratio of the standard error estimate to the true standard error as a function of $\beta_2$, which represents the degree of model misspecification. When $\beta_2=0$, the assumed model is correctly specified. As $\beta_2$ moves away from zero, the degree of misspecification becomes more serious. As shown in Figure \ref{Fig:s1}, local residual bootstrap performs well regardless of the degree of misspecification, while the other methods depend heavily on that degree.
When $\beta_2=0$, all the bootstrap methods provide accurate estimates with a ratio close to one. As $\beta_2$ moves away from zero, local residual bootstrap remains close to one, while the others deviate from one significantly. 
The same phenomenon appears in the other link functions.
The proposed method also outperforms in other scenarios in the appendix.

\begin{table}
\caption{Comparison of the coverage probabilities and average width (with standard error) for the bootstrap confidence interval and standard error estimation under SC1. 
The true standard errors of the pseudo-true estimations are in bold italics. `-' indicates that the Pearson residual is not well defined in the ordinal regression ($n=2000$).} \label{Tab:cover_ci_s1}
\centering
\begin{adjustbox}{width=1\textwidth,center}
\begin{tabular}{c c c c c c c c c} 
\toprule
&Confidence &Parametric &Pairwise &Wild &Multiplier &LRB- &LRB- &LRB- \\ 
&level &Bootstrap &Bootstrap &Bootstrap &Bootstrap &Pearson &SBS &Surrogate\\
\midrule
\multicolumn{9}{c}{Probit} \\
\hline
\multirow{6}{*}{$\mbox{CI}_{\mbox{nor}}$} 
&\multirow{2}{*}{0.95} &1.00  &1.00  &1.00  &1.00  &0.98  &0.98  &0.98  \\
& &0.032 (0.001) &0.029 (0.001) &0.031 (0.001) &0.029 (0.001) &0.004 (0.000) &0.004 (0.000) &0.004 (0.000) \\
&\multirow{2}{*}{0.90} &1.00  &1.00  &1.00  &1.00  &0.91  &0.91  &0.91  \\
& &0.027 (0.001) &0.025 (0.001) &0.026 (0.001) &0.024 (0.001) &0.004 (0.000) &0.004 (0.000) &0.004 (0.000) \\
&\multirow{2}{*}{0.75} &1.00  &1.00  &1.00  &1.00  &0.72  &0.73  &0.73  \\
& &0.019 (0.001) &0.017 (0.001) &0.018 (0.001) &0.017 (0.001) &0.003 (0.000) &0.003 (0.000) &0.003 (0.000) \\
\cline{2-9}
\multirow{6}{*}{$\mbox{CI}_{\mbox{per}}$} 
&\multirow{2}{*}{0.95} &1.00  &1.00  &0.00   &1.00  &0.93 &0.92  &0.95  \\
& &0.032 (0.001) &0.029 (0.001) &0.032 (0.002) &0.030 (0.001) &0.004 (0.000) &0.004 (0.000) &0.004 (0.000) \\
&\multirow{2}{*}{0.90} &1.00  &1.00  &0.00   &1.00  &0.88  &0.87  &0.88  \\
& &0.027 (0.001) &0.025 (0.001) &0.026 (0.001) &0.024 (0.001) &0.004 (0.000) &0.004 (0.000) &0.004 (0.000) \\
&\multirow{2}{*}{0.75} &1.00  &1.00  &0.00  &1.00  &0.71  &0.70  &0.73  \\
& &0.019 (0.001) &0.017 (0.001) &0.019 (0.001) &0.017 (0.001) &0.004 (0.000) &0.004 (0.000) &0.003 (0.000) \\
\cline{2-9}
Estimated SE $(\times 10^{-3})$ &\textbf{\emph{1.170}} &8.205 &7.476 &8.021 &8.582 &1.098 &1.097 &1.121 \\
(Estimated SE)/(true SE) &- &7.012 &6.389 &6.855 &7.334 &0.938 &0.937 &0.958 \\
\hline
\multicolumn{9}{c}{Logistic} \\
\hline
\multirow{6}{*}{$\mbox{CI}_{\mbox{nor}}$}  
&\multirow{2}{*}{0.95} &1.00  &1.00  &1.00  &1.00   &0.92  &0.92  &0.92  \\
& &0.052 (0.002) &0.045 (0.001) &0.050 (0.002) &0.045 (0.001) &0.009 (0.001) &0.009 (0.001) &0.009 (0.001) \\
&\multirow{2}{*}{0.90} &1.00  &1.00  &1.00  &1.00  &0.91  &0.89  &0.90 \\
& &0.043 (0.001) &0.038 (0.001) &0.042 (0.001) &0.038 (0.001) &0.008 (0.001) &0.008 (0.001) &0.008 (0.001) \\
&\multirow{2}{*}{0.75} &1.00  &1.00  &1.00  &1.00  &0.70  &0.69  &0.71  \\
& &0.030 (0.001) &0.027 (0.001) &0.029 (0.001) &0.027 (0.001) &0.005 (0.000) &0.005 (0.000) &0.005 (0.000) \\
\cline{2-9}
\multirow{6}{*}{$\mbox{CI}_{\mbox{per}}$} 
&\multirow{2}{*}{0.95} &1.00  &1.00  &0.00   &1.00  &0.93  &0.93  &0.93 \\
& &0.052 (0.002) &0.046 (0.002) &0.051 (0.002) &0.045 (0.002) &0.009 (0.001) &0.009 (0.001) &0.009 (0.001) \\
&\multirow{2}{*}{0.90} &1.00  &1.00  &0.00  &1.00   &0.90  &0.90  &0.89 \\
& &0.043 (0.002) &0.038 (0.001) &0.042 (0.002) &0.038 (0.001) &0.008 (0.001) &0.008 (0.001) &0.008 (0.001) \\
&\multirow{2}{*}{0.75} &1.00  &1.00  &0.00  &1.00   &0.70  &0.72  &0.70  \\
& &0.031 (0.001) &0.027 (0.001) &0.030 (0.001) &0.027 (0.001) &0.005 (0.000) &0.005 (0.000) &0.008 (0.001) \\
\cline{2-9}
Estimated SE $(\times 10^{-3})$ &\textbf{\emph{2.665}} &13.235 &11.537 &12.769 &12.808 &2.524 &2.513 &2.551 \\
(Estimated SE)/(true SE)  &- &4.966 &4.329 &4.791 &4.806 &0.947 &0.943 &0.957 \\
\hline
\multicolumn{9}{c}{Ordinal} \\
\hline
\multirow{6}{*}{$\mbox{CI}_{\mbox{nor}}$} 
&\multirow{2}{*}{0.95} &1.00  &1.00  &-  &-  &-  &0.95  &0.95 \\
& &0.057 (0.001) &0.063 (0.002) &- &-  &- &0.014 (0.001) &0.014 (0.001) \\
&\multirow{2}{*}{0.90} &1.00  &1.00  &-  &-  &-  &0.91  &0.91 \\
& &0.048 (0.001) &0.053 (0.002) &- &- &- &0.012 (0.000) &0.012 (0.000) \\
&\multirow{2}{*}{0.75} &0.98  &1.00  &-  &-  &-  &0.72  &0.73 \\
& &0.034 (0.001) &0.037 (0.001) &- &- &- &0.008 (0.000) &0.008(0.000) \\
\cline{2-9}
\multirow{6}{*}{$\mbox{CI}_{\mbox{per}}$}
&\multirow{2}{*}{0.95} &1.00  &1.00  &- &-  &-  &0.92  &0.92 \\
& &0.058 (0.002) &0.064 (0.003) &- &- &- &0.014 (0.001) &0.014(0.001) \\
&\multirow{2}{*}{0.90} &0.99  &1.00  &-  &-  &-  &0.89  &0.91 \\
& &0.048 (0.002) &0.053 (0.002) &- &- &- &0.012 (0.001) &0.012(0.001) \\
&\multirow{2}{*}{0.75} &0.92  &0.92  &-  &-  &-  &0.79  &0.79  \\
& &0.034 (0.001) &0.037 (0.002) &- &- &- &0.008 (0.000) &0.008(0.000) \\
\cline{2-9}
Estimated SE $(\times 10^{-3})$ &\textbf{\emph{5.752}} &13.924 &15.515 &- &- &- &6.328 &6.313 \\
(Estimated SE)/(true SE)  &- &2.421 &2.697 &- &- &- &1.100 &1.098\\
\bottomrule
\end{tabular}
\end{adjustbox}
\end{table}

\n

\subsection{Bootstrap Model Selection}\label{sec:simu2}

We assess the performance of the bootstrap model selection with various bootstrap methods in terms of the ability to rank the candidate models. We conduct local residual bootstrap with surrogate residuals along with the parametric bootstrap, pairwise bootstrap, wild bootstrap, and residual bootstrap with Pearson residuals and surrogate residuals. We then compare the true rank according to $L_n(\xi)$ (and $\Gamma_n(\xi)$) with the estimated rank of the candidate models. We also establish two criteria to evaluate ranking accuracy. One is the proportion of models whose ranks are correctly estimated, referred to as $\mbox{CR}_1$, that is, $\mbox{CR}_1=\sum_{\xi_i \in \mathcal{M}}I\{ \mbox{rank}(\xi_i) = \widehat{\mbox{rank}}(\xi_i) \}/|\mathcal{M}|$, where $|\mathcal{M}|$ is the cardinality of the candidate model set. The other is the proportion of all the candidate model pairs whose estimated orders are consistent with the true order, referred to as $\mbox{CR}_2=\sum_{i \neq j} I\{ (\mbox{rank}(\xi_i)-\mbox{rank}(\xi_j))(\widehat{\mbox{rank}}(\xi_i)-\widehat{\mbox{rank}}(\xi_j))>0 \}/\{|\mathcal{M}|(|\mathcal{M}|-1)\}$, which is similar to the C-index in survival analysis. The closer $\mbox{CR}_1$ and $\mbox{CR}_2$ are to one, the more accurate is the ranking estimate.

\begin{table}[t]
\caption{The true rank and estimated rank of each candidate model using the various bootstrap methods ($B=500$) under Cases I and II. The averages (standard deviations) are based on 50 replications.} \label{Tab:boot_rank_2}
\centering
\begin{adjustbox}{width=1\textwidth,center}
\begin{tabular}{c c c c c c c c c} 
\toprule
& \multirow{3}{*}{Model} &\multirow{3}{*}{True} &\multicolumn{6}{c}{Average of estimated rank} \\
\cline{4-9}
& & &Parametric &Pairwise &Wild &Pearson &Surrogate &LRB- \\ 
& &rank &Bootstrap &Bootstrap &Bootstrap &Bootstrap &Bootstrap &Surrogate\\
\midrule
\multicolumn{9}{c}{Case I} \\
\hline
\multirow{5}{*}{$L_n(\xi)$}
& $\{ x \}$ &3 &2.64 (0.53) &2.64 (0.53) &2.64 (0.53) &2.62 (0.53) &2.64 (0.53) &2.62 (0.57)\\
& $\{ x, \exp(x) \}$ &2 &1.60 (0.49) &1.60 (0.49) &1.60 (0.49) &1.60 (0.53) &1.60 (0.49) &1.64 (0.48)\\
& $\{ x, x^2, x^3 \}$ &1 &1.76 (0.94) &1.76 (0.94) &1.76 (0.94) &1.78 (0.93) &1.76 (0.94) &1.74 (0.94)\\
\cline{3-9}
& & $\mbox{CR}_1$ &0.61 (0.47) &0.61 (0.47) &0.61 (0.47) &0.59 (0.46) &0.61 (0.47) &0.63 (0.46) \\
& & $\mbox{CR}_2$ &0.74 (0.32) &0.74 (0.32) &0.74 (0.32) &0.73 (0.32) &0.74 (0.32) &0.74 (0.34) \\
\cline{2-9}
\multirow{5}{*}{$\Gamma_n(\xi)$}
& $\{ x \}$ &3 &2.24 (0.52) &2.52 (0.61) &2.60 (0.53) &2.20 (0.93) &2.18 (0.60) &2.66 (0.63)\\
& $\{ x, \exp(x) \}$ &2 &1.04 (0.20) &1.64 (0.48) &1.60 (0.49) &1.96 (0.53) &1.14 (0.35) &1.84 (0.47)\\
& $\{ x, x^2, x^3 \}$ &1 &2.72 (0.45) &1.84 (1.00) &1.80 (0.97) &1.84 (0.91) &2.68 (0.55) &1.50 (0.84)\\
\cline{3-9}
& & $\mbox{CR}_1$ &0.11 (0.16) &0.60 (0.48) &0.60 (0.48) &0.59 (0.42) &0.15 (0.24) &0.74 (0.41)\\
& & $\mbox{CR}_2$ &0.41 (0.17) &0.70 (0.36) &0.73 (0.33) &0.60 (0.43) &0.41 (0.23) &0.80 (0.33)\\
\hline
\multicolumn{9}{c}{Case II} \\
\hline
\multirow{6}{*}{$L_n(\xi)$}
& $\{ \mbox{probit}-(x_1, x_2)\}$ &4 &3.72 (0.45) &3.84 (0.37) &3.02 (0.14) &3.00 (0.00) &3.82 (0.39) &4.00 (0.00)\\
& $\{ \mbox{logit}-(x_1, x_2, x_1^2)\}$ &3 &3.28 (0.45) &3.16 (0.37) &3.98 (0.14) &4.00 (0.00) &3.18 (0.39) &3.00 (0.00)\\
& $\{ \mbox{logit}-(x_1, x_2, x_1x_2)\}$ &2 &2.00 (0.00) &1.50 (0.00) &2.00 (1.00) &2.00 (0.00) &2.00 (0.00) &2.00 (0.00)\\
& $\{ \mbox{probit}-(x_1, x_2, x_1x_2)\}$ &1 &1.00 (0.00) &1.50 (0.00) &1.00 (0.00) &1.00 (0.00) &1.00 (0.00) &1.00 (0.00)\\
\cline{3-9}
& & $\mbox{CR}_1$ &0.86 (0.23) &0.42 (0.19) &0.51 (0.07) &0.50 (0.00) &0.91 (0.19) &1.00 (0.00)\\
& & $\mbox{CR}_2$ &0.95 (0.08) &0.97 (0.06) &0.84 (0.02) &0.83 (0.00) &0.97 (0.06) &1.00 (0.00) \\
\cline{2-9}
\multirow{6}{*}{$\Gamma_n(\xi)$}
& $\{ \mbox{probit}-(x_1, x_2)\}$ &4 &4.00 (0.00) &3.84 (0.37) &3.00 (0.00) &3.00 (0.00) & 4.00(0.00) &3.96 (0.20) \\
& $\{ \mbox{logit}-(x_1, x_2, x_1^2)\}$ &3 &3.00 (0.00) &3.16 (0.37) &4.00 (0.00) &4.00 (0.00) &3.00 (0.00) &3.04 (0.20) \\
& $\{ \mbox{logit}-(x_1, x_2, x_1x_2)\}$ &2 &1.00 (0.00) &1.50 (0.00) &2.00 (0.00) &1.26 (0.44) &1.04 (0.20) &1.60 (0.49) \\
& $\{ \mbox{probit}-(x_1, x_2, x_1x_2)\}$ &1 &2.00 (0.00) &1.50 (0.00)  &1.00 (0.00) &1.74 (0.44) &1.96 (0.20) &1.40 (0.49) \\
\cline{3-9}
& & $\mbox{CR}_1$ &0.50 (0.00) &0.42 (0.19) &0.50 (0.00) &0.13 (0.22) &0.52 (0.10) &0.78 (0.25) \\
& & $\mbox{CR}_2$ &0.83 (0.00) &0.97 (0.06) &0.83 (0.00) &0.71 (0.07) &0.84 (0.03) &0.93 (0.08) \\
\bottomrule
\end{tabular}
\end{adjustbox}
\end{table}

We consider these two cases for generalized linear models. In Case I, the data $\{ y_i, x_i \}_{i=1}^n$ with $n=2000$ are generated from $\Phi^{-1}(Y_i=1)=\sin(2x_i-1) + 0.1\exp(x_i) + 0.5x^3$, where $x \sim \mbox{Unif}(-6,6)$. The candidate model set contains three models with misspecified predictors, $\mathcal{M}_1=\left\{ \{x\}, ~\{x, x^2, x^3\}, ~\{x, \exp(x)\} \right\}$.
In Case II, the data $\{ y_i, \textbf{x}_i \}_{i=1}^n$ with $n=2000$ are generated from $\Phi^{-1}(P(Y_i=1))=1+2x_{i1}-1.5x_{i2} + x_{i1}x_{i2} - (x_{i1}-x_{i2})^2$, where $x_1, x_2 \sim \mbox{Unif}(-6,6)$.
The candidate model set contains four models with a misspecified link function or mean structure, $\mathcal{M}_2=\left\{ \{\mbox{probit}\!-\!(x_1, x_2)\}, \{ \mbox{probit}\!-\!(x_1, x_2, x_1x_2) \}, \{ \mbox{logit}\!-\!(x_1, x_2, x_1^2)\}, \right. \\ \left.  \{ \mbox{logit}\!-\!(x_1, x_2, x_1x_2)\} \right\}$. 
Neither $\mathcal{M}_1$ nor $\mathcal{M}_2$ contains the true model. Table \ref{Tab:boot_rank_2} shows the results.

In Case I, compared with the alternatives, the proposed method shows only a slight improvement based on the in-sample average loss, but an obvious improvement based on the out-of-sample prediction. Only the proposed bootstrap can accurately estimate the rank of the candidate models and identify the least misspecified model, whereas the other bootstrap methods always make a mistake when ranking the models $\{ x, \exp(x) \}$ and $\{x, x^2, x^3 \}$. In Case II, model misspecification happens both on the mean structure and on the link function. A better mean structure with a misspecified link function may have less misspecification than a worse mean structure with a correct link function. Only the proposed method can identify the true rank by both $\hat{L}_n(\xi)$ and $\hat{\Gamma}_n(\xi)$.
\section{Real Data Analysis}\label{sec:realdata}

We use the proposed method to analyze a real data set on the fatal maiden voyage of the Titanic to provide insights into the likelihood of passengers surviving. 
This data set (in R package \texttt{titanic}) holds information on $n=714$ passengers, including their economic status (ticket price), gender, age, and survival. 
We explore the factors that affect survival probability by fitting a probit model and make inference for the coefficients. 
To accommodate the different age effects on survival for male and female, we include the interaction term in model I as follows:
$\Phi^{-1}\left\{ P(\mbox{Survival}=1) \right\} 
= \beta_0 + \beta_{\mbox{gender}} \mbox{Gender} + \beta_{\mbox{age}} \mbox{Age} + 
\beta_{\mbox{fare}} \mbox{Fare}
+ \beta_{\mbox{gender} \times \mbox{age}} \mbox{Gender} \times \mbox{Age}$.
``Gender" is coded one for male and zero for female, and other variables are continuous. If ``Gender" is zero, then ``Gender $\times$ Age" is zero for female, if ``Gender" is one, then ``Gender $\times$ Age" equals to the value of age for male. 
It leads to different models for male and female essentially, the coefficient $\beta_{\mbox{gender} \times \mbox{age}}$ represents the difference in the age effects on survival for male and female.

\begin{figure}[!t]
  \centering
  \subfigure[]{\includegraphics[width=0.3\textwidth]{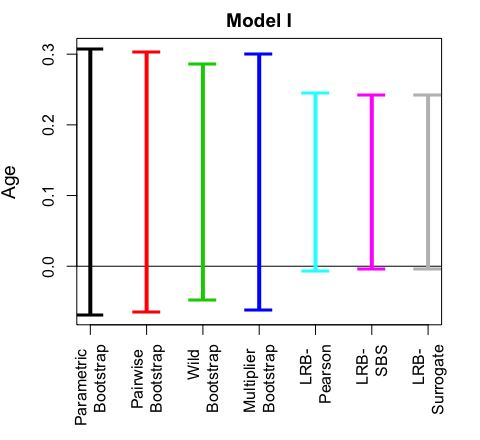}} 
  \subfigure[]{\includegraphics[width=0.3\textwidth]{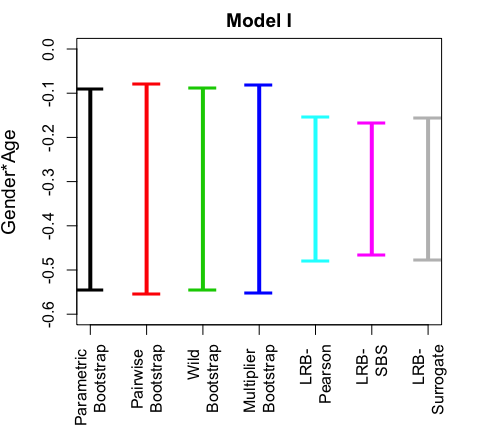}} 
  \subfigure[]{\includegraphics[width=0.3\textwidth]{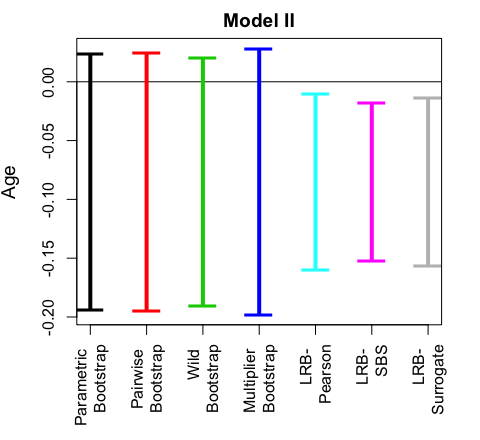}} 
  \caption{The confidence intervals of the different bootstrap methods. Panel (a) is for $\beta_{\mbox{age}}$ in model I, Panel (b) is for $\beta_{\mbox{gender} \times \mbox{age}}$ in model I, and Panel (c) is for $\beta_{\mbox{age}}$ in model II. The confidence intervals for the other coefficient estimates are omitted here.
}\label{Fig:titanic_ci}
\end{figure}

Table \ref{fit_coe} shows the coefficient estimates and Panels (a) and (b) of Figure \ref{Fig:titanic_ci} illustrate the confidence intervals for these coefficients from the various bootstrap procedures.
As we can see, the confidence intervals for $\beta_{\mbox{age}}$ contain zero, indicating that the age effect for female is not significant.
The confidence intervals for $\beta_{\mbox{gender} \times \mbox{age}}$ stay below zero, meaning the difference in the age effects for male and female is significant difference.  
The bootstrapped confidence intervals by local residual bootstrap are slightly narrower than those of the other bootstrap methods.
If the model is correctly specified, the confidence intervals for the parametric bootstrap should be the same as that of local residual bootstrap.
Therefore, this difference implies that even model I is mildly misspecified.

\begin{table}[!t]
\caption{Coefficient estimate of the probit model for models I and II.} \label{fit_coe}
\centering
\begin{tabular}{c c c c c c}
\toprule
&$\hat{\beta}_0$ & $\hat{\beta}_{\mbox{gender}}$ & $\hat{\beta}_{\mbox{age}}$ & $\hat{\beta}_{\mbox{fare}}$ & $\hat{\beta}_{\mbox{gender} \times \mbox{age}}$ \\
\midrule
Model I &0.692 &-1.476 &0.120 &0.384 &-0.316 \\
Model II &0.651 &-1.434 &-0.085 &0.389 &- \\
\bottomrule 
\end{tabular}
\end{table}

We now consider another misspecified model, model II, which omits the interaction term: 
$\Phi^{-1}\left\{ P(\mbox{Survival}=1) \right\} = \beta_0 + \beta_{\mbox{gender}} \mbox{Gender} + \beta_{\mbox{age}} \mbox{Age} + \beta_{\mbox{fare}} \mbox{Fare}$.
Since it ignores the different age effects for male and female, model II is even more misspecified than model I.
When selecting the bootstrap model using local residual bootstrap, the bootstrapped estimate for the in-sample average loss and out-of-sample prediction errors are $\hat{L}_n(\mbox{model I})=0.978$, $\hat{\Gamma}_n(\mbox{model I})=0.982$ and 
$\hat{L}_n(\mbox{model II})=0.984$, and $\hat{\Gamma}_n(\mbox{model II})=0.992$, indicating that model I is the better choice.

Suppose we mistakenly choose model II for the analysis. In this case, Table \ref{fit_coe} lists the estimated coefficients and Panel (c) of Figure \ref{Fig:titanic_ci} plots the corresponding confidence intervals, showing that age now has an estimated negative effect, which makes sense because the majority of the passengers are male.
However, according to the confidence intervals, such an effect is not significant under the parametric bootstrap, pairwise bootstrap, wild bootstrap, and multiplier bootstrap.
This clearly contradicts our understanding of the data since we have also shown the strong negative age effect for male and that the majority of the passengers are male.

On the contrary, the confidence intervals by local residual bootstrap do not contain zero and imply that such a negative effect is significant, ensuring a valid conclusion even though model II is misspecified.
Age is an important survival factor in this analysis and is justified in model I.
Older people have lower survival rates than younger people for a variety of reasons.
However, under the misspecified model II, the traditional bootstrap cannot identify the age effect, whereas the proposed bootstrap can still establish the significance of age.

Lastly, we examine hypothesis testing for $\beta_{\mbox{age}}$ and $\beta_{\mbox{gender} \times \mbox{age}}$ under models I and II. 
Table \ref{Tab:test_titanic_2} lists the bootstrapped p-values by the different methods. 
Under model I, all the bootstrap methods fail to reject the two-sided test $H_0: \beta_{\mbox{age}}=0$. 
All the methods except the wild bootstrap reject $H_0: \beta_{\mbox{gender} \times \mbox{age}} = 0$, suggesting that age has different effects on survival for male and female. 
Under the more misspecified model II, the parametric bootstrap, pairwise bootstrap, wild bootstrap, and multiplier bootstrap all fail to reject $H_0: \beta_{\mbox{age}}=0$, mistakenly suggesting age is not significant.
On the contrary, local residual bootstrap rejects both the two-sided test $H_0: \beta_{\mbox{age}}=0$ and the one-sided test $H_0: \beta_{\mbox{age}} \ge 0$, correctly suggesting that age has a negative effect on survival.
Therefore, local residual bootstrap can provide consistent and reasonable conclusions even when the model is misspecified.

\begin{table}[H]
\caption{Bootstrapped $p$-values for Titanic data.} \label{Tab:test_titanic_2}
\centering
\renewcommand\arraystretch{0.4}  
\resizebox{\linewidth}{!}{
\begin{tabular}{c c c c c c c c c }
\toprule
&\multirow{2}{*}{$H_0$} &Parametric &Pairwise &Wild &Multiplier &LRB- &LRB- &LRB- \\ 
& &Bootstrap &Bootstrap &Bootstrap &Bootstrap &Pearson &SBS &Surrogate\\
\midrule
\multirow{2}{*}{Model I} &$\beta_{\mbox{age}}=0$ 
&0.214 &0.213 &0.724 &0.192 &0.062 &0.054 &0.063 \\
&$\beta_{\mbox{gender} \times \mbox{age}}=0$ &0.005 &0.010 &0.414 &0.008 &0.000 &0.000 &0.002 \\
\cline{2-9}
\multirow{2}{*}{Model II} &$\beta_{\mbox{age}}=0$ 
&0.112 &0.138 &0.444 &0.133 &0.027 &0.009 &0.031 \\
&$\beta_{\mbox{age}} \ge 0$ &0.048 &0.077 &0.443 &0.071 &0.022 &0.001 &0.029 \\
\bottomrule
\end{tabular}}
\end{table}

\section{Conclusion}\label{sec:conclusions}

In this paper, we propose local residual bootstrap to assess estimation uncertainty, infer the pseudo-true parameter under model misspecification, and conduct bootstrap model selection.
The proposed method resamples the residuals under the neighborhood constraint to preserve the misspecification information and mimics the true sampling distribution.
A theoretical investigation and numerical studies are presented to confirm the validity of the proposed method, which can be applied to various models such as generalized linear models, linear models, and ordinal regressions. 

There are many directions for future research. 
We focus on the misspecification of model form in this article, whereas the serial correlation among observations in time series and longitudinal data is not handled.
It is an important yet challenging task to address in future studies.
Meanwhile, we focus more on estimation uncertainty than model selection uncertainty.
The model selection uncertainty is often exacerbated when the data-generating process is complex and all candidate models are somewhat misspecified.
Since statistical inference is usually conducted based on a selected model, 
assessing both the model selection and estimation uncertainty in the context of misspecification is an important task.
The interplay of the two sources of uncertainty is an interesting future topic.


\bibliographystyle{chicago}      
\bibliography{lrb}   

\begin{thebibliography}{}

\bibitem[\protect\citeauthoryear{Arteche and Orbe}{Arteche and
  Orbe}{2009}]{Arteche2009Bootstrap}
Arteche, J. and J.~Orbe (2009).
\newblock Bootstrap-based bandwidth choice for log-periodogram regression.
\newblock {\em Journal of Time Series Analysis\/}~{\em 30\/}(6), 591--617.

\bibitem[\protect\citeauthoryear{Arteche and Orbe}{Arteche and
  Orbe}{2017}]{Arteche2017Strategy}
Arteche, J. and J.~Orbe (2017).
\newblock A strategy for optimal bandwidth selection in local whittle
  estimation.
\newblock {\em Econometrics and Statistics\/}~{\em 4}, 3--17.

\bibitem[\protect\citeauthoryear{Biau, C{\'e}rou, and Guyader}{Biau
  et~al.}{2010}]{Biau2010Rate}
Biau, G., F.~C{\'e}rou, and A.~Guyader (2010).
\newblock On the rate of convergence of the bagged nearest neighbor estimate.
\newblock {\em Journal of Machine Learning Research\/}~{\em 11\/}(2), 687--712.

\bibitem[\protect\citeauthoryear{Biau and Devroye}{Biau and
  Devroye}{2010}]{Biau2010Layered}
Biau, G. and L.~Devroye (2010).
\newblock On the layered nearest neighbour estimate, the bagged nearest
  neighbour estimate and the random forest method in regression and
  classification.
\newblock {\em Journal of Multivariate Analysis\/}~{\em 101\/}(10), 2499--2518.

\bibitem[\protect\citeauthoryear{Bose and Chatterjee}{Bose and
  Chatterjee}{2003}]{Bose2003Generalized}
Bose, A. and S.~Chatterjee (2003).
\newblock Generalized bootstrap for estimators of minimizers of convex
  functions.
\newblock {\em Journal of Statistical Planning and Inference\/}~{\em 117\/}(2),
  225--239.

\bibitem[\protect\citeauthoryear{Chatterjee and Bose}{Chatterjee and
  Bose}{2005}]{Chatterjee2005Generalized}
Chatterjee, S. and A.~Bose (2005).
\newblock Generalized bootstrap for estimating equations.
\newblock {\em The Annals of Statistics\/}~{\em 33\/}(1), 414--436.

\bibitem[\protect\citeauthoryear{Claeskens, Aerts, and Molenberghs}{Claeskens
  et~al.}{2003}]{Claeskens2003A}
Claeskens, G., M.~Aerts, and G.~Molenberghs (2003).
\newblock A quadratic bootstrap method and improved estimation in logistic
  regression.
\newblock {\em Statistics \& Probability Letters\/}~{\em 61\/}(4), 383--394.

\bibitem[\protect\citeauthoryear{Efron}{Efron}{1983}]{Efron1983Estimating}
Efron, B. (1983).
\newblock Estimating the error rate of a prediction rule: Improvement on
  cross-validation.
\newblock {\em Journal of the American Statistical Association\/}~{\em
  78\/}(382), 316--331.

\bibitem[\protect\citeauthoryear{Fahrmexr}{Fahrmexr}{1990}]{Fahrmexr1990Maximum}
Fahrmexr, L. (1990).
\newblock Maximum likelihood estimation in misspecified generalized linear
  models.
\newblock {\em Statistics: A Journal of Theoretical and Applied Stats\/}~{\em
  21\/}(4), 487--502.

\bibitem[\protect\citeauthoryear{Freedman}{Freedman}{1981}]{Freedman1981Bootstrapping}
Freedman, D.~A. (1981).
\newblock Bootstrapping regression models.
\newblock {\em The Annals of Statistics\/}~{\em 9\/}(6), 1218--1228.

\bibitem[\protect\citeauthoryear{Friedl and Tilg}{Friedl and
  Tilg}{1997}]{Friedl1997Variance}
Friedl, H. and N.~Tilg (1997).
\newblock Variance estimates in logistic regression using the bootstrap.
\newblock {\em Communications in Statistics - Theory and Methods\/}~{\em
  24\/}(2), 473--486.

\bibitem[\protect\citeauthoryear{Gon{\c{c}}alves and White}{Gon{\c{c}}alves and
  White}{2004}]{Gonccalves2004Maximum}
Gon{\c{c}}alves, S. and H.~White (2004).
\newblock Maximum likelihood and the bootstrap for nonlinear dynamic models.
\newblock {\em Journal of Econometrics\/}~{\em 119\/}(1), 199--219.

\bibitem[\protect\citeauthoryear{Gon{\c{c}}alves and White}{Gon{\c{c}}alves and
  White}{2005}]{Goncalves2005Bootstrap}
Gon{\c{c}}alves, S. and H.~White (2005).
\newblock Bootstrap standard error estimates for linear regression.
\newblock {\em Journal of the American Statistical Association\/}~{\em
  100\/}(471), 970--979.

\bibitem[\protect\citeauthoryear{Gozalo}{Gozalo}{1997}]{Gozalo1997}
Gozalo, P.~L. (1997).
\newblock Nonparametric bootstrap analysis with applications to demographic
  effects in demand functions.
\newblock {\em Journal of Econometrics\/}~{\em 81\/}(2), 357--393.

\bibitem[\protect\citeauthoryear{Hall, Horowitz, and Jing}{Hall
  et~al.}{1995}]{Hall1995On}
Hall, P., J.~L. Horowitz, and B.~Y. Jing (1995).
\newblock On blocking rules for the bootstrap with dependent data.
\newblock {\em Biometrika\/}~{\em 82\/}(3), 561--574.

\bibitem[\protect\citeauthoryear{Hall and Jing}{Hall and
  Jing}{1996}]{Hall1996On}
Hall, P. and B.~Jing (1996).
\newblock On sample reuse methods for dependent data.
\newblock {\em Journal of the Royal Statistical Society. Series B:
  Methodological\/}~{\em 58\/}(4), 727--737.

\bibitem[\protect\citeauthoryear{Hsu, Ing, and Tong}{Hsu
  et~al.}{2019}]{Hsu2019Model}
Hsu, H.-L., C.-K. Ing, and H.~Tong (2019).
\newblock On model selection from a finite family of possibly misspecified time
  series models.
\newblock {\em The Annals of Statistics\/}~{\em 47\/}(2), 1061--1087.

\bibitem[\protect\citeauthoryear{Kline and Santos}{Kline and
  Santos}{2012}]{Kline2012Higher}
Kline, P. and A.~Santos (2012).
\newblock Higher order properties of the wild bootstrap under misspecification.
\newblock {\em Journal of Econometrics\/}~{\em 171\/}(1), 54--70.

\bibitem[\protect\citeauthoryear{Komunjer}{Komunjer}{2005}]{Komunjer2005Quasi}
Komunjer, I. (2005).
\newblock Quasi-maximum likelihood estimation for conditional quantiles.
\newblock {\em Journal of Econometrics\/}~{\em 128\/}(1), 137--164.

\bibitem[\protect\citeauthoryear{Lee}{Lee}{2016}]{Lee2016Asymptotic}
Lee, S. (2016).
\newblock Asymptotic refinements of a misspecification-robust bootstrap for
  {GEL} estimators.
\newblock {\em Journal of Econometrics\/}~{\em 192\/}(1), 86--104.

\bibitem[\protect\citeauthoryear{Li and Shepherd}{Li and
  Shepherd}{2012}]{Li2012A}
Li, C. and B.~E. Shepherd (2012).
\newblock A new residual for ordinal outcomes.
\newblock {\em Biometrika\/}~{\em 99\/}(2), 473--480.

\bibitem[\protect\citeauthoryear{Lin and Liu}{Lin and Liu}{2022}]{Lin2022Model}
Lin, Z. and D.~Liu (2022).
\newblock Model diagnostics of discrete data regression: a unifying framework
  using functional residuals.
\newblock {\em arXiv preprint arXiv:2207.04299\/}.

\bibitem[\protect\citeauthoryear{Liu and Zhang}{Liu and
  Zhang}{2018}]{Liu2018Residuals}
Liu, D. and H.~Zhang (2018).
\newblock Residuals and diagnostics for ordinal regression models: A surrogate
  approach.
\newblock {\em Journal of the American Statistical Association\/}~{\em
  113\/}(522), 845--854.

\bibitem[\protect\citeauthoryear{Liu}{Liu}{1988}]{Liu1988Bootstrap}
Liu, R.~Y. (1988).
\newblock Bootstrap procedures under some non-i.i.d. models.
\newblock {\em The Annals of Statistics\/}~{\em 16\/}(4), 1696--1708.

\bibitem[\protect\citeauthoryear{Lv and Liu}{Lv and Liu}{2014}]{Lv2014Model}
Lv, J. and J.~S. Liu (2014).
\newblock Model selection principles in misspecified models.
\newblock {\em Journal of the Royal Statistical Society. Series B:
  Methodological\/}~{\em 76\/}(1), 141--167.

\bibitem[\protect\citeauthoryear{MacKinnon}{MacKinnon}{2006}]{MacKinnon2006Bootstrap}
MacKinnon, J.~G. (2006).
\newblock Bootstrap methods in econometrics.
\newblock {\em Economic Record\/}~{\em 82}, S2--S18.

\bibitem[\protect\citeauthoryear{Mammen}{Mammen}{1993}]{Mammen1993Bootstrap}
Mammen, E. (1993).
\newblock Bootstrap and wild bootstrap for high dimensional linear models.
\newblock {\em The Annals of Statistics\/}~{\em 21\/}(1), 255--285.

\bibitem[\protect\citeauthoryear{Moulton and Zeger}{Moulton and
  Zeger}{1991}]{Moulton1991Bootstrapping}
Moulton, L.~H. and S.~L. Zeger (1991).
\newblock Bootstrapping generalized linear models.
\newblock {\em Computational Statistics \& Data Analysis\/}~{\em 11\/}(1),
  53--63.

\bibitem[\protect\citeauthoryear{Newey and Powell}{Newey and
  Powell}{1987}]{Newey1987Asymmetric}
Newey, W.~K. and J.~L. Powell (1987).
\newblock Asymmetric least squares estimation and testing.
\newblock {\em Econometrica: Journal of the Econometric Society\/}~{\em
  55\/}(4), 819--–847.

\bibitem[\protect\citeauthoryear{Paparoditis and Politis}{Paparoditis and
  Politis}{1999}]{Paparoditis1999Local}
Paparoditis, E. and D.~N. Politis (1999).
\newblock The local bootstrap for periodogram statistics.
\newblock {\em Journal of Time Series Analysis\/}~{\em 20\/}(2), 193--222.

\bibitem[\protect\citeauthoryear{Paparoditis and Politis}{Paparoditis and
  Politis}{2010}]{Paparoditis2010Residual}
Paparoditis, E. and D.~N. Politis (2010).
\newblock Residual-based block bootstrap for unit root testing.
\newblock {\em Econometrica\/}~{\em 71\/}(3), 813--855.

\bibitem[\protect\citeauthoryear{Rivers and Vuong}{Rivers and
  Vuong}{2002}]{Rivers2002Model}
Rivers, D. and Q.~Vuong (2002).
\newblock Model selection tests for nonlinear dynamic models.
\newblock {\em The Econometrics Journal\/}~{\em 5\/}(1), 1--39.

\bibitem[\protect\citeauthoryear{Scudilio and Pereira}{Scudilio and
  Pereira}{2020}]{Scudilio2020Adjusted}
Scudilio, J. and G.~H. Pereira (2020).
\newblock Adjusted quantile residual for generalized linear models.
\newblock {\em Computational Statistics\/}~{\em 35\/}(1), 399--421.

\bibitem[\protect\citeauthoryear{Shao}{Shao}{1996}]{Shao1996Bootstrap}
Shao, J. (1996).
\newblock Bootstrap model selection.
\newblock {\em Journal of the American Statistical Association\/}~{\em
  91\/}(434), 655--665.

\bibitem[\protect\citeauthoryear{Shao and Tu}{Shao and
  Tu}{2012}]{Shao2012Jackknife}
Shao, J. and D.~Tu (2012).
\newblock {\em The jackknife and bootstrap}.
\newblock Springer.

\bibitem[\protect\citeauthoryear{Shi}{Shi}{1991}]{Shi1991Local}
Shi, S.~G. (1991).
\newblock Local bootstrap.
\newblock {\em Annals of the Institute of Statistical Mathematics\/}~{\em
  43\/}(4), 667--676.

\bibitem[\protect\citeauthoryear{Silva, Franco, Reisen, and Cruz}{Silva
  et~al.}{2006}]{Silva2006local}
Silva, E., G.~C. Franco, V.~A. Reisen, and F.~R. Cruz (2006).
\newblock Local bootstrap approaches for fractional differential parameter
  estimation in arfima models.
\newblock {\em Computational Statistics \& Data Analysis\/}~{\em 51\/}(2),
  1002--1011.

\bibitem[\protect\citeauthoryear{Spokoiny and Zhilova}{Spokoiny and
  Zhilova}{2015}]{Spokoiny2015Bootstrap}
Spokoiny, V. and M.~Zhilova (2015).
\newblock Bootstrap confidence sets under model misspecification.
\newblock {\em The Annals of Statistics\/}~{\em 43\/}(6), 2653--2675.

\bibitem[\protect\citeauthoryear{Vuong}{Vuong}{1989}]{Vuong1989Likelihood}
Vuong, Q.~H. (1989).
\newblock Likelihood ratio tests for model selection and non-nested hypotheses.
\newblock {\em Econometrica\/}~{\em 57\/}(2), 307--333.

\bibitem[\protect\citeauthoryear{White}{White}{1982}]{White1982Maximum}
White, H. (1982).
\newblock Maximum likelihood estimation of misspecified models.
\newblock {\em Econometrica\/}~{\em 50\/}(1), 1--25.

\bibitem[\protect\citeauthoryear{Wu}{Wu}{1986}]{Wu1986Jackknife}
Wu, C.~F. (1986).
\newblock Jackknife, bootstrap and other resampling methods in regression
  analysis.
\newblock {\em Annals of Statistics\/}~{\em 14\/}(4), 1261--1295.

\bibitem[\protect\citeauthoryear{Yu, Zhang, and Yau}{Yu
  et~al.}{2018}]{Yu2018Asymptotic}
Yu, D., X.~Zhang, and K.~K. Yau (2018).
\newblock Asymptotic properties and information criteria for misspecified
  generalized linear mixed models.
\newblock {\em Journal of the Royal Statistical Society: Series B (Statistical
  Methodology)\/}~{\em 80\/}(4), 817--836.

\end{thebibliography}

\section{Appendix}

\subsection{Local Response Bootstrap}\label{sec:localresponse}

We introduce a modified version of the proposed method, termed local response bootstrap, by simply resampling response instead of residuals under neighborhood constraint.
It is a fast bootstrap method and can work well under low dimensional predictor cases. 
However, compared with local residual bootstrap, it often generates unsatisfactory results for the case of multivariate predictors as shown in the simulation.
The details of this method are in Algorithm \ref{Algorithm3}. 

\begin{spacing}{2.0}    
\begin{algorithm}[!htp]
\caption{Local response bootstrap procedure}\label{Algorithm3}
\LinesNumbered
\KwIn{data $\{ y_i, \textbf{x}_i \}_{i=1}^n$}
\KwOut{pseudo-true standard error estimate and confidence interval}
 Obtain the covariate distance matrix $\textbf{D}=[d_{ij}]_{n \times n}$ where $d_{ij}= \Vert \textbf{x}_i - \textbf{x}_j \Vert$.

\For{$b=1,\dots,B$}{

  \For{$i = 1, \dots, n$}{
   Generate the bootstrapped response $y_i^*$ by sampling from $\left\{ y_j: j \in N_i \right\}$ where $N_i=\left\{ j: d_{ij} \mbox{ are among the smallest } l \mbox{ distances of } \{ d_{ik} \}_{k=1}^n \right\}$; \\
    }
    Obtain the bootstrapped estimate $\hat{\boldsymbol{\beta}}_n^{*(b)}$ based on the bootstrap sample $\{ y_i^*, \textbf{x}_i \}_{i=1}^n$.
    }
   The pseudo-true standard error estimate is $\hat{\psi}=\sqrt{\sum_{b=1}^B (\hat{\boldsymbol{\beta}}_n^{*(b)}- \overline{\hat{\boldsymbol{\beta}}^*_n})^2/B}$, where $\overline{\hat{\boldsymbol{\beta}}}_n^* = \sum_{b=1}^B \hat{\boldsymbol{\beta}}_n^{*(b)}/B$. 
   
   The confidence interval for pseudo-true parameter is given by $\hat{\boldsymbol{\beta}}_n \pm \Phi^{-1}(1-\alpha/2) \hat{\psi}$. 
\end{algorithm}
\end{spacing}


\subsection{Additional Results for Simulations}\label{sec:add_simu}

Additional simulation scenarios can be founded in this section. SC2 - SC9 are for generalized linear models with various types of model misspecification, and SC10-SC12 are for linear regression models.  

\n
\begin{table}[!htp]
\caption{Comparison of the coverage probabilities and average width (standard error) for the bootstrap confidence interval and standard error estimation under SC1. The true standard errors of the pseudo-true estimations are in bold italics. `-' indicates that the Pearson residual is not well defined in the ordinal regression ($n=500$).} \label{Tab:cover_ci_s1_2}
\centering
\renewcommand\arraystretch{1.0}  
\resizebox{\linewidth}{!}{
\begin{tabular}{c c c c c c c c c} 
\toprule
&Confidence &Parametric &Pairwise &Wild &Multiplier &LRB- &LRB- &LRB- \\ 
&level &Bootstrap &Bootstrap &Bootstrap &Bootstrap &Pearson &SBS &Surrogate\\
\midrule
\multicolumn{9}{c}{Probit} \\
\hline
\multirow{6}{*}{$\mbox{CI}_{\mbox{nor}}$} 
&\multirow{2}{*}{0.95} &1.00 &1.00 &1.00 &1.00 &0.94 &0.94 &0.94\\
& &0.064 (0.002) &0.058 (0.002) &0.063 (0.002) &0.057 (0.002) &0.009 (0.001) &0.009 (0.001) &0.009 (0.001) \\
&\multirow{2}{*}{0.90} &1.00 &1.00 &1.00 &1.00 &0.89 &0.92 &0.91\\
& &0.053 (0.002) &0.048 (0.001) &0.053 (0.002) &0.048 (0.001) &0.007 (0.001) &0.007 (0.001) &0.008 (0.001) \\
&\multirow{2}{*}{0.75} &1.00 &1.00 &1.00 &1.00 &0.71 &0.76 &0.72\\
& &0.037 (0.001) &0.034 (0.001) &0.037 (0.001) &0.034 (0.001) &0.005 (0.001) &0.005 (0.001) &0.005 (0.001) \\
\cline{2-9}
\multirow{6}{*}{$\mbox{CI}_{\mbox{per}}$} 
&\multirow{2}{*}{0.95} &1.00 &1.00 &1.00 &1.00 &0.94 &0.94 &0.94\\
& &0.064 (0.003) &0.058 (0.002) &0.064 (0.003) &0.058 (0.002) &0.009 (0.001) &0.009 (0.001) &0.009 (0.001) \\
&\multirow{2}{*}{0.90} &1.00 &1.00 &1.00 &1.00 &0.92 &0.91 &0.91\\
& &0.053 (0.002) &0.048 (0.002) &0.053 (0.002) &0.048 (0.002) &0.007 (0.001) &0.007 (0.001) &0.008 (0.001) \\
&\multirow{2}{*}{0.75} &1.00 &1.00 &1.00 &1.00 &0.77 &0.77 &0.75\\
& &0.038 (0.001) &0.034 (0.001) &0.037 (0.001) &0.034 (0.001) &0.005 (0.001) &0.005 (0.001) &0.005 (0.001) \\
\cline{2-9}
Estimated SE $(\times 10^{-3})$ &\textbf{\emph{2.367}} &16.247 &14.740 &16.065 &14.686 &2.265 &2.261 &2.321 \\
(Estimated SE)/(true SE) &- &6.863 &6.227 &6.787 &6.204 &0.957 &0.955 &0.981 \\
\hline
\multicolumn{9}{c}{Logistic} \\
\hline
\multirow{6}{*}{$\mbox{CI}_{\mbox{nor}}$}  
&\multirow{2}{*}{0.95} &1.00 &1.00 &1.00  &1.00 &0.96 &0.95 &0.97 \\
& &0.102 (0.003) &0.088 (0.003) &0.101 (0.003) &0.089 (0.003) &0.018 (0.002) &0.018 (0.002) &0.018 (0.002) \\  
&\multirow{2}{*}{0.90} &1.00 &1.00 &1.00  &1.00 &0.90 &0.90 &0.90 \\
& &0.086 (0.003) &0.074 (0.002) &0.084 (0.003) &0.075 (0.002) &0.015 (0.002) &0.015 (0.002) &0.015 (0.002) \\
&\multirow{2}{*}{0.75} &1.00 &1.00 &1.00  &1.00 &0.72 &0.73 &0.73 \\
& &0.060 (0.002) &0.052 (0.002) &0.059 (0.002) &0.052 (0.002) &0.010 (0.001) &0.010 (0.001) &0.011 (0.001) \\
\cline{2-9}
\multirow{6}{*}{$\mbox{CI}_{\mbox{per}}$} 
&\multirow{2}{*}{0.95} &1.00 &1.00 &0.00  &1.00 &0.96 &0.95 &0.97\\
& &0.103 (0.004) &0.090 (0.004) &0.102 (0.005) &0.090 (0.004) &0.018 (0.002) &0.018 (0.002) &0.018 (0.002) \\
&\multirow{2}{*}{0.90} &1.00 &1.00 &0.00  &1.00 &0.92 &0.92 &0.93\\
& &0.086 (0.004) &0.074 (0.003) &0.084 (0.004) &0.074 (0.003) &0.015 (0.002) &0.015 (0.002) &0.015 (0.002) \\
&\multirow{2}{*}{0.75} &1.00 &1.00 &0.00  &1.00 &0.77 &0.75 &0.74\\
& &0.060 (0.002) &0.052 (0.002) &0.059 (0.002) &0.052 (0.002) &0.011 (0.001) &0.010 (0.001) &0.011 (0.001) \\
\cline{2-9}
Estimated SE $(\times 10^{-3})$ &\textbf{\emph{4.429}} &26.135 &22.589 &25.691 &22.715 &4.556 &4.527 &4.599 \\
(Estimated SE)/(true SE)  &- &5.901 &5.100 &5.801 &5.129 &1.029 &1.022 &1.039\\
\hline
\multicolumn{9}{c}{Ordinal} \\
\hline
\multirow{6}{*}{$\mbox{CI}_{\mbox{nor}}$} 
&\multirow{2}{*}{0.95} &1.00 &1.00 &- &- &- &0.92 &0.94\\
& &0.108 (0.004) &0.122 (0.004) &- &- &- &0.045 (0.003) &0.045 (0.003) \\
&\multirow{2}{*}{0.90} &0.99 &1.00 &- &- &- &0.89 &0.90\\
& &0.091 (0.003) &0.102 (0.003) &- &- &- &0.038 (0.002) &0.038 (0.002) \\
&\multirow{2}{*}{0.75} &0.97 &0.97 &- &- &- &0.80 &0.81\\
& &0.064 (0.002) &0.071 (0.002) &- &- &- &0.027 (0.002) &0.026 (0.002) \\
\cline{2-9}
\multirow{6}{*}{$\mbox{CI}_{\mbox{per}}$}
&\multirow{2}{*}{0.95} &1.00 &1.00 &- &- &- &0.94 &0.94\\
& &0.109 (0.005) &0.123 (0.005) &- &- &- &0.046 (0.003) &0.046 (0.003) \\
&\multirow{2}{*}{0.90} &0.99 &1.00 &- &- &- &0.90 &0.91\\
& &0.091 (0.004) &0.102 (0.004) &- &- &- &0.038 (0.002) &0.038 (0.003) \\
&\multirow{2}{*}{0.75} &0.97 &0.97 &- &- &- &0.82 &0.80\\
& &0.064 (0.002) &0.072 (0.003) &- &- &- &0.027 (0.002) &0.027 (0.002) \\
\cline{2-9}
Estimated SE $(\times 10^{-2})$ &\textbf{\emph{1.212}} &2.768 &3.104 &- &- &- &1.154 &1.152 \\
(Estimated SE)/(true SE)  &- &2.283 &2.560 &- &- &- &0.952 &0.950 \\
\bottomrule
\end{tabular}}
\end{table}

\n
{\bf SC2: Probit Models for Multivariate Predictors with Missing Term}

We consider the case of multivariate predictors by generating $n=2000$ observations from $\Phi^{-1} ( P(Y_i =1) )= 1+\sum_{j=1}^p (-1)^j x_{ij} - x_{i1}x_{i,10} + x_{i2}x_{i9}$ and $\textbf{x}_i \sim N_p(\textbf{0}, \boldsymbol{\Sigma})$. We set $\boldsymbol{\Sigma}= \textbf{I}$ for the independent case and $\Sigma_{ij} =0.5^{|i-j|}$ for the correlated case. The assumed model is $\Phi^{-1} ( P(Y_i =1) )= \beta_0 +\sum_{j=1}^p\beta_j x_{ij}$, which misses the interaction terms. 
Table \ref{Tab:cover_ci_s2} lists the ratio of standard error estimation and coverage rates.
As we can see, local residual bootstrap based on Pearson and surrogate residuals provides accurate standard error estimations and valid coverage probabilities, while local residual bootstrap based on SBS residuals and the local response bootstrap underestimate the standard error. The other bootstrap methods overestimate the standard errors and provide conservative confidence intervals. 

\begin{table}[!htp]
	\caption{Comparison of the coverage probabilities and average width (standard error) for the bootstrap confidence interval and standard error estimation under SC2. The true standard errors of the pseudo-true estimations are in bold italics. } \label{Tab:cover_ci_s2}
	\centering
\renewcommand\arraystretch{1.0}  
\resizebox{\linewidth}{!}{
		\begin{tabular}{c c c c c c c c c c} 
			\toprule
			&Confidence &Parametric &Pairwise &Wild &Multiplier &Local Response &LRB- &LRB- &LRB- \\ 
			&level &Bootstrap &Bootstrap &Bootstrap &Bootstrap &Bootstrap &Pearson &SBS &Surrogate\\
			\midrule
			\multicolumn{10}{c}{Independent predictors} \\
			\hline
			\multirow{6}{*}{$\mbox{CI}_{\mbox{nor}}$} 
			&\multirow{2}{*}{0.95} &1.00 &1.00 &1.00 &1.00 &0.82 &0.98 &0.83 &0.98 \\
                 & &0.179 (0.008) &0.201 (0.011) &0.171 (0.007) &0.207 (0.011) &0.111 (0.004) &0.143 (0.008) &0.112 (0.004) &0.145 (0.007) \\ 
			&\multirow{2}{*}{0.90} &1.00 &1.00 &0.98 &1.00 &0.73 &0.86 &0.73 &0.86 \\
                 & &0.150 (0.006) &0.169 (0.009) &0.144 (0.006) &0.174 (0.010) &0.093 (0.003) &0.120 (0.006) &0.094 (0.003) &0.122 (0.006) \\
			&\multirow{2}{*}{0.75} &0.83 &0.88 &0.79 &0.93 &0.59 &0.69 &0.59 &0.70 \\
                & &0.105 (0.004) &0.118 (0.006) &0.101 (0.004) &0.122 (0.007) &0.065 (0.002) &0.084 (0.004) &0.066 (0.002) &0.085 (0.004) \\
			\cline{2-10}
			Estimated SE $(\times 10^{-2})$ &\textbf{\emph{3.384}} &4.568 &5.129 &4.376 &5.295 &2.835 &3.656 &2.858 &3.705  \\ 
			(Estimated SE)/(true SE) &- &1.350 &1.516 &1.293 &1.565 &0.838 &1.080 &0.845 &1.095 \\
			\hline
			\multicolumn{10}{c}{Correlated predictors} \\
			\hline
			\multirow{6}{*}{$\mbox{CI}_{\mbox{nor}}$} 
			&\multirow{2}{*}{0.95} &0.99 &0.99 &0.99 &1.00 &0.85 &0.93 &0.74 &0.94 \\
              & &0.172 (0.006) &0.188 (0.008) &0.158 (0.006) &0.193 (0.009) &0.121 (0.003) &0.150 (0.006) &0.120 (0.005) &0.154 (0.007) \\
			&\multirow{2}{*}{0.90} &0.95 &0.99 &0.94 &0.99 &0.77 &0.90 &0.69 &0.89 \\
              & &0.144 (0.005) &0.158 (0.007) &0.133 (0.005) &0.162 (0.007) &0.102 (0.003) &0.126 (0.005) &0.101 (0.004) &0.129 (0.006) \\
			&\multirow{2}{*}{0.75} &0.82 &0.90 &0.82 &0.84 &0.62 &0.74 &0.50 &0.72 \\
                & &0.101 (0.004) &0.110 (0.005) &0.093 (0.004) &0.113 (0.005) &0.071 (0.002) &0.088 (0.004) &0.071 (0.003) &0.090 (0.004) \\
			\cline{2-10}
			Estimated SE $(\times 10^{-2})$ &\textbf{\emph{3.796}} &4.380 &4.804 &4.043 &4.930 &3.096 &3.839 &3.075 &3.931 \\ 
			(Estimated SE)/(true SE) &- &1.154 &1.265 &1.065 &1.299 &0.816 &1.011 &0.810 &1.036 \\
			\bottomrule
		\end{tabular}}
\end{table}

In this setting, we have introduced a variation of the proposed method, local response bootstrap, which locally resamples the responses instead of the residuals. 
The detailed algorithm is given in the appendix. 
However,  local residual bootstrap still compares favorably to local response bootstrap because the resampling responses require similar response distributions in the neighborhood, whereas resampling residuals only require a similar type of misspecification in the neighborhood. 
The response distribution is affected by many factors such as $\boldsymbol{\beta}$ and ${\bf x}_i$ and is unstable when $p$ is large.
Meanwhile, misspecification is much less sensitive in that regard.
Hence, resampling residuals introduce less bias than resampling responses.

To illustrate this point, we simulate the data according to $\Phi^{-1}(P(Y_i=1)) = {\bf z}_i^T \boldsymbol{\beta}$, where $\textbf{z}_i=(1,\textbf{x}_i^T,x_{i1}^2)^T$ and $\textbf{x}_i \sim N(\textbf{0}, \textbf{I}_{8 \times 8})$, and set $\boldsymbol{\beta} = (-2,1,-0.2,-0.1,0.1,-0.2,0.3,-0.5,-0.2,0.9)^T$. 
We fit both $\Phi^{-1}(P(Y_i=1)) = {\bf z}_i^T \boldsymbol{\beta}$ and $\Phi^{-1}(P(Y_i=1)) = {\bf x}_i^T \boldsymbol{\beta}^{\dagger}$ to the data. 

\begin{figure}[!htp]
	\centering
	\subfigure[Local response bootstrap under the correctly specified case]{
		\includegraphics[width=0.3\textwidth]{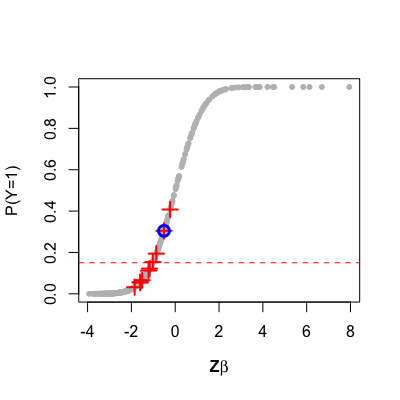}
	}
	\subfigure[Local residual bootstrap under the correctly specified case]{
		\includegraphics[width=0.3\textwidth]{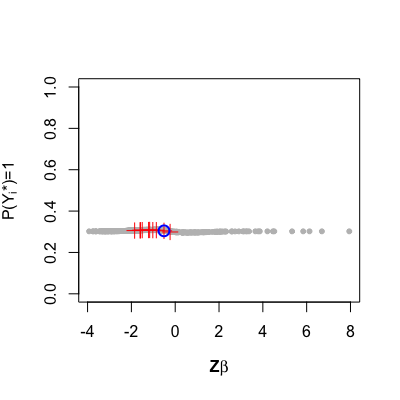}
	} 
	\subfigure[Local residual bootstrap under the misspecified case]{
		\includegraphics[width=0.3\textwidth]{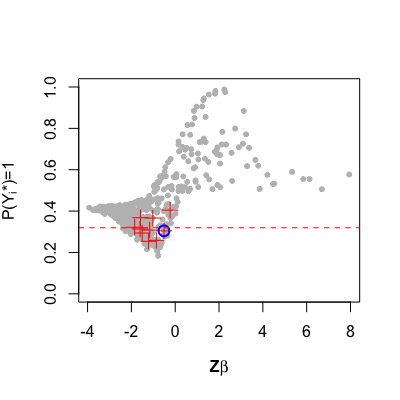}
	} 
	\caption{Comparison of local residual bootstrap with the local response bootstrap.
		Panel (a) shows $P(Y^*_{k}(j)=1)$ against ${\bf z}_j\boldsymbol{\beta}$ for the local response bootstrap under the correctly specified case. 
		Panels (b) and (c) show $P(Y^*_{k}(R_j)=1)$ against ${\bf z}_j\boldsymbol{\beta}$ for local residual bootstrap under the correctly specified model and misspecified model, respectively.
	}
	\label{Fig:Multi_change}
\end{figure}

Let us focus the $k$-th bootstrapped response $Y^*_{k}$ and consider the response and residuals as random variables.
For the local response bootstrap, we write $Y^*_{k}$ as $Y^*_{k}(j)$ if it is sampled as $Y_j$.
Figure \ref{Fig:Multi_change}a plots $P(Y^*_{k}(j)=1)$ against ${\bf z}_j\boldsymbol{\beta}$ for $j=1,...,n$. 
The blue circle is $(P(Y_{k}=1), {\bf z}_k\beta)$ and the red `+'s are $\{(P(Y^*_{k}(j)=1), {\bf z}_j\beta) \}_{j \in N_k}$.
The dashed line is $P(Y^*_{k}=1)=\sum_{j \in N_{k}} P(Y^*_{k}(j)=1)/l$.
As we can see, $P(Y^*_{k}=1)$ deviates from $P(Y_{k}=1)$ significantly, meaning the local response bootstrap cannot mimic the original response.
By contrast, for local residual bootstrap, we write $Y^*_{k}$ as $Y^*_{k}(r_j)$ if it is generated using $r_j$ (again as a random variable).
Figure \ref{Fig:Multi_change}b plots $P(Y^*_{k}(r_j)=1)$ against $\textbf{z}_j \boldsymbol{\beta}$ under the correctly specified case and Figure \ref{Fig:Multi_change}c shows that under the misspecified case.
The blue circle is still $(P(Y_{k}=1), {\bf z}_k\beta)$. The red `+'s are $\{(P(Y^*_{k}(r_j)=1), {\bf z}_j\beta) \}_{j \in N_k}$.
The dashed line is $P(Y^*_{k}=1)=\sum_{j \in N_{k}} P(Y^*_{k}(r_j)=1)/l$.
When the model is correctly specified, all the residuals follow the same distribution, so $P(Y^*_{k}=1) = P(Y_{k}=1)$.
When the model is misspecified, $P(Y^*_{k}(r_j)=1)$ is still close to $P(Y_{k}=1)$.
Therefore, local residual bootstrap does mimic the original response well.
Overall, local residual bootstrap performs better than the local response bootstrap under both the correctly specified and the misspecified cases.

\n
{\bf SC3--SC6: Missing Predictors, Misspecified Mean Structures, Mixed Populations, and Categorical Predictors} 

SC3 considers the case of missing predictors. We generate data $\{ y_i, \textbf{x}_i \}_{i=1}^n$ with $n=2000$ according to $\Phi^{-1} ( P(Y_i=1) ) = -1+2x_{i1}+2x_{i2}$, where $(x_{i1},v_i) \sim N({\bf 0}, \boldsymbol{\Sigma})$, and $x_{i2}=\exp(v_i)$. We consider the case of independent predictors with $\boldsymbol{\Sigma}=\textbf{I}$ and that of correlated predictors with $\Sigma_{12}=0.7$. The assumed model is $\Phi^{-1} ( P(Y_i=1) ) = \beta_0^\dagger +\beta_1^\dagger x_{i1}$, which misses $x_{i2}$. 
Table \ref{Tab:cover_ci_s3_2} lists the ratio of standard error estimations and coverage probabilities.

\begin{table}[!htp]
	\caption{Comparison of the coverage probabilities and average width (standard error) for the bootstrap confidence interval and standard error estimation under SC3, SC4, SC5, and SC6. The true standard errors of the pseudo-true estimations are in bold italics. }\label{Tab:cover_ci_s3_2}
	\centering
\renewcommand\arraystretch{1.0}  
\resizebox{\linewidth}{!}{
		\begin{tabular}{c c c c c c c c c} 
			\toprule
			&Confidence &Parametric &Pairwise &Wild &Multiplier &LRB- &LRB- &LRB- \\ 
			&level &Bootstrap &Bootstrap &Bootstrap &Bootstrap &Pearson &SBS &Surrogate\\
			\midrule
			\multicolumn{9}{c}{SC3: Independent} \\
			\hline
			\multirow{6}{*}{$\mbox{CI}_{\mbox{nor}}$} 
			&\multirow{2}{*}{0.95} &1.00 &1.00 &0.97 &1.00  &1.00 &1.00 &1.00\\
               & &0.179 (0.007) &0.178 (0.007) &0.143 (0.006) &0.180 (0.008) &0.160 (0.007) &0.161 (0.007) &0.161 (0.007) \\
			&\multirow{2}{*}{0.90} &0.98 &0.97 &0.94 &0.99  &0.96 &0.97 &0.96\\
              & &0.150 (0.006) &0.149 (0.006) &0.120 (0.005) &0.151 (0.007) &0.135 (0.006) &0.135 (0.006) &0.135 (0.006) \\
			&\multirow{2}{*}{0.75} &0.90 &0.91 &0.82 &0.92  &0.88 &0.88 &0.88\\
               & &0.105 (0.004) &0.104 (0.004) &0.084 (0.003) &0.106 (0.005) &0.094 (0.004) &0.095 (0.004) &0.094 (0.004) \\
			\cline{2-9}
			Estimated SE $(\times 10^{-2})$ &\textbf{\emph{3.076}} &4.559 &4.537 &3.664 &4.607 &4.093 &4.122 &4.107  \\
			(Estimated SE)/(true SE) &-  &1.482 &1.475 &1.191 &1.498 &1.331 &1.340 &1.335 \\
			\hline
			\multicolumn{9}{c}{SC3: Correlated} \\
			\hline
			\multirow{6}{*}{$\mbox{CI}_{\mbox{nor}}$} 
			&\multirow{2}{*}{0.95} &0.95 &0.96 &0.91 &0.96  &0.94 &0.97 &0.94\\
                & &0.336 (0.018) &0.339 (0.024) &0.277 (0.019) &0.343 (0.026) &0.307 (0.021) &0.326 (0.064) &0.309 (0.021) \\
			&\multirow{2}{*}{0.90} &0.92 &0.93 &0.86 &0.93  &0.89 &0.94 &0.91\\
                & &0.282 (0.015) &0.285 (0.020) &0.232 (0.016) &0.287 (0.022) &0.258 (0.017) &0.273 (0.053) &0.260 (0.018) \\
			&\multirow{2}{*}{0.75} &0.80 &0.80 &0.73 &0.78  &0.75 &0.84 &0.76\\
                & &0.197 (0.011) &0.199 (0.014) &0.162 (0.011) &0.201 (0.015) &0.180 (0.012) &0.191 (0.037) &0.182 (0.012) \\
			\cline{2-9}
			Estimated SE $(\times 10^{-2})$ &\textbf{\emph{7.823}} &8.589 &8.660 &7.067 &8.747 &7.844 &8.313 &7.904 \\
			(Estimated SE)/(true SE) &- &1.098 &1.107 &0.903 &1.118 &1.003 &1.063 &1.010\\
			\hline
			\multicolumn{9}{c}{SC4: Exponential} \\
			\hline
			\multirow{3}{*}{Coverage rate of $\mbox{CI}_{\mbox{nor}}$} 
			&\multirow{2}{*}{0.95} &0.76 &0.95 &0.92  &0.96  &0.94 &0.94 &0.93 \\
            & &0.168 (0.021) &0.295 (0.059) &0.235 (0.046) &0.444 (0.251)   &0.276 (0.056) &0.275 (0.057) &0.278 (0.057) \\
			&\multirow{2}{*}{0.90} &0.67 &0.92 &0.84  &0.93  &0.90 &0.90 &0.91 \\
            & &0.141 (0.018) &0.248 (0.050) &0.197 (0.038) &0.373 (0.210)   &0.232 (0.047) &0.231 (0.048) &0.233 (0.048) \\
			&\multirow{2}{*}{0.75} &0.48 &0.79 &0.64  &0.86  &0.73 &0.73 &0.73 \\
            & &0.099 (0.013) &0.173 (0.035) &0.138 (0.027) &0.261 (0.147)  &0.162 (0.033) &0.162 (0.033) &0.163 (0.033) \\
			\cline{2-9}
			Estimated SE $(\times 10^{-2})$ &\textbf{\emph{7.145}} &4.186 &7.535 &5.997  &11.336  &7.045  &7.031 &7.087  \\
			(Estimated SE)/(true SE) &- &0.600 &1.055 &0.839 &1.530 &0.986 &0.984 &0.992 \\
			\hline
			\multicolumn{9}{c}{SC4: Sine} \\
			\hline
			\multirow{3}{*}{Coverage rate of $\mbox{CI}_{\mbox{nor}}$} 
			&\multirow{2}{*}{0.95} &1.00 &1.00 &1.00 &1.00  &0.93 &0.92 &0.92\\
            & &0.035 (0.001) &0.029 (0.001) &0.027 (0.001) &0.029 (0.001)   &0.012 (0.001) &0.012 (0.001) &0.012 (0.001) \\
			&\multirow{2}{*}{0.90} &1.00 &1.00 &1.00 &1.00  &0.83 &0.82 &0.84\\
            & &0.029 (0.001) &0.024 (0.001) &0.022 (0.001) &0.024 (0.001)  &0.010 (0.001) &0.010 (0.001) &0.010 (0.001) \\
			&\multirow{2}{*}{0.75} &0.99 &0.99 &0.97 &1.00  &0.66 &0.68 &0.69\\
            & &0.021 (0.001) &0.017 (0.001) &0.016 (0.001) &0.017 (0.001)  &0.007 (0.000) &0.007 (0.000) &0.007 (0.000) \\ 
			\cline{2-9}
			Estimated SE $(\times 10^{-3})$ &\textbf{\emph{3.514}} &8.959 &7.281 &6.808 &8.354 &3.038 &3.050 &3.100 \\
			(Estimated SE)/(true SE) &-  &2.549 &2.072 &1.937 &2.377 &0.864 &0.868 &0.882\\
			\hline
			\multicolumn{9}{c}{SC5: Slope} \\
			\hline
			\multirow{3}{*}{Coverage rate of $\mbox{CI}_{\mbox{nor}}$} 
			&\multirow{2}{*}{0.95} &1.00 &1.00 &1.00  &1.00  &0.95 &0.93 &0.94 \\
            & &0.032 (0.001) &0.033 (0.001) &0.034 (0.001) &0.033 (0.001)  &0.007 (0.000) &0.007 (0.000) &0.007 (0.001) \\
			&\multirow{2}{*}{0.90} &1.00 &1.00 &1.00  &1.00  &0.90 &0.90 &0.91 \\
             & &0.027 (0.001) &0.028 (0.001) &0.029 (0.001) &0.028 (0.001)  &0.006 (0.000) &0.006 (0.000) &0.006 (0.000) \\
			&\multirow{2}{*}{0.75} &1.00 &1.00 &1.00  &1.00  &0.75 &0.75 &0.76 \\
            & &0.019 (0.001) &0.019 (0.001) &0.020 (0.001) &0.019 (0.001)   &0.004 (0.000) &0.004 (0.000) &0.004 (0.000) \\
			\cline{2-9}
			Estimated SE $(\times 10^{-3})$ &\textbf{\emph{2.045}} &8.287 &8.404 &8.752 &8.746 &1.845 &1.850 &1.860  \\
			(Estimated SE)/(true SE) &- &4.053 &4.110 &4.280 &4.277  &0.960 &0.960 &0.962 \\
			\hline
			\multicolumn{9}{c}{SC5: Intercept and slope} \\
			\hline
			\multirow{3}{*}{Coverage rate of $\mbox{CI}_{\mbox{nor}}$} 
			&\multirow{2}{*}{0.95} &1.00 &1.00 &1.00 &1.00  &0.93 &0.93 &0.94\\
            & &0.032 (0.001) &0.032 (0.001) &0.032 (0.001) &0.032 (0.001)   &0.004 (0.000) &0.004 (0.000) &0.004 (0.000) \\
			&\multirow{2}{*}{0.90} &1.00 &1.00 &1.00 &1.00  &0.91 &0.90 &0.90\\
            &  &0.027 (0.001) &0.027 (0.001) &0.027 (0.001) &0.027 (0.001)   &0.004 (0.000) &0.004 (0.000) &0.004 (0.000) \\
			&\multirow{2}{*}{0.75} &1.00 &1.00 &1.00 &1.00  &0.73 &0.71 &0.73\\
            & &0.019 (0.001) &0.019 (0.001) &0.019 (0.001) &0.019 (0.000)  &0.003 (0.000) &0.003 (0.000) &0.003 (0.000) \\
			\cline{2-9}
			Estimated SE $(\times 10^{-3})$ &\textbf{\emph{1.168}} &8.147 &8.121 &8.073 &7.639 &1.101 &1.102 &1.103 \\
			(Estimated SE)/(true SE) &- &6.976 &6.954 &6.912 &6.540 &0.942 &0.943 &0.945\\
			\hline
			\multicolumn{9}{c}{SC6: Categorical predictors} \\
			\hline
			\multirow{3}{*}{Coverage rate of $\mbox{CI}_{\mbox{nor}}$} 
			&\multirow{2}{*}{0.95} &1.00 &0.99 &1.00 &1.00 &0.95 &0.95 &0.95\\
            & &0.368 (0.025) &0.265 (0.011) &0.334 (0.015) &0.269 (0.010)  &0.184 (0.014) &0.185 (0.014) &0.185 (0.014) \\
			&\multirow{2}{*}{0.90} &0.99 &0.97 &0.99 &1.00 &0.90 &0.90 &0.89\\
            & &0.309 (0.013) &0.223 (0.010) &0.280 (0.012) &0.226 (0.008)  &0.155 (0.012) &0.155 (0.012) &0.155 (0.012) \\
			&\multirow{2}{*}{0.75} &0.96 &0.90 &0.95 &0.93 &0.75 &0.79 &0.75\\
            & &0.216 (0.009) &0.156 (0.007) &0.196 (0.009) &0.158 (0.006)  &0.108 (0.008) &0.109 (0.008) &0.108 (0.008) \\
			\cline{2-9}
			Estimated SE $(\times 10^{-2})$ &\textbf{\emph{4.835}} &9.394 &6.777 &8.533 &7.547 &4.711 &4.724 &4.715 \\
			(Estimated SE)/(true SE) &-  &1.943 &1.402 &1.765 &1.561  &0.974 &0.977 &0.975\\
			\bottomrule
		\end{tabular}}
\end{table}

SC4 considers the case of the misspecified mean structure. We generate data with $n=2000$ under two non-linear mean structures: exponential and sine. 
Under the exponential case, the data-generating process is $\Phi^{-1} ( P(Y_i=1) ) = -2+4 \exp(x_i)$, where $x_i \sim \mbox{Unif}(-6,6)$.
Under the sine case, the data-generating process is $\Phi^{-1} ( P(Y_i=1) ) = 5 \sin(x_i)$.
In both cases, we fit the model $\Phi^{-1} ( P(Y_i=1) ) = \beta_0 +\beta_1 x_i$. 
Table \ref{Tab:cover_ci_s3_2} lists the results.

SC5 considers the case of multiple subpopulations within which the response depends on the predictors differently. We generate data with $n=2000$ under two cases. In the case of different slopes, the data-generating process is $\Phi^{-1} ( P(Y_i=1|u_i) )= -2+x_i$ if $u_i=0$ and $\Phi^{-1} ( P(Y_i=1|u_i) )= -2-x_i$ if $u_i=1$.
In the case of different slopes and intercepts, the data-generating process is $\Phi^{-1} ( P(Y_i=1|u_i) )= -1+x_i$ if $u_i=0$ and $\Phi^{-1} ( P(Y_i=1|u_i) )= 1-x_i$ if $u_i=1$, where $x_i \sim \textup{Unif}(-6,6)$ and $u_i \sim \textup{Bern}(0.5)$.
In these two cases, the assumed model is $\Phi^{-1} ( P(Y_i=1) )=\beta_0 + \beta_1 x_i + \beta_2 u_i$, which misses the interaction term $x_iu_i$ in the latter case and  misses the interaction term but has the extra predictor $u_i$ in the former case. Table \ref{Tab:cover_ci_s3_2} lists the results. 

SC6 considers that all the predictors are categorical. We generate the data $n=2000$ according to $\Phi^{-1} ( P(Y_i=1) ) = 1 + x_{i1} - x_{i2} -4 x_{i1} x_{i2}$, where $x_{i1} \sim \mbox{Bern}(0.2)$ and $x_{i2} \sim \mbox{Bern}(0.8)$. The assumed model is $\Phi^{-1} ( P(Y_i=1) ) = \beta_0^\dagger + \beta_1^\dagger x_{i1} + \beta_2^\dagger x_{i2}$, which misses the interaction term. Table \ref{Tab:cover_ci_s3_2} lists the results. 

As shown in Table \ref{Tab:cover_ci_s3_2}, the proposed bootstrap method performs better than the others in most simulation settings with the ratio closer to one and coverage probabilities for the confidence intervals closer to the nominal level. However, the alternative methods underestimate the standard errors and cannot contain the pseudo-true value with the given confidence levels in the exponential case in SC4; moreover, they overestimate the standard errors and provide conservative confidence intervals in the sine case in SC4, SC5, and SC6. In SC3, all the methods overestimate the standard errors and provide wide confidence intervals in the independent case. 
Performance improves in the correlated case, especially for local residual bootstrap, which indicates that the correlation between $x_1$ and $x_2$ helps preserve the information of $x_2$ in the bootstrap sample. 

\n
{\bf SC7: Extra-binomial Variation}

In this setting, we consider the presence of extra-binomial variation. The setting is similar to the data generation in \cite{Friedl1997Variance}. We generate data $\{ y_i, x_i \}_{i=1}^n$ with $n=200$ from $Y_i \sim \mbox{Binomial}(m, p_i)$, where $m=10$ and $p_i \sim \mbox{Beta}(2\mu_i, 2(1-\mu_i))$. With the logit link, $\mu_i=\exp(-2+2x_i)/(1+\exp(-2+2x_i))$, where $x \sim \mbox{Unif}(0,2)$. In terms of the correlated-binomial model, this setting gives a common correlation $\rho=1/3$, and the global dispersion factor is $\sigma^2=4$ with the consideration of overdispersion. 
We fit a binary logistic model for this data.
The ratio of standard error estimation for binary model and binomial model are listed in Table \ref{Tab:cover_ci_ss3}.
Compared with the alternatives, the proposed bootstrap can a better performance with ratio of standard error closer to 1 and confidence interval closer to the nominal level.

\begin{table}[!htp]
\caption{Comparison of the coverage probabilities and average width (standard error) for the bootstrap confidence interval and standard error estimation under SC7. The true standard errors of the pseudo-true estimations are in bold italics.} \label{Tab:cover_ci_ss3}
\centering
\renewcommand\arraystretch{1.0}  
\resizebox{\linewidth}{!}{
\begin{tabular}{c c c c c c c c c} 
\toprule
&Confidence &Parametric &Pairwise &Wild &Multiplier &LRB- &LRB- &LRB- \\ 
&level &Bootstrap &Bootstrap &Bootstrap &Bootstrap &Pearson &SBS &Surrogate\\
\midrule
\multirow{5}{*}{Coverage rate of $\mbox{CI}_{\mbox{nor}}$} 
&\multirow{2}{*}{0.99} &0.99 &0.99 &0.95 &0.99 &0.96 &0.96 &0.95\\
& &0.504 (0.019) &0.514 (0.019) &0.373 (0.015) &0.513 (0.018) &0.399 (0.019) &0.397 (0.019) &0.401 (0.020) \\
&\multirow{2}{*}{0.95} &0.97 &0.96 &0.92 &0.97 &0.94 &0.93 &0.93\\
& &0.384 (0.015) &0.391 (0.015) &0.284 (0.011) &0.391 (0.013) &0.304 (0.015) &0.302 (0.014) &0.305 (0.015) \\
&\multirow{2}{*}{0.90} &0.93 &0.93 &0.85 &0.94 &0.87 &0.88 &0.88\\
& &0.322 (0.012) &0.328 (0.012) &0.238 (0.010) &0.328 (0.011) &0.255 (0.012) &0.254 (0.012) &0.256 (0.013) \\
&\multirow{2}{*}{0.80} &0.87 &0.87 &0.76 &0.87 &0.78 &0.78 &0.79\\
& &0.251 (0.010) &0.256 (0.010) &0.185 (0.007) &0.255 (0.009) &0.199 (0.010) &0.198 (0.009) &0.200 (0.010) \\
&\multirow{2}{*}{0.75} &0.82 &0.82 &0.69 &0.82 &0.74 &0.73 &0.73\\
& &0.225 (0.009) &0.230 (0.009) &0.166 (0.007) &0.229 (0.008) &0.178 (0.009) &0.177 (0.008) &0.179 (0.009) \\
\cline{2-9}
\multirow{5}{*}{Coverage rate of $\mbox{CI}_{\mbox{per}}$} 
&\multirow{2}{*}{0.99} &0.99 &0.99 &0.92 &0.98 &0.97 &0.97 &0.98\\
& &0.517 (0.035) &0.529 (0.038) &0.386 (0.028) &0.525 (0.033) &0.414 (0.032) &0.409 (0.031) &0.414 (0.032) 
\\
&\multirow{2}{*}{0.95} &0.97 &0.96 &0.81 &0.97 &0.94 &0.93 &0.93\\
& &0.387 (0.019) &0.395 (0.018) &0.287 (0.014) &0.396 (0.016) &0.305 (0.016) &0.306 (0.017) &0.307 (0.016) \\
&\multirow{2}{*}{0.90} &0.93 &0.93 &0.75 &0.93 &0.87 &0.88 &0.88\\
& &0.322 (0.015) &0.327 (0.015) &0.237 (0.012) &0.328 (0.013) &0.254 (0.013) &0.255 (0.012) &0.256 (0.013) 
\\
&\multirow{2}{*}{0.80} &0.87 &0.87 &0.62 &0.86 &0.77 &0.78 &0.78\\
& &0.251 (0.012) &0.256 (0.012) &0.185 (0.009) &0.254 (0.011) &0.199 (0.011) &0.198 (0.010) &0.200 (0.010) \\
&\multirow{2}{*}{0.75} &0.80 &0.85 &0.59 &0.81 &0.75 &0.74 &0.72\\
& &0.226 (0.010) &0.231 (0.011) &0.167 (0.008) &0.229 (0.011) &0.179 (0.010) &0.178 (0.010) &0.180 (0.011) \\
\cline{2-9}
Estimated SE $(\times 10^{-2})$ &\textbf{\emph{8.624}} &9.794 &9.990 &7.243 &9.972 &7.756 &7.721 &7.792\\
(Estimated SE)/(true SE) &- &1.136 &1.158 &0.839 &1.156 &0.899 &0.895 &0.904\\
\bottomrule
\end{tabular}}
\end{table}

\n
{\bf SC8: Poisson Models with Missing Term}

In this setting, we consider the case where miss the quadratic term in the poisson model. The data-generating process and the assumed model are as follows:
\[
\begin{array}{l l}
\mbox{Data-generating process} & \log( E(Y_i) ) = 4+2x_i-x_i^2, \quad x_i \sim \mbox{Unif}(-6, 6)\\
\mbox{Assumed model} & \log( E(Y_i) ) = \beta_0^\dagger+\beta_1^\dagger x_i.
\end{array}
\]
We conduct LRB procedure based on Pearson residuals since the surrogate residuals and SBS residuals are not available in the poisson model.
The ratio of standard error estimation and coverage probabilities are listed in Table \ref{Tab:cover_ci_poisson}.
In this case, local residual bootstrap based on Pearson residuals provides accurate standard error estimations and valid coverage probabilities, whereas other bootstrap methods overestimate the standard errors and provide conservative confidence intervals, especially for pairwise bootstrap, wild bootstrap and multiplier bootstrap.

\begin{table}[!htp]
\caption{Comparison of the coverage probabilities and average width (standard error) for the bootstrap confidence interval and standard error estimation under SC8. The true standard errors of the pseudo-true estimations are in bold italics.} \label{Tab:cover_ci_poisson}
\centering
\renewcommand\arraystretch{1.0}  
\resizebox{\linewidth}{!}{
\begin{tabular}{c c c c c c c} 
\toprule
&Confidence &Parametric &Pairwise &Wild &Multiplier &LRB- \\ 
&level &Bootstrap &Bootstrap &Bootstrap &Bootstrap &Pearson \\
\midrule
\multirow{5}{*}{Coverage rate of $\mbox{CI}_{\mbox{nor}}$} 
&\multirow{2}{*}{0.99} &1.00 &1.00 &1.00 &1.00 &1.00\\
& &0.016 (0.001) &0.078 (0.002) &0.077 (0.003) &0.077 (0.003) &0.009 (0.000) \\
&\multirow{2}{*}{0.95} &1.00 &1.00 &1.00 &1.00 &0.94\\
& &0.012 (0.000) &0.059 (0.002) &0.059 (0.002) &0.059 (0.002) &0.007 (0.000) \\
&\multirow{2}{*}{0.90} &1.00 &1.00 &1.00 &1.00 &0.88\\
& &0.010 (0.000) &0.050 (0.002) &0.049 (0.002) &0.049 (0.002) &0.006 (0.000) \\
&\multirow{2}{*}{0.80} &0.96 &1.00 &1.00 &1.00 &0.81\\
& &0.008 (0.000) &0.039 (0.001) &0.038 (0.001) &0.038 (0.001) &0.005 (0.000) \\
&\multirow{2}{*}{0.75} &0.93 &1.00 &1.00 &1.00 &0.76\\
& &0.007 (0.000) &0.035 (0.001) &0.035 (0.001) &0.035 (0.001) &0.004 (0.000) \\
\cline{2-7}
\multirow{5}{*}{Coverage rate of $\mbox{CI}_{\mbox{per}}$} 
&\multirow{2}{*}{0.99} &1.00 &1.00 &1.00 &1.00 &0.98\\
& &0.016 (0.001) &0.081 (0.005) &0.079 (0.005) &0.080 (0.005) &0.010 (0.001) \\
&\multirow{2}{*}{0.95} &1.00 &1.00 &1.00 &1.00 &0.94\\
& &0.012 (0.001) &0.059 (0.002) &0.059 (0.002) &0.059 (0.003) &0.007 (0.000) \\
&\multirow{2}{*}{0.90} &0.99 &1.00 &1.00 &1.00 &0.85\\
& &0.010 (0.000) &0.049 (0.002) &0.050 (0.002) &0.049 (0.002) &0.006 (0.000) \\
&\multirow{2}{*}{0.80} &0.96 &1.00 &1.00 &1.00 &0.76\\
& &0.008 (0.000) &0.039 (0.001) &0.039 (0.001) &0.038 (0.002) &0.005 (0.000) \\
&\multirow{2}{*}{0.75} &0.92 &1.00 &1.00 &1.00 &0.72\\
& &0.007 (0.000) &0.035 (0.001) &0.035 (0.001) &0.035 (0.001) &0.004 (0.000) \\
\cline{2-7}
Estimated SE $(\times 10^{-3})$ &\textbf{\emph{1.973}} &3.062 &15.093 &15.032 &15.027 &1.820 \\
(Estimated SE)/(true SE) &- &1.552 &7.651 &7.621 &7.618 &0.923 \\
\bottomrule
\end{tabular}}
\end{table}

\n
{\bf SC9: Gamma Models with Missing Term}

In this setting, we consider the case where miss the quadratic term in the gamma model. The data-generating process and the assumed model are as follows, and the inverse link is used here, that is, $h(t)=1/t$:
\[
\begin{array}{l l}
\mbox{Data-generating process} &  E(Y_i) = h(\exp(-2 - x_i + x_i^2)), \quad x_i \sim \mbox{Unif}(0, 1)\\
\mbox{Assumed model} & E(Y_i) = h( \beta_0^\dagger+\beta_1^\dagger x_i).
\end{array}
\]
The ratio of standard error estimation and coverage probabilities are listed in Table \ref{Tab:cover_ci_gamma}.
In this case, local residual bootstrap based on Pearson residuals provides accurate standard error estimations and valid coverage probabilities, whereas parameter bootstrap, pairwise bootstrap and multiplier bootstrap slightly overestimate the standard errors and provide conservative confidence intervals. 
Wild bootstrap has accurate estimate of standard errors but biased estimate of distributions.

\begin{table}[!htp]
\caption{Comparison of the coverage probabilities and average width (standard error) for the bootstrap confidence interval and standard error estimation under SC9. The true standard errors of the pseudo-true estimations are in bold italics.} \label{Tab:cover_ci_gamma}
\centering
\renewcommand\arraystretch{1.0}  
\resizebox{\linewidth}{!}{
\begin{tabular}{c c c c c c c} 
\toprule
&Confidence &Parametric &Pairwise &Wild &Multiplier &LRB- \\ 
&level &Bootstrap &Bootstrap &Bootstrap &Bootstrap &Pearson \\
\midrule
\multirow{5}{*}{Coverage rate of $\mbox{CI}_{\mbox{nor}}$} 
&\multirow{2}{*}{0.99} &1.00 &1.00 &0.99 &1.00 &0.99\\
& &0.025 (0.001) &0.028 (0.001) &0.018 (0.001) &0.028 (0.001) &0.018 (0.001) \\
&\multirow{2}{*}{0.95} &0.99 &1.00 &0.93 &1.00 &0.93\\
& &0.019 (0.001) &0.021 (0.001) &0.014 (0.001) &0.021 (0.001) &0.014 (0.001) \\
&\multirow{2}{*}{0.90} &0.95 &0.99 &0.92 &0.99 &0.91\\
& &0.016 (0.001) &0.018 (0.001) &0.011 (0.000) &0.018 (0.001) &0.011 (0.001) \\
&\multirow{2}{*}{0.80} &0.93 &0.94 &0.80 &0.94 &0.79\\
& &0.013 (0.000) &0.014 (0.001) &0.009 (0.000) &0.014 (0.001) &0.009 (0.001) \\
&\multirow{2}{*}{0.75} &0.92 &0.92 &0.77 &0.92 &0.75\\
& &0.011(0.000) &0.012(0.001) &0.008(0.000) &0.012(0.001) &0.008(0.000) \\
\cline{2-7}
\multirow{5}{*}{Coverage rate of $\mbox{CI}_{\mbox{per}}$} 
&\multirow{2}{*}{0.99} &1.00 &1.00 &0.24 &1.00 &1.00\\
& &0.026 (0.002) &0.028 (0.002) &0.019 (0.001) &0.029 (0.002) &0.019 (0.001) \\
&\multirow{2}{*}{0.95} &1.00 &1.00 &0.05 &1.00 &0.95\\
& &0.019 (0.001) &0.021 (0.001) &0.014 (0.001) &0.021 (0.001) &0.014 (0.001) \\
&\multirow{2}{*}{0.90} &0.97 &1.00 &0.01 &0.99 &0.87\\
& &0.016 (0.001) &0.018 (0.001) &0.011 (0.001) &0.018 (0.001) &0.011 (0.001) \\
&\multirow{2}{*}{0.80} &0.94 &0.96 &0.00 &0.96 &0.76\\
& &0.013 (0.001) &0.014 (0.001) &0.009 (0.000) &0.014 (0.001) &0.009 (0.001) \\
&\multirow{2}{*}{0.75} &0.88 &0.94 &0.00 &0.95 &0.71\\
& &0.011 (0.000) &0.012 (0.001) &0.008 (0.000) &0.012 (0.001) &0.008 (0.001) \\
\cline{2-7}
Estimated SE $(\times 10^{-3})$ &\textbf{\emph{3.672}} &4.893 &5.354 &3.471 &5.389 &3.474 \\
(Estimated SE)/(true SE) &- &1.333 &1.458 &0.945 &1.467 &0.946 \\
\bottomrule
\end{tabular}}
\end{table}

\n
{\bf SC10: Linear Models with Missing Interaction Term}

In this setting, we consider multivariate regression model. We generate the data $\{ y_i, x_i \}_{i=1}^n$ with $n=2000$. We adopt the setting from \citet{Kline2012Higher}. The data-generating process and the assumed model are as follows:
\[ 
\begin{array}{l l}
\mbox{Data-generating process} & y_i = x_{i1} + x_{i2} + x_{i3} + 0.5x_{i1}x_{i2} + \epsilon_i, \\
\mbox{Assumed model} & y_i = \beta_0^\dagger + \beta_1^\dagger x_{i1} + \beta_2^\dagger x_{i2} + \beta_3^\dagger x_{i3} + \epsilon_i.
\end{array}
\]
where $(v, x_2, x_3)$ follows multivariate normal with an equal correlation of 0.2, and $x_1 = \frac{\exp(v)-\mathbb{E}(\exp(v))}{\var(\exp(v))^{1/2}}$. The error $\epsilon$ is generated independently of regressors as the mixture of a $N(-\frac{1}{9}, 1)$ variable with probability 0.9 and a $N(1,4)$ variable with probability 0.1. The assumed model omits the interaction term. The results are listed in Table \ref{Tab:cover_ci_ss5}. The proposed bootstrap can give effective standard error estimation and coverage probabilities. Parametric bootstrap slightly overestimate the standard errors and the confidence intervals is conservative. The other bootstrap methods overestimate the standard errors more significantly and provide wider bootstrap confidence intervals.

\begin{table}[!htp]
\caption{Comparison of the coverage probabilities and average width (standard error) for the bootstrap confidence interval and standard error estimation under SC10. The true standard errors of the pseudo-true estimations are in bold italics.} \label{Tab:cover_ci_ss5}
\centering
\renewcommand\arraystretch{1.0}  
\resizebox{\linewidth}{!}{
\begin{tabular}{c c c c c c c}
\toprule
&Confidence &Parametric &Pairwise &Wild &Multiplier &\multirow{2}{*}{LRB} \\
&level &Bootstrap &Bootstrap &Bootstrap &Bootstrap  & \\
\midrule
\multirow{5}{*}{Coverage rate of $\mbox{CI}_{\mbox{nor}}$} 
&\multirow{2}{*}{0.99}  &1.00 &1.00 &1.00 &1.00 &0.99\\
& &0.153 (0.007) &0.299 (0.019) &0.301 (0.018) &0.307 (0.023) &0.132 (0.011) \\
&\multirow{2}{*}{0.95}  &0.98 &1.00 &1.00 &1.00 &0.92\\
& &0.116 (0.006) &0.228 (0.015) &0.229 (0.014) &0.234 (0.018) &0.100 (0.008) \\
&\multirow{2}{*}{0.90}  &0.94 &1.00 &1.00 &1.00 &0.89\\
& &0.098 (0.005) &0.191 (0.012) &0.192 (0.012) &0.196 (0.015) &0.084 (0.007) \\
&\multirow{2}{*}{0.80}  &0.86 &1.00 &1.00 &1.00 &0.78\\
& &0.076 (0.004) &0.149 (0.009) &0.150 (0.009) &0.153 (0.012) &0.065 (0.005) \\
&\multirow{2}{*}{0.75}  &0.81 &0.99 &0.99 &0.99 &0.75\\
& &0.068 (0.003) &0.134 (0.009) &0.135 (0.008) &0.137 (0.010) &0.059 (0.005) \\
\cline{2-7}
\multirow{5}{*}{Coverage rate of $\mbox{CI}_{\mbox{per}}$} 
&\multirow{2}{*}{0.99}  &1.00 &1.00 &1.00 &1.00 &0.98\\
& &0.149 (0.011) &0.289 (0.026) &0.284 (0.022) &0.301 (0.029) &0.126 (0.013) \\
&\multirow{2}{*}{0.96}  &0.97 &1.00 &1.00 &1.00 &0.95\\
& &0.116 (0.007) &0.224 (0.016) &0.225 (0.015) &0.233 (0.021) &0.099 (0.009) \\
&\multirow{2}{*}{0.90}  &0.93 &1.00 &1.00 &1.00 &0.89\\
& &0.098 (0.006) &0.191 (0.013) &0.193 (0.013) &0.196 (0.016) &0.084 (0.007) \\
&\multirow{2}{*}{0.80}  &0.85 &0.99 &1.00 &0.99 &0.80\\
& &0.076 (0.004) &0.149 (0.011) &0.151 (0.010) &0.152 (0.012) &0.066 (0.006) \\
&\multirow{2}{*}{0.75}  &0.83 &0.99 &1.00 &0.98 &0.75\\
& &0.068 (0.004) &0.133 (0.010) &0.135 (0.010) &0.135 (0.011) &0.059 (0.005) \\
\cline{2-7}
Estimated SE $(\times 10^{-2})$ &\textbf{\emph{2.548}} &2.975 &5.819 &5.860 &5.975 &2.558\\
(Estimated SE)/(true SE) &- &1.168 &2.284 &2.300 &2.345 &1.004\\
\bottomrule
\end{tabular}}
\end{table}

\n
{\bf SC11: Linear Models with Heteroscedasticity}

In this setting, we generate the data $\{ y_i, x_i \}_{i=1}^n$ with $n=2000$. The data-generating process and the assumed model are as follows:
\[ 
\begin{array}{l l}
\mbox{Data-generating process} & y_i = 1 + x_i + x_i^2  + \epsilon_i,  \quad x_i \sim \mbox{Unif}(0,1), \\
& \epsilon_i \sim N(0,\sigma_i^2), \sigma_i = (x_i-0.5)^2 \\
\mbox{Assumed model} & y_i = \beta_0^\dagger + \beta_1^\dagger x_i + \beta_2^\dagger x_i^2 + \epsilon_i.
\end{array}
\]
The assumed model is misspecified by ignoring the heteroscedasticity, since it corresponds to $\sigma_i\epsilon_i \sim N(0,1)$. The results are listed in Table \ref{Tab:cover_ci_ss6}. Under heteroscedastic case, parametric bootstrap underestimate the standard error and both of asymptotic normal confidence intervals and bootstrap percentile confidence intervals cannot contain the pseudo-true value with the given confidence levels. The other bootstrap methods can give accurate standard error estimation and valid confidence intervals. The wild bootstrap and the multiplier bootstrap are designed for this heteroscedasticity, however, the proposed method still performs slightly better in this case.

\begin{table}[!htp]
\caption{Comparison of the coverage probabilities and average width (standard error) for the bootstrap confidence interval and standard error estimation under SC11. The true standard errors of the pseudo-true estimations are in bold italics.} \label{Tab:cover_ci_ss6}
\centering
\renewcommand\arraystretch{1.0}  
\resizebox{\linewidth}{!}{
\begin{tabular}{c c c c c c c}
\toprule
&Confidence &Parametric &Pairwise &Wild &Multiplier &\multirow{2}{*}{LRB} \\
&level &Bootstrap &Bootstrap &Bootstrap &Bootstrap  & \\
\midrule
\multirow{5}{*}{Coverage rate of $\mbox{CI}_{\mbox{nor}}$} 
&\multirow{2}{*}{0.99}  &0.93 &1.00 &1.00 &1.00 &0.99\\
& &0.177 (0.007) &0.257 (0.013) &0.256 (0.013) &0.256 (0.012) &0.232 (0.010) \\
&\multirow{2}{*}{0.95}  &0.87 &0.94 &0.93 &0.94 &0.93\\
& &0.135 (0.005) &0.196 (0.010) &0.194 (0.010) &0.195 (0.009) &0.177 (0.007) \\
&\multirow{2}{*}{0.90}  &0.78 &0.92 &0.92 &0.92 &0.88\\
& &0.113 (0.005) &0.164 (0.008) &0.163 (0.008) &0.164 (0.008) &0.148 (0.006) \\
&\multirow{2}{*}{0.80}  &0.70 &0.85 &0.83 &0.84 &0.79\\
& &0.088 (0.004) &0.128 (0.007) &0.127 (0.006) &0.128 (0.006) &0.116 (0.005) \\
&\multirow{2}{*}{0.75}  &0.64 &0.81 &0.80 &0.80 &0.77\\
& &0.079 (0.003) &0.115 (0.006) &0.114 (0.006) &0.115 (0.005) &0.104 (0.004) \\
\cline{2-7}
\multirow{5}{*}{Coverage rate of $\mbox{CI}_{\mbox{per}}$} 
&\multirow{2}{*}{0.99}  &0.93 &1.00 &1.00 &1.00 &1.00\\
& &0.184 (0.012) &0.264 (0.019) &0.261 (0.019) &0.264 (0.018) &0.238 (0.016) \\
&\multirow{2}{*}{0.95}  &0.87 &0.96 &0.94 &0.96 &0.96\\
& &0.137 (0.006) &0.197 (0.011) &0.195 (0.011) &0.197 (0.011) &0.178 (0.010) \\
&\multirow{2}{*}{0.90}  &0.79 &0.91 &0.91 &0.92 &0.92\\
& &0.113 (0.005) &0.164 (0.009) &0.163 (0.009) &0.164 (0.008) &0.148 (0.008) \\
&\multirow{2}{*}{0.80}  &0.70 &0.87 &0.84 &0.84 &0.84\\
& &0.088 (0.004) &0.129 (0.007) &0.127 (0.007) &0.128 (0.007) &0.116 (0.005) \\
&\multirow{2}{*}{0.75}  &0.63 &0.81 &0.82 &0.79 &0.80\\
& &0.079 (0.004) &0.116 (0.007) &0.115 (0.006) &0.115 (0.006) &0.105 (0.005) \\
\cline{2-7}
Estimated SE $(\times 10^{-2})$ &\textbf{\emph{4.569}} &3.443 &5.003 &4.965 &4.982 &4.513\\
(Estimated SE)/(true SE) &-  &0.753 &1.095 &1.087 &1.090 &0.988\\
\bottomrule
\end{tabular}}
\end{table}

\n
{\bf SC12: Linear Models with Mixed Populations}

In this setting, we consider mixed populations for linear regression, that is, the data points are from two different linear models. We generate the data $\{ y_i, x_i \}_{i=1}^n$ with $n=2000$. The data-generating process and the assumed model are as follows:
\[
\begin{array}{l l}
\mbox{Data-generating process} & 
y_i|u_i = \left\{
\begin{array}{lcr}
-2+2x_i+\epsilon_i, & & \mbox{if } u_i=0\\
-2-2x_i+\epsilon_i,  & &  \mbox{if } u_i=1,
\end{array} \right. \\
& x_i \sim N(0,1), u_i \sim \mbox{Bern}(0.5), \epsilon_i \sim N(0,1) \\
\mbox{Assumed model} & y_i= \beta_0^\dagger + \beta_1^\dagger x_i + \beta_2^\dagger u_i + \epsilon_i. 
\end{array}
\]
We fit one model ignoring the mixed population. The results are listed in Table \ref{Tab:cover_ci_ss7}. The local residual bootstrap can give effective standard error estimation and confidence intervals, while other bootstrap methods still overestimate standard errors and give wide confidence intervals.

\begin{table}[!htp]
\caption{Comparison of the coverage probabilities and average width (standard error) for the bootstrap confidence interval and standard error estimation under SC12. The true standard errors of the pseudo-true estimations are in bold italics.} \label{Tab:cover_ci_ss7}
\centering
\renewcommand\arraystretch{1.0}  
\resizebox{\linewidth}{!}{
\begin{tabular}{c c c c c c c}
\toprule
&Confidence &Parametric &Pairwise &Wild &Multiplier &\multirow{2}{*}{LRB} \\
&level &Bootstrap &Bootstrap &Bootstrap &Bootstrap  & \\
\midrule
\multirow{5}{*}{Coverage rate of $\mbox{CI}_{\mbox{nor}}$} 
&\multirow{2}{*}{0.99}  &1.00 &1.00 &1.00 &1.00 &0.99\\
& &0.257 (0.009) &0.409 (0.014) &0.409 (0.015) &0.408 (0.015) &0.117 (0.005) \\
&\multirow{2}{*}{0.95}  &1.00 &1.00 &1.00 &1.00 &0.96\\
& &0.196 (0.007) &0.311 (0.011) &0.311 (0.011) &0.311 (0.011) &0.089 (0.004) \\
&\multirow{2}{*}{0.90}  &1.00 &1.00 &1.00 &1.00 &0.90\\
& &0.164 (0.006) &0.261 (0.009) &0.261 (0.009) &0.261 (0.01) &0.075 (0.003) \\
&\multirow{2}{*}{0.80}  &1.00 &1.00 &1.00 &1.00 &0.80\\
& &0.128 (0.004) &0.203 (0.007) &0.204 (0.007) &0.203 (0.007) &0.058 (0.003) \\
&\multirow{2}{*}{0.75}  &1.00 &1.00 &1.00 &1.00 &0.78\\
& &0.115 (0.004) &0.183 (0.006) &0.183 (0.007) &0.182 (0.007) &0.052 (0.002) \\
\cline{2-7}
\multirow{5}{*}{Coverage rate of $\mbox{CI}_{\mbox{per}}$} 
&\multirow{2}{*}{0.99}  &1.00 &1.00 &1.00 &1.00 &1.00\\
& &0.263 (0.018) &0.423 (0.030) &0.426 (0.028) &0.420 (0.026) &0.120 (0.009) \\
&\multirow{2}{*}{0.95}  &1.00 &1.00 &1.00 &1.00 &0.96\\
& &0.198 (0.009) &0.314 (0.015) &0.312 (0.015) &0.314 (0.015) &0.09 (0.005) \\
&\multirow{2}{*}{0.90}  &1.00 &1.00 &1.00 &1.00 &0.89\\
& &0.165 (0.007) &0.261 (0.011) &0.262 (0.011) &0.260 (0.012) &0.075 (0.004) \\
&\multirow{2}{*}{0.80}  &1.00 &1.00 &1.00 &1.00 &0.81\\
& &0.129 (0.006) &0.203 (0.008) &0.204 (0.009) &0.203 (0.008) &0.058 (0.003) \\
&\multirow{2}{*}{0.75}  &1.00 &1.00 &1.00 &1.00 &0.77\\
& &0.116 (0.005) &0.184 (0.008) &0.184 (0.008) &0.183 (0.008) &0.053 (0.003) \\
\cline{2-7}
Estimate SE $(\times 10^{-2})$ &\textbf{\emph{2.181}} &4.996 &7.944 &7.950 &7.932 &2.272 \\
(Estimated SE)/(true SE) &- &2.223 &3.643 &3.645 &3.637 &1.042 \\
\bottomrule
\end{tabular}}
\end{table}

\subsection{Proofs}

\n
\textbf{Proof of Theorem 1}

\begin{proof}

Suppose the true data generating process is $h^{-1}(m_i)=g_0(\tx_i, \bbeta_0)$ with the expectation $\mathbb{E}_g(Y_i)=m_i$ and the variance $\mbox{Var}_g(Y_i)=\sigma_i$. Suppose the true distribution function of $Y_i$ is $G_i(\cdot)$.
Let the assumed model be $h^{\prime-1}(\mu_i)=\tx_i\bbeta$, which may be misspecified with the pseudo-true parameter $\bbeta_{n,0}^\dagger$. 

For generalized linear models with surrogate residuals defined (other residuals have the similar argument),
we first consider $R_i$ as the surrogate residual under the pseudo-true model, which is available in the case of categorical response with a cumulative link.
The residual variable $R_i=S_i -\tx_i\bbetad$, where $S_i$ is the surrogate random variable following the truncated distribution of $Z_i$. 
The latent variable $Z_i$ follows the distribution related to the link function with mean $\tx_i\bbetad$.
For example, for the logit link, $Z_i$ follows the logistic distribution.
For the probit link, then $Z_i$ follows the normal distribution. 
Denote its distribution function as $F_i(\cdot)$.
If $Y_i=j$, it indicates that latent variable $\alpha_{j-1} < Z_i \leq \alpha_j$, $j=1,\dots,J$ and $-\infty = \alpha_0 < \alpha_1 < \cdots < \alpha_{j-1} < \alpha_j <\cdots <\alpha_J =\infty$, $S_i \sim Z_i |\alpha_{j-1} < Z_i \leq \alpha_j$.
 For any arbitrary but fixed $c$, suppose $\alpha_{k-1}-\tx_i\bbetad \leq c \leq  \alpha_k-\tx_i\bbetad$ where $1 \leq k \leq J$,
 the distribution of $R_i$ is
\begin{equation}\label{eq:part1disr}
\begin{split}
  P(R_i \leq c) 
& = P(S_i \leq c+\tx_i\bbetad) =\sum_{j=1}^J P(S_i \leq c+\tx_i\bbetad |Y_i=j)P(Y_i=j) \\
& = P(Y_i \leq k-1) + P(S_i \leq c+\tx_i\bbetad|Y_i=k)P(Y_i=k) \\
& = P(Y_i \leq k-1)+\frac{P(\alpha_{k-1} < Z_i \leq c+\tx_i\bbetad)}{P(\alpha_{k-1} < Z \leq \alpha_k)}P(Y_i=k) \\
& =G_i(k-1) + \frac{F_i(c + \tx_i\bbetad)-F_i(\alpha_{k-1})}{F_i(\alpha_k)-F_i(\alpha_{k-1})}(G_i(k)-G_i(k-1)).
\end{split}
\end{equation}
Let $\hat{R}_i$ be the residual under the estimated model as a random variable, $R_i^*$ be the bootstrapped residual under the estimated model as a random variable. 
Here, since we treat them as random variables, we use uppercase $R$ instead of lowercase $r$.
Note that the distribution functions of $R_i$ and $R_i^*$ satisfy
\begin{equation}
\| P(R_i^* \leq c ) - P(R_i \leq c) \| 
\leq
\| P(\hat{R}_i \leq c) - P(R_i \leq c) \| + \| P(R_i^* \leq c) - P(\hat{R}_i \leq c) \|.
\end{equation}

As for $P(\hat{R}_i \leq c)-P(R_i \leq c)$, the distribution function of $\hat{R}_i$ can be obtained by plugging in the parameter estimation $\hat{\alpha}$ and $\hat{\bbeta}_n$. 
Since given $\tx_i$, $F_i(\cdot)$ is a continuous function, and $\hat{\bbeta}_n \rightarrow \bbetad$ \citep{Lv2014Model}, thus $P(\hat{R}_i \leq c) \rightarrow P(R_i \leq c)$ and $\| P(\hat{R}_i \leq c)-P(R_i \leq c)\| \rightarrow 0$ in probability.

As for $P(\hat{R}_i \leq c)-P(R_i^* \leq c)$, $R_i^*$ is resampled uniformly from $\left\{ \hat{R}_k: k \in N_i\right\}$, and the distribution of $R_i^*$ is the mixture of the distribution of $\hat{R}_k$, thus, $P(R^*_i \leq c) $ can be seen as the average of the $ P(\hat{R}_k \leq c)$ for $k \in N_i$. Therefore, it can be rewritten as
\[
\begin{split}
\| P(\hat{R}_i \leq c)-P(R_i^* \leq c) \| 
& = \| \frac{1}{l_n}\sum_{k \in N_i} P(\hat{R}_k \leq c) - P(\hat{R}_i \leq c) \|   \\
& \leq \frac{1}{l_n} \sum_{k \in N_i} \|P(\hat{R}_k \leq c) - P(\hat{R}_i \leq c) \|.
\end{split}
\]
Fixed $\hat{\bbeta}_n$, $P(R_i \leq c)$ can be viewed as a continuous function of $\tx_i$. 
Since $\Vert \tx_k -\tx_i\Vert_2 \leq \delta_n$, where $\delta_n \rightarrow 0$ when $l_n/n \rightarrow 0$ and $n\rightarrow 0$. Thus, for some positive constant $c_1$,
\[
\begin{split}
 \| P(\hat{R}_i \leq c)-P(R_i^* \leq c) \| 
&  \leq \frac{1}{l_n} \sum_{k \in N_i} \|P(\hat{R}_k \leq c) - P(\hat{R}_i \leq c) \| \\
& \leq \frac{c_1}{l_n} \sum_{k \in N_i} \Vert\tx_k -\tx_i \Vert_2 
\leq \frac{c_1}{l_n} \sum_{k \in N_i} \delta_n \rightarrow 0.
 \end{split}
\]
This completes the proof. 

\end{proof}

\n
\textbf{ Proof of Theorem 2}

\begin{proof}
The proof process consists three steps. 
Step 1: prove that the bootstrapped $Y^*_i$ has the same distribution as $Y_i$ asymptotically. 
Step 2: derive the asymptotic distribution of QMLE $\hat{\boldsymbol{\beta}}_n$ under the assumed model, following \citet{Fahrmexr1990Maximum} and \citet{Lv2014Model}. Step 3: derive the asymptotic distribution of $\hat{\boldsymbol{\beta}}^*_n$ and show the resampling variance of $\sqrt{n}
(\hat{\boldsymbol{\beta}}^*_n - \hat{\boldsymbol{\beta}}_n)$ is consistent for the asymptotic variance of $\sqrt{n}(\hat{\boldsymbol{\beta}}_n - \boldsymbol{\beta}_{n,0}^\dagger)$.

\noindent
{\bf Step 1:}
In general case, we first consider $R_i$ as surrogate residual. The bootstrapped response $Y_i^*=k$ if $\hat{\alpha}_{k-1} < S_i^* \leq \hat{\alpha}_k$, and $S_i^*=R_i^* + \tx_i\hat{\bbeta}_n$.
\[
\begin{split}
& \|P(Y_i^*=k) - P(Y_i=k) \|
= \| P(\hat{\alpha}_{k-1} < S_i^* \leq \hat{\alpha}_k) - P(Y_i=k) \| \\
& = \| P(\hat{\alpha}_{k-1}-\tx_i\hat{\bbeta}_n < R_i^* \leq \hat{\alpha}_k-\tx_i\hat{\bbeta}_n) -P(Y_i=k) \| \\
& \leq \| P(\hat{\alpha}_{k-1}-\tx_i\hat{\bbeta}_n < R_i^* \leq \hat{\alpha}_k-\tx_i\hat{\bbeta}_n) - P(\alpha_{k-1}-\tx_i\bbeta_{n,0}^\dagger < R_i \leq \alpha_k-\tx_i\bbeta_{n,0}^\dagger) \|  \\
& \quad + \| P(\alpha_{k-1}-\tx_i\bbeta_{n,0}^\dagger < R_i \leq \alpha_k-\tx_i\bbeta_{n,0}^\dagger)- P(Y_i=k)\| \\
\end{split}
\]
The first term is $o(1)$ according to the consistency of $\hat{\alpha}$ and $\hat{\bbeta}_n$ as well as Theorem 1. In the second term, we have $P(\alpha_{k-1}-\tx_i\bbeta_{n,0}^\dagger < R_i \leq \alpha_k-\tx_i\bbeta_{n,0}^\dagger)=P(R_i \leq \alpha_k-\tx_i\bbeta_{n,0}^\dagger)-P(R_i \leq \alpha_{k-1}-\tx_i\bbeta_{n,0}^\dagger)$, substitute it into equation \eqref{eq:part1disr}, we can obtain $P(\alpha_{k-1}-\tx_i\bbeta_{n,0}^\dagger < R_i \leq \alpha_k-\tx_i\bbeta_{n,0}^\dagger)=G_i(k)-G_i(k-1)=P(Y_i=k)$. Therefore, we have $\|P(Y_i^*=k) - P(Y_i=k) \|=o(1)$.
Under the bootstrap version, define the true expectation $\mathbb{E}(Y_i^*)=m_i^*$ and $\var(Y_i^*)=\sigma_i^*$.
Since $Y_i^* \overset{d}{\rightarrow} Y_i$ and $Y_i^*$ has bounded second moment and bounded third moment,
we have $m_i^* \rightarrow m_i$ and $\sigma_i^* \rightarrow \sigma_i$.

\noindent 
{\bf Step 2:}
The score function is
\bg\label{solu_betahat}
\Psi_n(\boldsymbol{\beta}) = \partial \mathcal{L}_n/\partial \boldsymbol{\beta} 
=\sum_{i=1}^n \textbf{x}_i D_i(\boldsymbol{\beta})V_i^{-1}(\boldsymbol{\beta})\( y_i-h_i(\boldsymbol{\beta}) \).
\ed
Noted that $\mu_i=h(\textbf{x}_i^T \boldsymbol{\beta}) \overset{\wedge}{=} h_i(\boldsymbol{\beta})$, where $h^{-1}(\cdot)$ is the link function in generalized linear models.
Variance function $V(\mu_i) \overset{\wedge}{=} V_i(\boldsymbol{\beta})$ and $D_i(\boldsymbol{\beta}) = \partial h_i / \partial \eta_i$, where $\eta_i=h^{-1}(\mu_i)$. The pseudo-true value $\boldsymbol{\beta}_{n,0}^\dagger$ maximizes the expected log-likelihood, that is, is the root of the expected score function
\bg\label{expectscore}
\mathbb{E}\( \Psi_n(\boldsymbol{\beta}_{n,0}^\dagger) \) = \sum_{i=1}^n \textbf{x}_i D_i(\boldsymbol{\beta}_{n,0}^\dagger) V_i^{-1}(\boldsymbol{\beta}_{n,0}^\dagger)\( m_i-h_i(\boldsymbol{\beta}_{n,0}^\dagger) \)=0.
\ed
Denote $D_i=D_i(\boldsymbol{\beta}_{n,0}^\dagger)$, $V_i=V_i^{-1}(\boldsymbol{\beta}_{n,0}^\dagger)$, and $h_i=h_i(\boldsymbol{\beta}_{n,0}^\dagger)$. From the above equation, we have $\sum_{i=1}^n \textbf{x}_i D_i V_i^{-1} m_i = 
\sum_{i=1}^n \textbf{x}_i D_i V_i^{-1}h_i$, and further, 
$\Psi_n(\boldsymbol{\beta}_{n,0}^\dagger) 
= \sum_{i=1}^n \textbf{x}_i D_i V_i^{-1} \( y_i-h_i \) 
= \sum_{i=1}^n \textbf{x}_i D_i V_i^{-1} (y_i-m_i)$. Therefore, we have
\begin{equation}\label{Vtildebeta}
\tilde{\textbf{B}}_n
=\mathbb{E}\( \Psi_n(\boldsymbol{\beta}_{n,0}^\dagger) \Psi_n(\boldsymbol{\beta}_{n,0}^\dagger) ^T \) 
=\sum_{i=1}^n \textbf{x}_i D_i V_i^{-1} \sigma_i V_i^{-1} D_i \textbf{x}_i^T.
\end{equation}
The second derivatives of log-likelihood 
\[
\partial^2 \mathcal{L}_n/\partial\boldsymbol{\beta} \partial \boldsymbol{\beta}^T
= \sum_{i=1}^n \textbf{x}_i W_i(\boldsymbol{\beta}) \textbf{x}_i^T \(y_i -h_i(\boldsymbol{\beta}) \) 
- \sum_{i=1}^n \textbf{x}_i D_i(\boldsymbol{\beta}) V_i^{-1}(\boldsymbol{\beta})D_i(\boldsymbol{\beta}) \textbf{x}_i^T,
\]
where $W_i = \partial^2 \theta_i/ \partial \eta_i^2$ and define $W_i=W_i(\boldsymbol{\beta}_{n,0}^\dagger)$. 
Thus, 
\bg\label{Htildebeta}
\tilde{\textbf{A}}_n
= - \mathbb{E}\left( \partial^2 \mathcal{L}_n/\partial\boldsymbol{\beta} \partial \boldsymbol{\beta}^T \Big{|}_{\boldsymbol{\beta}_{n,0}^\dagger} \right)  
=  \sum_{i=1}^n \textbf{x}_i D_i V_i^{-1}D_i \textbf{x}_i^T
- \sum_{i=1}^n \textbf{x}_i W_i \textbf{x}_i^T ( m_i -h_i ).
\ed
According to \citet{Lv2014Model}, $\hat{\boldsymbol{\beta}}_n$ is weakly consistent for $\boldsymbol{\beta}_{n, 0}^\dagger$ under condition C1 and C2, 
and $\textbf{C}_n(\hat{\boldsymbol{\beta}}_n - \boldsymbol{\beta}_{n, 0}^\dagger)$ is asymptotically normal $N(0, \textbf{I}_p)$ under condition C1-C4, where $\textbf{C}_n = \tilde{\textbf{B}}_n^{-1/2} \tilde{\textbf{A}}_n$, and $\tilde{\textbf{B}}_n$ and $\tilde{\textbf{A}}_n$ is the form of (\ref{Vtildebeta}) and (\ref{Htildebeta}), respectively. 

\noindent 
Step 3:

\noindent
When we obtain the bootstrapped data $\left\{y_i^*, \textbf{x}_i \right\}_{i=1}^n$, $\hat{\boldsymbol{\beta}}_n^*$ maximizes the log-likelihood $\mathcal{L}_n^*(\boldsymbol{\beta})$ with the same assumed parametric fit, that is, $\hat{\boldsymbol{\beta}}_n^*$ is the root of the score function
\bg\label{solu_betahatstar}
\Psi_n^*(\boldsymbol{\beta}) = \partial \mathcal{L}_n^* / \partial \boldsymbol{\beta} 
= \sum_{i=1}^n \textbf{x}_i D_i(\boldsymbol{\beta})V_i^{-1}(\boldsymbol{\beta}) \( y_i^*-h_i(\boldsymbol{\beta}) \). 
\ed
Similar as step 2, we have 
\begin{equation}\label{Vtildebetastar}
\textbf{B}_n^*(\boldsymbol{\beta})
=  \sum_{i=1}^n \textbf{x}_i D_i(\boldsymbol{\beta})V_i^{-1}(\boldsymbol{\beta}) \sigma_i^* V_i^{-1}(\boldsymbol{\beta}) D_i(\boldsymbol{\beta}) \textbf{x}_i^T
\end{equation}
\begin{equation}\label{Htildebetastar}
\textbf{A}_n^*(\boldsymbol{\beta})
= \sum_{i=1}^n \textbf{x}_i D_i(\boldsymbol{\beta})V_i^{-1}(\boldsymbol{\beta})D_i(\boldsymbol{\beta}) \textbf{x}_i^T
- \sum_{i=1}^n \textbf{x}_i W_i(\boldsymbol{\beta}) \textbf{x}_i^T \( m_i^* -h_i(\boldsymbol{\beta}) \).
\end{equation}
We first prove the consistency of $\hat{\boldsymbol{\beta}}_n^* \rightarrow \hat{\boldsymbol{\beta}}_n$, conditional on $\hat{\boldsymbol{\beta}}_n$. Define $\hat{\textbf{B}}_n=\textbf{B}_n^*(\hat{\boldsymbol{\beta}}_n)$, $\hat{\textbf{A}}_n=\textbf{A}_n^*(\hat{\boldsymbol{\beta}}_n)$.

The quasi-likelihood $\mathcal{L}_n^*(\textbf{y}^*, \boldsymbol{\beta})$ can be expanded about $\hat{\boldsymbol{\beta}}_n$ by Taylor's theorem, we have, for any $\boldsymbol{\beta}$,
\[
\mathcal{L}_n^*(\textbf{y}^*, \boldsymbol{\beta}) = \mathcal{L}_n^*(\textbf{y}^*, \hat{\boldsymbol{\beta}}_n) 
+ (\boldsymbol{\beta} - \hat{\boldsymbol{\beta}}_n)^T \Psi^*_n(\hat{\boldsymbol{\beta}}_n) 
- \frac{1}{2} (\boldsymbol{\beta} - \hat{\boldsymbol{\beta}}_n)^T \textbf{A}_n^*(\boldsymbol{\beta}^\prime) (\boldsymbol{\beta} - \hat{\boldsymbol{\beta}}_n)^T ,
\]
where $\boldsymbol{\beta}^\prime$ is on the lime segment between $\boldsymbol{\beta}$ and $\hat{\boldsymbol{\beta}}_n$. Setting $\textbf{u}=\delta^{-1} \hat{\textbf{B}}_n^{1/2} (\boldsymbol{\beta}- \hat{\boldsymbol{\beta}}_n)$, it follows:
\[
\mathcal{L}_n^*(\textbf{y}^*, \boldsymbol{\beta}) - \mathcal{L}_n^*(\textbf{y}^*, \hat{\boldsymbol{\beta}}_n) 
= \delta \textbf{u}^T \hat{\textbf{B}}_n^{-1/2} \Psi^*_n(\hat{\boldsymbol{\beta}}_n)
- \frac{1}{2} \delta^2 \textbf{u}^T \hat{\textbf{T}}_n(\boldsymbol{\beta}^\prime) \textbf{u}. 
\]
where $\hat{\textbf{T}}_n(\boldsymbol{\beta}^\prime) = \hat{\textbf{B}}_n^{-1/2} \textbf{A}_n^*(\boldsymbol{\beta}^\prime) \hat{B}_n^{-1/2}$. 
Denote $\partial \hat{H}_n(\delta)=\{ \boldsymbol{\beta} \in \mathbb{R}^p: \Vert \hat{\textbf{B}}_n^{1/2} (\boldsymbol{\beta} - \hat{\boldsymbol{\beta}}_n) \Vert = \delta \}$ as the boundary of the closed set $\hat{H}_n(\delta)=\{ \boldsymbol{\beta} \in \mathbb{R}^p: \Vert \hat{\textbf{B}}_n^{1/2} (\boldsymbol{\beta} - \hat{\boldsymbol{\beta}}_n) \Vert \leq \delta \}$.  Note that $\boldsymbol{\beta} \in \partial \hat{H}_n(\delta)$ if and only if $\Vert \textbf{u} \Vert = 1$, and that $\boldsymbol{\beta} \in \partial \hat{H}_n(\delta)$ indicates that $\boldsymbol{\beta}^\prime \in \hat{H}_n(\delta)$ by the convexity of $\hat{H}_n(\delta)$. We have
$\max_{\Vert \textbf{u} \Vert = 1} \textbf{u}^T \hat{\textbf{B}}_n^{-1/2} \Psi^*_n(\hat{\boldsymbol{\beta}}_n) 
= \Vert \hat{\textbf{B}}_n^{-1/2} \Psi^*_n(\hat{\boldsymbol{\beta}}_n) \Vert$, 
and according to the second part of condition C2, we have
$\min_{\Vert \textbf{u} \Vert = 1} \textbf{u}^T \hat{\textbf{T}}_n(\boldsymbol{\beta}^\prime) \textbf{u}
\ge \min_{\boldsymbol{\beta} \in  \hat{H}_n(\delta)}\lambda_{\min} \{ \hat{\textbf{T}}_n(\boldsymbol{\beta}) \} \ge c$.
Thus we have 
\[
\max_{\boldsymbol{\beta} \in \partial \hat{H}_n(\delta)} \mathcal{L}_n^*(\textbf{y}^*, \boldsymbol{\beta}) - \mathcal{L}_n^*(\textbf{y}^*, \hat{\boldsymbol{\beta}}_n) 
\leq  \delta\( \Vert \hat{\textbf{B}}_n^{-1/2} \Psi^*_n(\hat{\boldsymbol{\beta}}_n) \Vert - \frac{1}{2} c \delta \),
 \]
which entails that according to Markov's inequality 
\[
\begin{split}
& P\( \max_{\boldsymbol{\beta} \in \partial \hat{H}_n(\delta)} \mathcal{L}_n^*(\textbf{y}^*, \boldsymbol{\beta}) < \mathcal{L}_n^*(\textbf{y}^*, \hat{\boldsymbol{\beta}}_n) \) 
\ge P\( \Vert \hat{\textbf{B}}_n^{-1/2} \Psi^*_n(\hat{\boldsymbol{\beta}}_n) \Vert^2 \leq \frac{1}{4} c^2 \delta^2 \) \\
& \ge 1 - \frac{\mathbb{E}\(  \Vert \hat{\textbf{B}}_n^{-1/2} \Psi^*_n(\hat{\boldsymbol{\beta}}_n) \Vert^2 \)}{c^2 \delta^2/4} 
 \ge 1- \frac{4p}{c^2 \delta^2},
\end{split}
\]
along with $\mathbb{E}\(  \Vert \hat{\textbf{B}}_n^{-1/2} \Psi^*_n(\hat{\boldsymbol{\beta}}_n) \Vert^2 \) = \mbox{tr}\{ \mathbb{E}(\Psi^*_n(\hat{\boldsymbol{\beta}}_n) \Psi^*_n(\hat{\boldsymbol{\beta}}_n)^T) \hat{\textbf{B}}_n^{-1}\} = \mbox{tr}(\textbf{I}_p)=p$.
Thus, for any $0<\eta<1$, there exist $\delta = 2p^{1/2}/(c \eta^{1/2}) >0$, such that 
\[
P\( \max_{\boldsymbol{\beta} \in \partial \hat{H}_n(\delta)} \mathcal{L}_n^*(\textbf{y}^*, \boldsymbol{\beta}) < \mathcal{L}_n^*(\textbf{y}^*, \hat{\boldsymbol{\beta}}_n) \) 
\ge 1-\eta.
\]
In fact, if $\{ \max_{\boldsymbol{\beta} \in \partial \hat{H}_n(\delta)} \mathcal{L}_n^*(\textbf{y}^*, \boldsymbol{\beta}) < \mathcal{L}_n^*(\textbf{y}^*, \hat{\boldsymbol{\beta}}_n) \}$, the continuous function $\mathcal{L}_n^*(\textbf{y}^*, \cdot)$ has a local maximum in the interior of $\hat{H}_n(\delta)$, and since the condition C1 which implies that $\mathcal{L}_n^*(\textbf{y}^*, \cdot)$ is strictly concave, this maximum must be located at $\hat{\boldsymbol{\beta}}_n^*$. Thus,
\[
P(\hat{\boldsymbol{\beta}}_n^* \in \hat{H}_n(\delta) ) = P(\Vert \hat{\textbf{B}}_n^{1/2} (\hat{\boldsymbol{\beta}}_n^* - \hat{\boldsymbol{\beta}}_n) \Vert \leq \delta) 
\ge 1-\eta,
\]
for any $\eta \in (0,1)$, together with the first part of condition C2, proves that $\hat{\boldsymbol{\beta}}_n^* - \hat{\boldsymbol{\beta}}_n = o_p(1)$.



The target ``true" parameter of bootstrap model $\boldsymbol{\beta}_{n,0}^*$ is the root of the expected score function
\bg\label{boot_expectscore}
\mathbb{E}\( \Psi_n^*(\boldsymbol{\beta}_{n,0}^*) \) = \sum_{i=1}^n \textbf{x}_i D_i(\boldsymbol{\beta}_{n,0}^*)V_i^{-1}(\boldsymbol{\beta}_{n,0}^*) \( m_i^*-h_i(\boldsymbol{\beta}_{n,0}^*) \)=0.
\ed
Define $\textbf{C}_n^*=\textbf{B}_n^{*-1/2} \textbf{A}_n^*$, where $\textbf{A}_n^*=\textbf{A}^*_n(\boldsymbol{\beta}_{n,0}^*)$ and $ \textbf{B}_n^*=\textbf{B}^*_n(\boldsymbol{\beta}_{n,0}^*)$, thus, $\textbf{C}_n^*(\hat{\boldsymbol{\beta}}_n^* - \boldsymbol{\beta}_{n,0}^*) \stackrel{d}{\rightarrow} N(0, \textbf{I}_p)$ according to asymptotic normality for QMLE. Observe that
\[
\textbf{C}_n^*(\hat{\boldsymbol{\beta}}_n^* - \hat{\boldsymbol{\beta}}_n)
= \textbf{C}_n^*(\hat{\boldsymbol{\beta}}_n^* - \boldsymbol{\beta}_{n,0}^*)
+ \textbf{C}_n^*(\boldsymbol{\beta}_{n,0}^* - \hat{\boldsymbol{\beta}}_n).
\]
We have $\hat{\boldsymbol{\beta}}_n^* \rightarrow \hat{\boldsymbol{\beta}}_n$, and $\hat{\boldsymbol{\beta}}_n^* \rightarrow \boldsymbol{\beta}_{n,0}^*$ according to the consistency of QMLE, thus, we can obtain that $|\boldsymbol{\beta}_{n,0}^* - \hat{\boldsymbol{\beta}}_n| \rightarrow 0$ conditional on $\hat{\boldsymbol{\beta}}_n$.
According to Slutsky's lemma, $\textbf{C}_n^*(\hat{\boldsymbol{\beta}}_n^* - \hat{\boldsymbol{\beta}}_n) \stackrel{d}{\rightarrow} N(0, \textbf{I}_p)$. 
Next we illustrate the consistency of $\textbf{C}^*_n \rightarrow \textbf{C}_n$. 

Note that the objective function $\mathcal{L}_n(\boldsymbol{\beta})= \sum_{i=1}^n \( y_i \theta_i - b(\theta_i) \)$. Specifically, in probit binary model, both of $\theta_i=\log(\frac{\mu_i}{1-\mu_i})$ and $b(\theta_i)=\log(1+e^{\theta_i})$ are the functions of only $\boldsymbol{\beta}$ give $\textbf{x}_i$. The pseudo-true parameter $\boldsymbol{\beta}_{n,0}^\dagger$ is to maximum the $\mathbb{E}(\mathcal{L}_n(\boldsymbol{\beta}))$, that is, 
\[
\boldsymbol{\beta}_{n,0}^\dagger = \arg\max_{\boldsymbol{\beta}} \mathbb{E}( \mathcal{L}_n(\boldsymbol{\beta})) =  \arg\max_{\boldsymbol{\beta}} \left\{\sum_{i=1}^n \( m_i \theta_i - b(\theta_i) \) \right\}.
\]
While $\boldsymbol{\beta}_{n,0}^*$ is to maximum the $\mathbb{E}(\mathcal{L}_n^*(\boldsymbol{\beta}))$, and 
\[
\boldsymbol{\beta}_{n,0}^* =  \arg\max_{\boldsymbol{\beta}} \mathbb{E}(\mathcal{L}_n^*(\boldsymbol{\beta})) =  \arg\max_{\boldsymbol{\beta}} \left\{ \sum_{i=1}^n \( m_i^* \theta_i - b(\theta_i) \) \right\}.
\]
Because  $m_i^* \rightarrow m_i$, for any $\boldsymbol{\beta}$ in parameter space, we have $\mathbb{E}(\mathcal{L}_n^*(\boldsymbol{\beta})) \rightarrow \mathbb{E}( \mathcal{L}_n(\boldsymbol{\beta}))$, according to Lemma A in \citet{Newey1987Asymmetric}, $\boldsymbol{\beta}_{n,0}^* \rightarrow \boldsymbol{\beta}_{n,0}^\dagger$.
We also have $m_i^* \rightarrow m_i$, and $\sigma_i^* \rightarrow \sigma_i$, the results follows immediately,
$\textbf{A}_n^* \rightarrow \tilde{\textbf{A}}_n$, $ \textbf{B}_n^* \rightarrow \tilde{\textbf{B}}_n$, 
thus, we have $\textbf{C}_n^* \rightarrow \textbf{C}_n$.

\end{proof}

\n
\textbf{ Proof of Theorem 3}

\begin{proof}

Using Taylor expansion, we have $
\textbf{0}=\Psi_n^*(\hat{\boldsymbol{\beta}}_n^*)
=\Psi_n^*(\boldsymbol{\beta}_{n,0}^*) + \textbf{A}_n^*(\boldsymbol{\beta}_{n,0}^*)(\hat{\boldsymbol{\beta}}_n^*-\boldsymbol{\beta}_{n,0}^*)
+ R_n(\hat{\boldsymbol{\beta}}_n^*)$.
We can show that $\mathbb{E}\left( \textbf{A}_n^*(\boldsymbol{\beta}_{n,0}^*)(\boldsymbol{\beta}_{n,0}^*-\hat{\boldsymbol{\beta}}_n^*) \right)=\mathbb{E}\left( \Psi_n^*(\boldsymbol{\beta}_{n,0}^*) \right) + \mathbb{E}\left( R_n(\hat{\boldsymbol{\beta}}_n^*) \right)$, and
\[
\mathbb{E}\left( \textbf{A}_n^*(\boldsymbol{\beta}_{n,0}^*)(\hat{\boldsymbol{\beta}}_n^*-\boldsymbol{\beta}_{n,0}^*) 
(\hat{\boldsymbol{\beta}}_n^*-\boldsymbol{\beta}_{n,0}^*)^T \textbf{A}_n^*(\boldsymbol{\beta}_{n,0}^*)^T \right)
=\mathbb{E} \left( \Psi_n^*(\boldsymbol{\beta}_{n,0}^*)^T \Psi_n^*(\boldsymbol{\beta}_{n,0}^*) \right) + \mathbb{E} \left( R_n^2(\hat{\boldsymbol{\beta}}_n^*) \right),
\]
where $R_n(\hat{\boldsymbol{\beta}}_n^*)=o(\hat{\boldsymbol{\beta}}_n^*-\boldsymbol{\beta}_{n,0}^*) = o(1)$ and $\mathbb{E}\left( \Psi_n^*(\boldsymbol{\beta}_{n,0}^*) \right)=\textbf{0}$.
\[
\begin{split}
\mbox{Var}\left( \hat{\boldsymbol{\beta}}_n^* \right)
& = \mathbb{E} \left( (\hat{\boldsymbol{\beta}}_n^* - \mathbb{E}(\hat{\boldsymbol{\beta}}_n^*))(\hat{\boldsymbol{\beta}}_n^* - \mathbb{E}(\hat{\boldsymbol{\beta}}_n^*))^T \right) \\
& \leq \mathbb{E} \left( (\hat{\boldsymbol{\beta}}_n^* - \boldsymbol{\beta}_{n,0}^*)(\hat{\boldsymbol{\beta}}_n^* - \boldsymbol{\beta}_{n,0}^*)^T \right) + (\mathbb{E}(\hat{\boldsymbol{\beta}}_n^*) - \boldsymbol{\beta}_{n,0}^*)(\mathbb{E}(\hat{\boldsymbol{\beta}}_n^*) - \boldsymbol{\beta}_{n,0}^*)^T \\
& \leq \textbf{A}_n^{*-1}\textbf{B}_n^* \textbf{A}_n^{*-1} + \textbf{A}_n^{*-1}\mathbb{E}( R_n^2(\hat{\boldsymbol{\beta}}_n^*))\textbf{A}_n^{*-1} + \textbf{A}_n^{*-1}(\mathbb{E}(R_n(\hat{\boldsymbol{\beta}}_n^*)))^2\textbf{A}_n^{*-1}.
\end{split}
\]
Thus we have $\mbox{Var}( \hat{\boldsymbol{\beta}}_n^*) \rightarrow \textbf{A}_n^{*-1}\textbf{B}_n^* \textbf{A}_n^{*-1}$, according to the step 3 in the proof of Theorem 2, $ \textbf{A}_n^{*-1}\textbf{B}_n^* \textbf{A}_n^{*-1} \rightarrow \tilde{\textbf{A}}_n^{-1} \tilde{\textbf{B}}_n \tilde{\textbf{A}}_n^{-1}$. This completes the proof.

\end{proof}

\end{document}